\documentclass[twocolumn,
superscriptaddress,
showpacs,
altaffilletter,
nofootinbib,
amsmath,
 amssymb,
aps,
prd,
]{revtex4-2}

\usepackage{graphicx}\usepackage{dcolumn}\usepackage{bm}\usepackage{hyperref}\hypersetup{colorlinks=true, citecolor=blue, urlcolor=blue, linkcolor=blue}

\usepackage[alsoload=hep]{siunitx} 
\usepackage{xcolor} \usepackage{booktabs}

\usepackage[ampersand]{easylist} 

\usepackage{amsmath}\usepackage{xspace}

\usepackage{scalefnt}

\begin{document}

\newcommand{\gm}{\ensuremath{(g-2)}\xspace}
\newcommand{\wa}{\ensuremath{\omega_{a}}\xspace}
\newcommand{\wam}{\ensuremath{\omega_{a}^m}\xspace}
\newcommand{\way}{\ensuremath{\omega_{a_y}}\xspace}
\newcommand{\waz}{\ensuremath{\omega_{a_z}}\xspace}
\newcommand{\vecwa}{\ensuremath{\vec \omega_{a}}\xspace}
\renewcommand{\wp}{\ensuremath{\omega_{p}}\xspace}
\newcommand{\opprime}{\ensuremath{\omega^{\prime}_p}\xspace}
\newcommand{\opprimetilde}{\ensuremath{\tilde{\omega}^{\prime}_p}\xspace}
\newcommand{\optilde}{\ensuremath{\tilde{\omega}^{}_p}\xspace}
\renewcommand{\amu}{\ensuremath{a_{\mu}}\xspace}
\newcommand{\gmtwo}{\ensuremath{g\!-\!2}\xspace}
\newcommand{\rmagic}{\ensuremath{R_{0}}\xspace}
\newcommand{\cmagic}{\ensuremath{c_{0}}\xspace}
\newcommand{\pmagic}{\ensuremath{p_{0}}\xspace}
\newcommand{\gammamagic}{\ensuremath{\gamma_{0}}\xspace}
\newcommand{\betamagic}{\ensuremath{\beta_{0}}\xspace}
\newcommand{\dipbfield}{\ensuremath{B_{0}}\xspace}

\newcommand{\ringsim}{{\scalefont{0.76}GM2RINGSIM}\xspace}
\newcommand{\COSY}{{\scalefont{0.76}COSY}\xspace}
\newcommand{\COSYINFINITY}{{\scalefont{0.76}COSY INFINITY}\xspace}
\newcommand{\BMAD}{{\scalefont{0.76}BMAD}\xspace}
\newcommand{\OPERA}{{\scalefont{0.76}OPERA}\xspace}
\newcommand{\OPERATHREED}{{\scalefont{0.76}OPERA-3D}\xspace}
\newcommand{\GEANT}{{\scalefont{0.76}GEANT4}\xspace}
\newcommand{\GBeamline}{{\scalefont{0.76}G4BEAMLINE}\xspace}
\newcommand{\COULOMB}{{\scalefont{0.76}COULOMB}\xspace}
\newcommand{\INTEGRATED}{{\scalefont{0.76}INTEGRATED}\xspace}
\newcommand{\CADmesh}{CAD{\scalefont{0.76}MESH}\xspace}

\def\bmbetahat{{\hat{\bm\beta}}}
\def\bfs{{\vec {S}}}
\def\bfhx{{\hat{\bf x}}}
\def\bfhy{{\hat{\bf y}}}
\def\bfhz{{\hat{\bf z}}}
\def\bfB{{\bf B}}
\def\bfs{{\vec {S}}}
\def\vev#1{{\langle{#1}\rangle}}
\def\onehalf{{\frac{1}{2}}}
\def\bfE{{\bf E}}
\def\mE{{\mathcal{E}}}

\newcommand{\runone}{Run-1\xspace}
\newcommand{\runtwo}{Run-2\xspace}
\newcommand{\runthree}{Run-3\xspace}
\newcommand{\runfour}{Run-4\xspace}
\newcommand\runonea{Run-1a\xspace}
\newcommand\runoneb{Run-1b\xspace}
\newcommand\runonec{Run-1c\xspace}
\newcommand\runoned{Run-1d\xspace}
\newcommand\precession{precession-run1}
\newcommand\field{field-run1}
\newcommand\BD{BD-run1}

\graphicspath{{./fig/}}

\title{Beam dynamics corrections to the \runone\ measurement of the muon anomalous magnetic moment at Fermilab}
\affiliation{Argonne National Laboratory, Lemont, Illinois, USA}
\affiliation{Boston University, Boston, Massachusetts, USA}
\affiliation{Brookhaven National Laboratory, Upton, New York, USA}
\affiliation{Budker Institute of Nuclear Physics, Novosibirsk, Russia}
\affiliation{Center for Axion and Precision Physics (CAPP) / Institute for Basic Science (IBS), Daejeon, Republic of Korea}
\affiliation{Cornell University, Ithaca, New York, USA}
\affiliation{Fermi National Accelerator Laboratory, Batavia, Illinois, USA}
\affiliation{INFN Gruppo Collegato di Udine, Sezione di Trieste, Udine, Italy}
\affiliation{INFN, Laboratori Nazionali di Frascati, Frascati, Italy}
\affiliation{INFN, Sezione di Napoli, Naples, Italy}
\affiliation{INFN, Sezione di Pisa, Pisa, Italy}
\affiliation{INFN, Sezione di Roma Tor Vergata, Rome, Italy}
\affiliation{INFN, Sezione di Trieste, Trieste, Italy}
\affiliation{Istituto Nazionale di Ottica - Consiglio Nazionale delle Ricerche, Pisa, Italy}
\affiliation{Department of Physics and Astronomy, James Madison University, Harrisonburg, Virginia, USA}
\affiliation{Institute of Physics and Cluster of Excellence PRISMA+, Johannes Gutenberg University Mainz, Mainz, Germany}
\affiliation{Joint Institute for Nuclear Research, Dubna, Russia}
\affiliation{Department of Physics, Korea Advanced Institute of Science and Technology (KAIST), Daejeon, Republic of Korea}
\affiliation{Lancaster University, Lancaster, United Kingdom}
\affiliation{Michigan State University, East Lansing, Michigan, USA}
\affiliation{North Central College, Naperville, Illinois, USA}
\affiliation{Northern Illinois University, DeKalb, Illinois, USA}
\affiliation{Regis University, Denver, Colorado, USA}
\affiliation{Scuola Normale Superiore, Pisa, Italy}
\affiliation{School of Physics and Astronomy, Shanghai Jiao Tong University, Shanghai, China}
\affiliation{Tsung-Dao Lee Institute, Shanghai Jiao Tong University, Shanghai, China}
\affiliation{Institut f\"ur Kern—und Teilchenphysik, Technische Universit\"at Dresden, Dresden, Germany}
\affiliation{Universit\`a del Molise, Campobasso, Italy}
\affiliation{Universit\`a di Cassino e del Lazio Meridionale, Cassino, Italy}
\affiliation{Universit\`a di Napoli, Naples, Italy}
\affiliation{Universit\`a di Pisa, Pisa, Italy}
\affiliation{Universit\`a di Roma Tor Vergata, Rome, Italy}
\affiliation{Universit\`a di Trieste, Trieste, Italy}
\affiliation{Universit\`a di Udine, Udine, Italy}
\affiliation{Department of Physics and Astronomy, University College London, London, United Kingdom}
\affiliation{University of Illinois at Urbana-Champaign, Urbana, Illinois, USA}
\affiliation{University of Kentucky, Lexington, Kentucky, USA}
\affiliation{University of Liverpool, Liverpool, United Kingdom}
\affiliation{Department of Physics and Astronomy, University of Manchester, Manchester, United Kingdom}
\affiliation{Department of Physics, University of Massachusetts, Amherst, Massachusetts, USA}
\affiliation{University of Michigan, Ann Arbor, Michigan, USA}
\affiliation{University of Mississippi, University, Mississippi, USA}
\affiliation{University of Rijeka, Rijeka, Croatia}
\affiliation{Department of Physics, University of Texas at Austin, Austin, Texas, USA}
\affiliation{University of Virginia, Charlottesville, Virginia, USA}
\affiliation{University of Washington, Seattle, Washington, USA}
\author{T.~Albahri}  \affiliation{University of Liverpool, Liverpool, United Kingdom}
\author{A.~Anastasi} \thanks{Deceased.} \affiliation{INFN, Sezione di Pisa, Pisa, Italy}
\author{K.~Badgley}  \affiliation{Fermi National Accelerator Laboratory, Batavia, Illinois, USA}
\author{S.~Bae{\ss}ler} \altaffiliation[Also at ]{Oak Ridge National Laboratory, Oak Ridge, Tennessee, USA.}  \affiliation{University of Virginia, Charlottesville, Virginia, USA}
\author{I.~Bailey} \altaffiliation[Also at ]{The Cockcroft Institute of Accelerator Science and Technology, Daresbury, United Kingdom.}  \affiliation{Lancaster University, Lancaster, United Kingdom}
\author{V.~A.~Baranov}  \affiliation{Joint Institute for Nuclear Research, Dubna, Russia}
\author{E.~Barlas-Yucel}  \affiliation{University of Illinois at Urbana-Champaign, Urbana, Illinois, USA}
\author{T.~Barrett}  \affiliation{Cornell University, Ithaca, New York, USA}
\author{F.~Bedeschi}  \affiliation{INFN, Sezione di Pisa, Pisa, Italy}
\author{M.~Berz}  \affiliation{Michigan State University, East Lansing, Michigan, USA}
% ...
\collaboration{Muon \gmtwo Collaboration} \noaffiliation
\vskip 0.25cm

\date{\today}

\begin{abstract}
\medskip

This paper presents the beam dynamics systematic corrections and their uncertainties for the \runone dataset of the Fermilab Muon \gmtwo Experiment. Two corrections to the measured muon precession frequency \wam are associated with well-known effects owing to the use of electrostatic quadrupole (ESQ) vertical focusing in the storage ring.  An average vertically oriented motional magnetic field is felt by relativistic muons passing transversely through the radial electric field components created by the ESQ system.   The correction depends on the stored momentum distribution and the tunes of the ring, which has relatively weak vertical focusing.   Vertical betatron motions imply that the muons do not orbit the ring in a plane exactly orthogonal to the vertical magnetic field direction.  A correction is necessary to account for an average pitch angle associated with their trajectories.  A third small correction is necessary, because muons that escape the ring during the storage time are slightly biased in initial spin phase compared to the parent distribution.    Finally, because  two high-voltage resistors in the ESQ network had longer than designed $RC$ time constants, the vertical and horizontal centroids and envelopes of the stored muon beam drifted slightly, but coherently, during each storage ring fill.  This led to the discovery of an important phase-acceptance relationship that requires a correction.  The sum of the corrections to \wam is $0.50 \pm 0.09$\,ppm; the uncertainty is small compared to the 0.43\,ppm statistical precision of \wam.

\end{abstract}
 
\maketitle 

\tableofcontents

\section{\label{sec:intro} Introduction}

The measurement of the muon magnetic anomaly\footnote{\amu is a scalar quantity and thus not technically the ``anomalous magnetic moment'' that is ubiquitous in the literature.} $ \amu \equiv (g_\mu - 2)/2$, where $g_\mu$ is the factor describing the relationship of the muon magnetic moment to its spin, has undergone significant development since the late 1960s when the idea of using a magnetic storage ring for the measurement was first introduced~\cite{Bailey:1972eu}.  Two storage ring experiments at CERN~\cite{Bailey:1972eu,Bailey:1978mn} and the Brookhaven National Laboratory (BNL) Experiment (E821)~\cite{Bennett:2006fi} have increasingly refined the technique, leading to a determination of \amu to a precision of 0.54\,ppm~\cite{Bennett:2006fi}.  These experiments determine \amu by measuring the muon spin precession frequency relative to the momentum vector while a muon beam is confined in a storage ring. In an ideal case with muons orbiting in a horizontal plane with a uniform vertical magnetic field, the anomalous precession frequency \wa is given by the difference between the spin ($s$) and  cyclotron ($c$) frequencies, $\wa = \omega_{s} - \omega_{c}$.  The observable \wa is proportional to \amu.

A highly precise determination of \amu is motivated by the fact that the standard model (SM) prediction~\cite{Aoyama:2020ynm} is known to 0.37\,ppm and thus a direct comparison with the experiment serves as an excellent and sensitive test of its completeness.
The E821 \amu measurement~\cite{Bennett:2006fi} is larger than the SM prediction; the level of significance is 3.7 standard deviations $(\sigma)$. This difference can be interpreted as a  hint of new physics or an indication of unaccounted-for systematic errors in the experiment or theory.
The Fermilab Muon \gmtwo Experiment (E989) is designed to test the validity of the BNL result and to go further by improving on the experimental precision.

The principle of an \amu measurement in a storage ring lies in the expression
\begin{equation}
\begin{aligned}
  \amu = \frac{\wa}{\opprimetilde(T_{r})} \frac{\mu^{\prime}_p(T_{r})}{\mu_e(H)} \frac{{\mu_e(H)}}{\mu_e} \frac{m_{\mu}}{m_e} \frac{g_e}{2},
  \label{eq:amueq}
\end{aligned}
\end{equation}
where the experiment measures the two quantities in the ratio $\wa / \opprimetilde(T_{r})$, and the
fundamental constants---the proton-to-electron magnetic moment ratio~\cite{Phillips:1977}, the QED factor $\mu^{}_e(H)/\mu_e$~\cite{CODATA:2018}, the muon-to-electron mass ratio~\cite{CODATA:2018,Liu:1999iz}, and the electron $g_e$ factor~\cite{Hanneke:2010au}---are obtained from external measurements and calculations.

Reference~\cite{E989PRL-1}---a companion to this paper---reports the first results from the analysis of the \runone data collected in 2018:
\begin{equation*}
  a_\mu({\rm FNAL}) = 116\,592\,040(54) \times  10^{-11} ~ (\text{0.46\,ppm}).
\end{equation*}
The statistical (0.43\,ppm), systematic (0.15\,ppm), and external fundamental constant (0.03\,ppm) uncertainties are combined in quadrature.
The result is in good agreement with the BNL result and, when the two measurements are combined and compared to the SM, the significance of the difference between experiment and theory rises to $4.2\,\sigma$.
In order to obtain this result, a number of beam dynamics effects had to be accounted for in the analysis.  This paper reports the methods used to quantify the impact of these effects and their uncertainties.  These effects were part of a larger analysis, which can be summarized briefly as follows.

In Eq.~\ref{eq:amueq}, term $\opprimetilde(T_{r})$ represents the magnetic field sampled by the muons and measured using pulsed proton NMR. The magnitude of the field is calibrated in terms of the equivalent precession frequency $\opprime(T_{r})$ of a proton shielded in a spherical sample of water at a reference temperature $T_{r}=34.7^{\circ}$C.
The detailed analysis of the magnetic field determination for this experiment is reported in a second companion paper~\cite{\field}.

The  anomalous  precession frequency \wa in the numerator of Eq.~\ref{eq:amueq} is extracted from a fit to the time and energy spectrum of positrons, which encodes a modulation of its intensity at the anomalous precession frequency.
Importantly, the frequency extracted by fitting the data is the {\em measured} quantity \wam, and not the desired quantity \wa needed to obtain \amu in Eq.~\ref{eq:amueq}.
Details of the measured precession frequency \wam analysis, including statistical and systematic uncertainties, are reported in the final companion paper to this report~\cite{\precession}.

The muons in the experiment do not orbit in a perfectly horizontal plane in a homogeneous vertical magnetic field.
An electrostatic quadrupole (ESQ) system is required to provide relatively weak vertical focusing.
The resulting presence of an electric field and vertical betatron oscillations, as well as muons lost during the measurement period, and a faulty ESQ component that led to a time-dependent optical lattice necessitate ``beam dynamics'' corrections that must be applied to \wam in order to obtain \wa, the quantity needed to determine the magnetic anomaly.

Two corrections are a result of a physical reduction in the spin precession frequency of the muons. The electric-field correction $C_e$ is a result of muons traveling orthogonally through the radial electric field from the ESQ system and experiencing a motional vertical magnetic field.
The pitch correction $C_p$ compensates for the vertical betatron motion that causes, on average, a slightly nonorthogonal relation between the muon momentum and the vertically aligned magnetic field.
While these were known from previous experiments, two additional corrections must be made to compensate for a dynamic change to the stored muon ensemble's average spin phase during each injection cycle of muons into the ring.  The $C_{ml}$ correction is associated with muons lost during storage that have a slightly different spin phase compared to those that remain stored.   Finally, because  two high-voltage resistors in the ESQ network had longer than designed $RC$ time constants, the vertical and horizontal centroids and envelopes of the stored muon beam drifted slightly, but coherently, during each storage ring fill.  This led to the discovery of an important phase-acceptance relationship that requires a correction $C_{pa}$.
The four corrections are applied in a linear combination\footnote{Cross terms between corrections are neglected here which is more than sufficient.} to \wam,
\begin{equation}
  \wa \approx \wam \left( 1 + C_e + C_p + C_{ml} + C_{pa} \right) \,.
  \label{eq:wamdef}
\end{equation}

The total shift to the measured \wam value is $(+0.50 \pm 0.09)$\,ppm.  Although these small corrections are all at the sub-ppm level, the net correction exceeds the statistical uncertainty on \wam, and thus the corrections must be scrutinized carefully and their uncertainties determined precisely.
A thorough discussion
is presented in this paper.

Section~\ref{sec:bd_overview} includes background information on the formal experimental principle, \runone dataset characteristics, injection and storage information, and descriptions of the experimental instrumentation, magnetic field, and the simulation tools.  The electric-field correction is evaluated using the stored muon momentum distribution that is obtained via a data-driven procedure described in Sec.~\ref{sec:fastrotation}.  The $C_e$ correction and its uncertainty can be found in Sec.~\ref{sec:efield}.  The pitch correction $C_p$ is discussed in Sec.~\ref{sec:pitch}.

The time-dependent nature of the detected oscillation phase which compels the final two corrections was exacerbated during \runone due to damage to 2 out of 32 high-voltage resistors in the ESQ charging system.  Section~\ref{sec:time_changing_fields} examines the consequences of this unexpected configuration for the time-dependent electric quadrupole field and the dynamics of the stored beam.  Section~\ref{sec:muon_loss_phase} discusses how this slow change of the steering field leads to larger than expected muon losses and evaluates the resulting correction $C_{ml}$ due to a correlation between muon momentum and initial average spin phase.  Section~\ref{sec:phase_acc} decribes the phase-acceptance correction $C_{pa}$ that is introduced because the average spatial distribution of the beam drifts during storage as a result of the same ESQ resistor issue.

Finally, we summarize and conclude in Sec.~\ref{sec:conc} and provide four appendices covering details of electric field and pitch corrections, the generation of phase, asymmetry, and acceptance maps needed for the $C_{pa}$ correction, and a typical  fitting function used to determine \wam.
 \section{\label{sec:bd_overview} Experimental Details}

\subsection{Experimental principle}
\label{sc:experiment}
Polarized positive muons are injected into a 14.2\,m-diameter storage ring having a highly uniform ${\sim}1.45$\,\SI{}{\tesla} vertical magnetic field; each individual injection sequence is termed a ``fill''.  For muons of momentum $p_\mu$ orbiting in a horizontal plane that is perpendicular to a perfectly uniform magnetic field ($\vec{B} \cdot \vec{p}_\mu = 0$), the magnitude of the cyclotron frequency is $\omega_c =|q| B/ m \gamma$, where $q$, $m$, and $\gamma$ are the muon charge, mass, and Lorentz factor respectively.
The torque on the magnetic moment, together with the Thomas precession, rotates the muon spin at the frequency $\omega_s = g_\mu |q|B / 2 m + (1-\gamma) |q| B / \gamma m$~\cite{Bargmann:1959gz}.
The anomalous precession frequency \wa\ can then be described as the difference
\begin{equation}
 \wa \equiv  \omega_s -  \omega_c = \left(\frac{g_\mu-2}{2} \right) \frac{|q| B}{m} = \amu \frac{|q|B}{m}.
\label{eq:simpleomega}
\end{equation}
Because of parity violation in the $\mu^+$ weak decay, positrons are emitted with an energy and an angular distribution that are each highly correlated to the muon spin direction in its rest frame. Ignoring effects from beam dynamics, the detectors that are distributed uniformly around the inside of the storage ring see a positron count rate versus time in fill $t$ and positron energy $E$ with the following functional form:
\begin{equation}
N \left(t,E\right)=N_0(E) e^{-t / \gamma\tau_\mu} \left\{ 1 + A(E)\cos\left[ \wa t+\varphi_0(E) \right] \right\}.
\label{eq:wiggle_func}
\end{equation}
The normalization, time-dilated muon lifetime,  asymmetry, anomalous precession frequency, and phase constant at the time of injection are represented in Eq.~\ref{eq:wiggle_func} by $N_0(E)$, $\gamma\tau_\mu$, $A(E)$, \wa and $\varphi_0(E)$, respectively.
The energy dependence of $N_0$ and $A$ has its origin in the Lorentz-boosted muon decay (Michel) spectrum;  the actual values depend on the detector acceptance.
The  asymmetry also depends on the muon ensemble average polarization magnitude ${\mathcal P}_\mu$, which is $\approx 95\%$ in this experiment.
In practice,  \wa\ is determined from a fit to a  positron time spectrum having a range of positron energies.
The detector design is optimized to accept higher-energy positrons for which the figure of merit $NA^2$ is maximized.

 We emphasize that,  throughout this paper, the variable $\varphi_0$ represents the phase constant in the time-dependent phase angle of the cosine in Eq.~\ref{eq:wiggle_func}.
Its value is determined by the fitting procedure and need not, and cannot, be determined with sufficient precision {\it{a priori}} to the fit of the spin precession data sample.
The phase constant $\varphi_0$ is dependent on the stored muon momentum $p_\mu$, the decay positron energy $E$, and the transverse decay coordinates\footnote{Our coordinate system is with respect to the center of the storage volume at radius $R_0$, with $x$ or $r$ radially outward, $y$ vertically up, and $\phi$ increasing clockwise when viewed from above.} $(x,y)$ inside the storage ring.
The $N_0$ and $A$ terms in Eq.~\ref{eq:wiggle_func} are also functions of the same quantities, but they couple much more weakly to \wam.
The physical interpretation of these dependencies of the phase constant and their implications to systematic uncertainties in the determination of \wa\ are most important in the determination of the $C_{ml}$ and $C_{pa}$ correction factors discussed in Secs.~\ref{sec:muon_loss_phase} and \ref{sec:phase_acc}, respectively.

The relevant observable for the anomalous precession frequency is the oscillation in the quantity $\hat{\beta}\cdot\vec{S}$, where $\vec{S}$ is the spin vector of the muon.
Following the Thomas-Bargmann–Michel–Telegdi (Thomas-BMT) equation~\cite{jackson-classical-1999} for spin and the Lorentz equation for momentum, one observes that, in the presence of an electric field $\vec{E}$,
\begin{equation}
{d(\hat{\beta}\cdot\vec{S}) \over dt} = -{q \over m }\vec{S}_{T}\cdot\left[a_\mu\hat{\beta}\times\vec{B} + \beta \left(a_\mu - \frac{1}{\gamma^{2}-1}\right){\vec{E} \over c} \right],
\label{eq:trueomega}
\end{equation}
where $\vec{S}_T$ is the component of $\vec{S}$ perpendicular to $\hat{\beta}$ and we have ignored a possible term owing to a nonzero muon electric dipole moment, which has been determined to be negligible~\cite{Bennett:2008dy}.
This expression corresponds exactly to the muon spin precession in the ring.
The electric field term in Eq.~\ref{eq:trueomega} vanishes for muons having the ``magic'' momentum   \pmagic\ = 3.094\,GeV/$c$ ($\gamma \sim 29.3$). The experiment is therefore designed around injection and storage of muons centered on \pmagic.

\subsection{The \runone dataset}
The muon delivery and experimental apparatus were commissioned from June
2017  to March 2018. The \runone\ data-taking period began on
March 26, 2018 and concluded on July 7, 2018. The datasets in this
analysis comprise 14.13\,M muon fills. Approximately 5000 muons are stored
in each fill, at an average of 11.4 fills per second.

Vertical focusing in the storage ring is achieved using a suite of ESQ plates that occupy 43\% of the ring circumference.
The field index $n$, responsible for the relatively weak focusing in the vertical direction, is defined by
\begin{equation}
n = {R_0 \over vB_0 } {\partial E_y \over \partial y},
\label{eq:field-index}
\end{equation}
where $R_0$ is the central orbit radius, $v$ is the muon velocity, $B_0$ is the magnetic field, and
the gradient in the effective vertical electric field
is determined from the plate voltages and geometry.
Four distinct datasets, referred to hereafter as 1a, 1b, 1c, and 1d, have been separately analyzed corresponding to four different combinations of
kicker magnet and ESQ voltages (see Table~\ref{tb:mastertable}). The separate analyses of these datasets yield consistent results for \amu.
The 18.3\,kV ($n$~=~0.108) and 20.4\,kV ($n$~=~0.120) values were chosen to avoid storage ring betatron resonance conditions that would lead to large muon losses.

\begin{table}[htbp]
\begin{ruledtabular}
\begin{tabular}{ccccc}
  Dataset & ~$\delta$\wam (stat)~  &~ESQ~& ~Effective Field~  & ~Kicker~   \\
              &   ppb            &  kV    &  Index             &   kV  \\ \hline
\runonea & 1206           & 18.3 & 0.108       & 130   \\
\runoneb & 1024          & 20.4 & 0.120       & 137 \\
\runonec & 825             & 20.4 & 0.120      & 130  \\
\runoned & 676$^a$  & 18.3 & 0.107       & 125 \\
\end{tabular}
\end{ruledtabular}
\caption{The  \runone\ dataset characteristics: The columns indicate the statistical uncertainty of \wam based on the asymmetry-weighted analysis method~\cite{\precession}; the demand ESQ voltage; the measured field index $n$ which is affected by the damaged ESQ resistors (Sec.~\ref{sec:injtostorage}); and the typical kicker strength in kilovolts used to deflect the incoming beam into the storage volume.}
\label{tb:mastertable}
$^a$ The precession fit start time for \runoned was delayed to \SI{50}{\micro\second}, in contrast to \SI{30}{\micro\second} for the other data groups.
\end{table}

The different field indices lead to differing beam frequencies, since the
horizontal and vertical betatron tunes for a uniform set of quadrupole fields that occupy
the full azimuth of the storage ring are given by
\begin{equation}
\nu_x =  \sqrt{1-n} ~~~~ \text{and}~~~~  \nu_y = \sqrt{n},
\end{equation}
with corresponding  betatron frequencies
\begin{equation}
\omega_x =  \omega_{c}\sqrt{1-n} ~~~~ \text{and}~~~~  \omega_y =  \omega_{c}\sqrt{n}.
\end{equation}
These expressions are sufficiently accurate for our purposes. Calculations of the frequencies for the different field indices
are shown in Table~\ref{tab:beamfreq}.
Coherent radial and vertical betatron oscillations of the centroid and width of the stored beam are driven by a combination of the mismatch between the beam line admittance and the storage ring acceptance, the intrinsic divergence of the incoming beam, and the strength of the storage ring kicker system.
The radial coherent betatron motion of the muon ensemble introduces an oscillatory time dependence to the $N_0$, $A$, and $\varphi_0$ terms in Eq.~\ref{eq:wiggle_func}.
Since $\omega_x > \omega_c / 2$, the observed frequency at each detector is the aliased frequency $\omega_\text{CBO} = \omega_c - \omega_x$.
This coherent betatron oscillation (CBO) occurs inside the storage volume but becomes imprinted on the positron count spectrum $N(t)$ because of the radial dependence of the detector acceptance.  The radial mean of the muon distribution is modulated at frequency $\omega_\text{CBO}$ and the radial width (rms) at $2\omega_\text{CBO}$ and also at  $\omega_\text{CBO}$ if the stored beam is not centered in the aperture.
A smaller but similar effect exists for the vertical oscillations, where $\omega_y$ is observed without aliasing, but the width oscillations at $2\omega_y$ are aliased to $\omega_{VW} = \omega_c - 2\omega_y$. Because of the symmetric nature of the vertical detector acceptance, the $\omega_{VW}$ effect is stronger than that from $\omega_y$.  The calculated values of $\omega_\text{CBO}$ and $\omega_{VW}$ are also presented in Table~\ref{tab:beamfreq}.

These effects impact $N_0$, $A$, and $\varphi_0$ and are accounted for by additional terms in the fitted positron decay time spectrum (see Appendix~\ref{ap:fitfunction} and Ref.~\cite{\precession}).  The field indices have been chosen such that the modulations do not couple strongly to \wam.   If the ESQ system does not maintain stable voltages within a fill, $\omega_\text{CBO}$ and $\omega_{VW}$ vary as a function of time in fill, which was the case in \runone but fixed thereafter.

\begin{table}[htbp]
\begin{ruledtabular}
\begin{tabular}{cccc}
Physical            & Calculated             & \multicolumn{2}{c}{Frequency (\SI{}{\radian / \micro\second})}\\
frequency           & expression             & $n = 0.108$  & $n = 0.120$ \\ \hline
$\omega_c$          & $v/R_0$                & 42.15        & 42.15       \\
$\omega_x$          & $\sqrt{1-n}\omega_c$   & 39.81        & 39.54       \\
$\omega_y$          & $\sqrt{n}\omega_c$     & 13.85        & 14.60       \\
$\omega_\text{CBO}$ & $\omega_c - \omega_x$  & 2.34         & 2.61        \\
$\omega_{VW}$       & $\omega_c - 2\omega_y$ & 14.45        & 12.95       \\
$\omega_{a}$        & $e a_\mu B / m $       & 1.44         & 1.44        \\
\end{tabular}
\end{ruledtabular}
\caption{Characteristic frequencies for the \runone field indices. The physical frequencies are described in the text. All values are calculated from the simple expressions shown and are similar to the measured values.}
\label{tab:beamfreq}
\end{table}

\subsection{From muon injection to muon storage}
\label{sec:injtostorage}
Muons from the Fermilab accelerator complex are created as follows.  Bunches of 8-GeV protons strike a target in the AP0 building of the Muon Campus~\cite{Stratakis:2019sun}. The average number of protons incident on the target per muon fill was $9.84 \times 10^{11}$. Positive 3.1\,GeV/$c$ particles are extracted and transported through the 279-m-long M2/M3 FODO\footnote{Alternating focusing and defocusing quadrupole magnets.} beam line.  Roughly 80\% of the pions decay to muons in this beam line section, which is optimized to produce a muon beam with an average longitudinal polarization of approximately 95\%. The particles then enter the 505-m-circumference Delivery Ring, where they are allowed to circulate for four full turns.  The accompanying 3.1\,GeV/$c$ protons from the target station lag behind the faster muons  and are swept out by an in-ring fast-kicker magnet.  The purified and polarized muon bunch is then directed along the M4 and M5 beam lines and into the storage ring. The net $2463$\,m path length from the target to the storage ring reduces the initial pion intensity by a factor $1.5\times10^{6}$, an important feature that eliminates the hadronically induced flash at injection, which was challenging in the E821 experiment~\cite{Bennett:2006fi}.

\begin{figure}[htbp]
	\centering
        \includegraphics[width=\columnwidth]{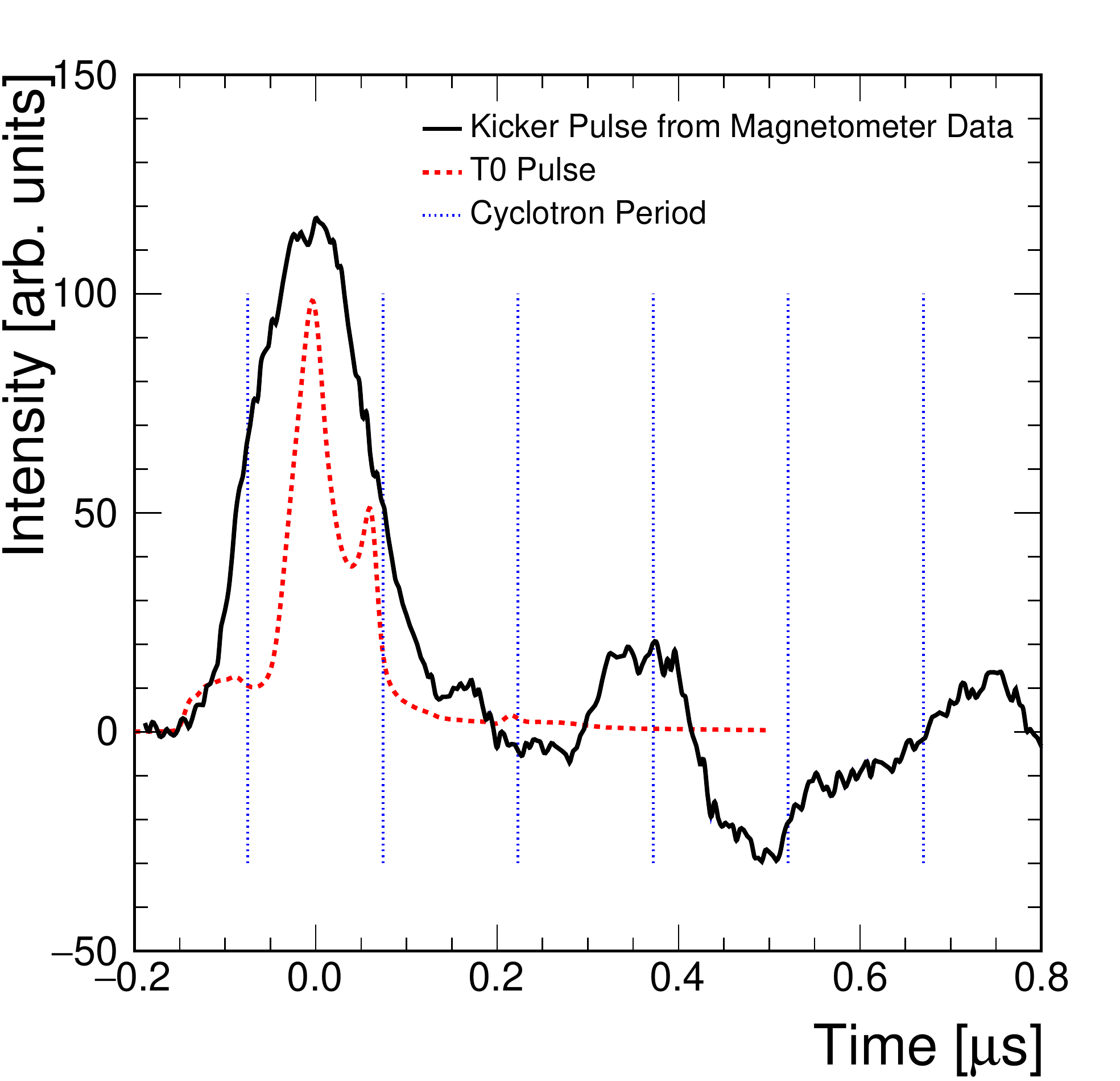}
\caption{The solid black trace represents the shape of the kicker magnetic field. The ringing is owing to an unavoidable impedance mismatch.  The dashed red trace represents the average injected muon bunch temporal intensity distribution as measured by the T0 detector. The vertical dotted lines are separated by the 149\,ns cyclotron period. The overlay between kicker and beam shapes shows that not all entering muons experience the same magnetic deflection.  It is clear from the ringing that muons experience multiple kicks, both inward and outward.}\label{fig:kicker-T0}
\end{figure}

Sixteen individual bunches of muons are injected every 1.4\,s cycle in two sequences of eight with 10\,ms separation and 267\,ms between the start of each sequence.  The average temporal intensity distribution of muons at the entrance to the storage ring is shown as a dashed red line in Fig.~\ref{fig:kicker-T0}.  The shape of this intensity-time distribution varies slightly for each of the eight fills in a group.

\begin{figure}[htbp]
\includegraphics[width=\columnwidth]{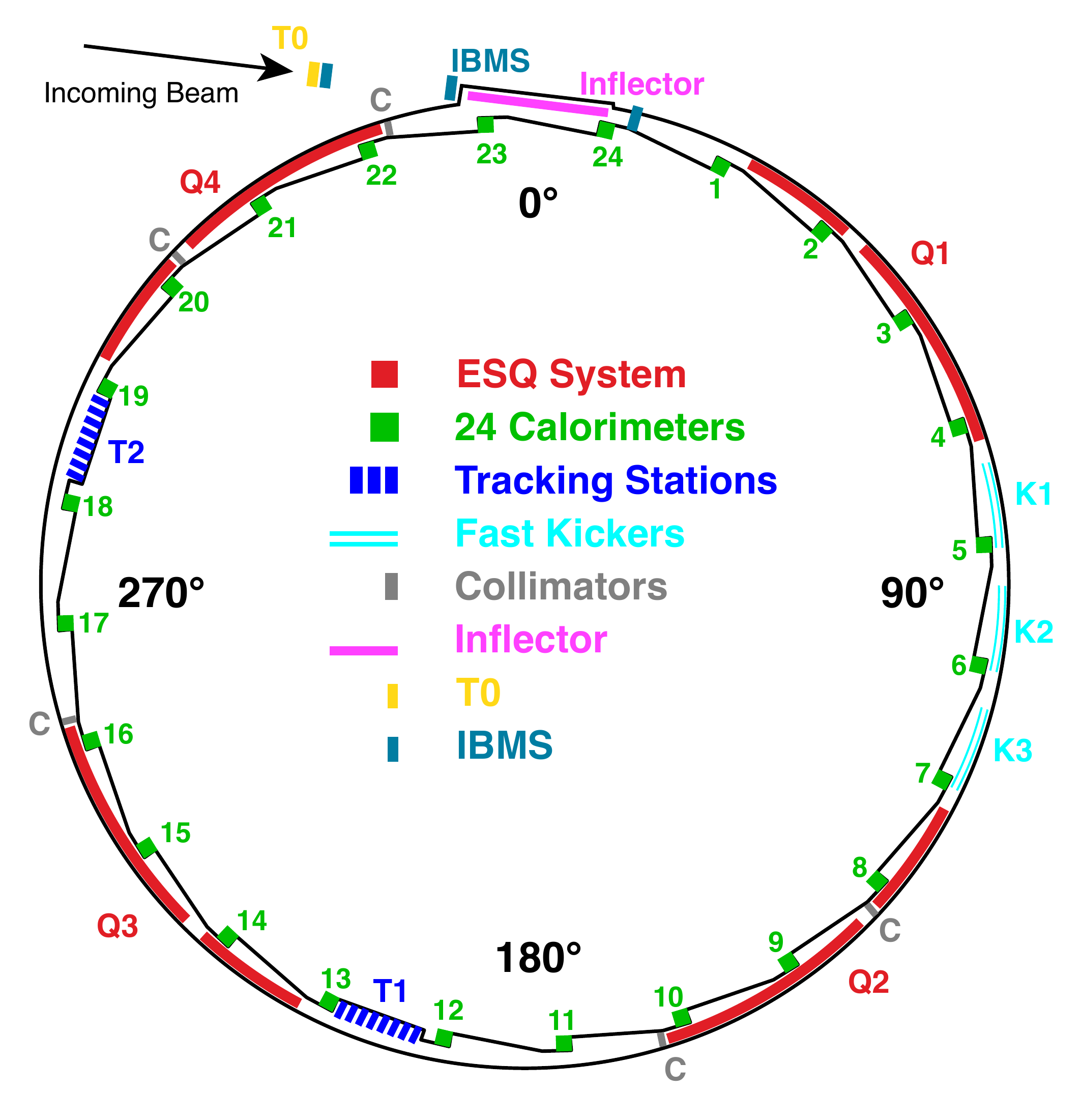}
\caption{Plan view of the \gmtwo storage ring vacuum chamber and instrumentation highlighting the important components discussed in this paper. The beam enters through a field-free corridor provided by the inflector magnet.  Three kicker stations operate in concert to deflect the beam onto a stable orbit during the first turn. Four ESQ stations, each consisting of a short and long section, provide vertical containment. The T0 and IBMS detectors monitor the injected beam intensity, time profile, and spatial distribution. }\label{fig:ring}
\end{figure}

Figure~\ref{fig:ring} shows a plan view of the main components and detectors in the storage ring. Not shown for clarity in the figure is the ``C-shaped'' superconducting storage ring magnet~\cite{Danby:2001eh}.  It was transported from BNL to Fermilab, rebuilt, and reshimmed by our Collaboration to a roughly threefold improved uniformity.  The NMR probes used to measure the absolute and relative magnetic field are all new, but they closely follow designs developed for E821~\cite{Prigl:1996fpk}. The superconducting inflector magnet~\cite{Yamamoto:2002bb}, the ESQ hardware~\cite{Semertzidis:2003zs}, and the vacuum chambers are used ``as is" from E821.  The ESQ power supply system and its controls, the entire storage ring kicker system, and all detectors and associated electronics were custom designed for E989.

The storage ring is designed to accept muons in a narrow momentum range around \pmagic; however, the incoming beam has a comparatively wide momentum spread of $\pm1.6\%$ and no upstream dispersive focus.  Only a few percent of the particles are stored with a momentum width less than $\pm0.15\%$.
The muon bunch enters the ring through a nearly field-free corridor---18\,mm wide $\times$ 56\,mm high $\times$ 1700\,mm long---provided by the superconducting inflector magnet~\cite{Yamamoto:2002bb}. Three kicker stations (K1, K2, and K3) create a vertical magnetic field that opposes the main storage ring field and, thus, deflects the muons passing through them outward by ${\sim}10$\,mrad. Ideally, this transient field is turned off before the muons complete a full turn in about 149\,ns.  A challenge in this experiment is creating such a transient magnetic kick and timing it with respect to each muon bunch.  Figure~\ref{fig:kicker-T0} also shows a sample of the measured kicker-induced magnetic field overlaid with the muon bunch.  The kicker pulse shape and magnetic field strength are discussed later in the context of the stored muon momentum distribution within the ring.  The kick is not uniform across the full temporal extent of the incoming muon bunch, a fact that impacts the total storage efficiency and the momentum distribution of the stored beam.

Four ESQ stations are symmetrically placed around the ring. Each consists of a  long (L) and a  short (S) section spanning $26^\circ$ and $13^\circ$, respectively.  Thus Q1 in Fig.~\ref{fig:ring} has sections Q1L and Q1S, each with four plates that are connected to power supplies through individual high-voltage resistors.  The plates are raised from ground to operating voltage prior to each fill with $RC$ charging time constants of ${\sim}5$\,\SI{}{\micro\second}. They are returned to ground at the end of the 700-\SI{}{\micro\second}-long fill.  Plates are charged using either one-step or two-step power supplies. Those plates connected to the two-step supplies rise to a preset voltage that is  set to be ${\sim}$5--7\,kV below their final designed voltage. After a programmable delay of approximately \SI{7}{\micro\second}, they are raised to the full set point voltage. This procedure, known as scraping, initially displaces the beam vertically and horizontally with respect to the central closed orbit.    When a muon's horizontal and vertical oscillations conspire to exceed a 45 mm-radius with respect to the quadrupole center, it will likely strike a collimator, scatter, and lose energy.  Such muons  leave the storage volume in a few turns. When the voltages are symmetrized, the muon distribution relaxes back to the nominal center, where muon losses are minimized. This scraping process is designed to be completed by \SI{30}{\micro\second} after injection, before the nominal measurement start time.

\begin{figure}[htbp]
\includegraphics[width=\columnwidth]{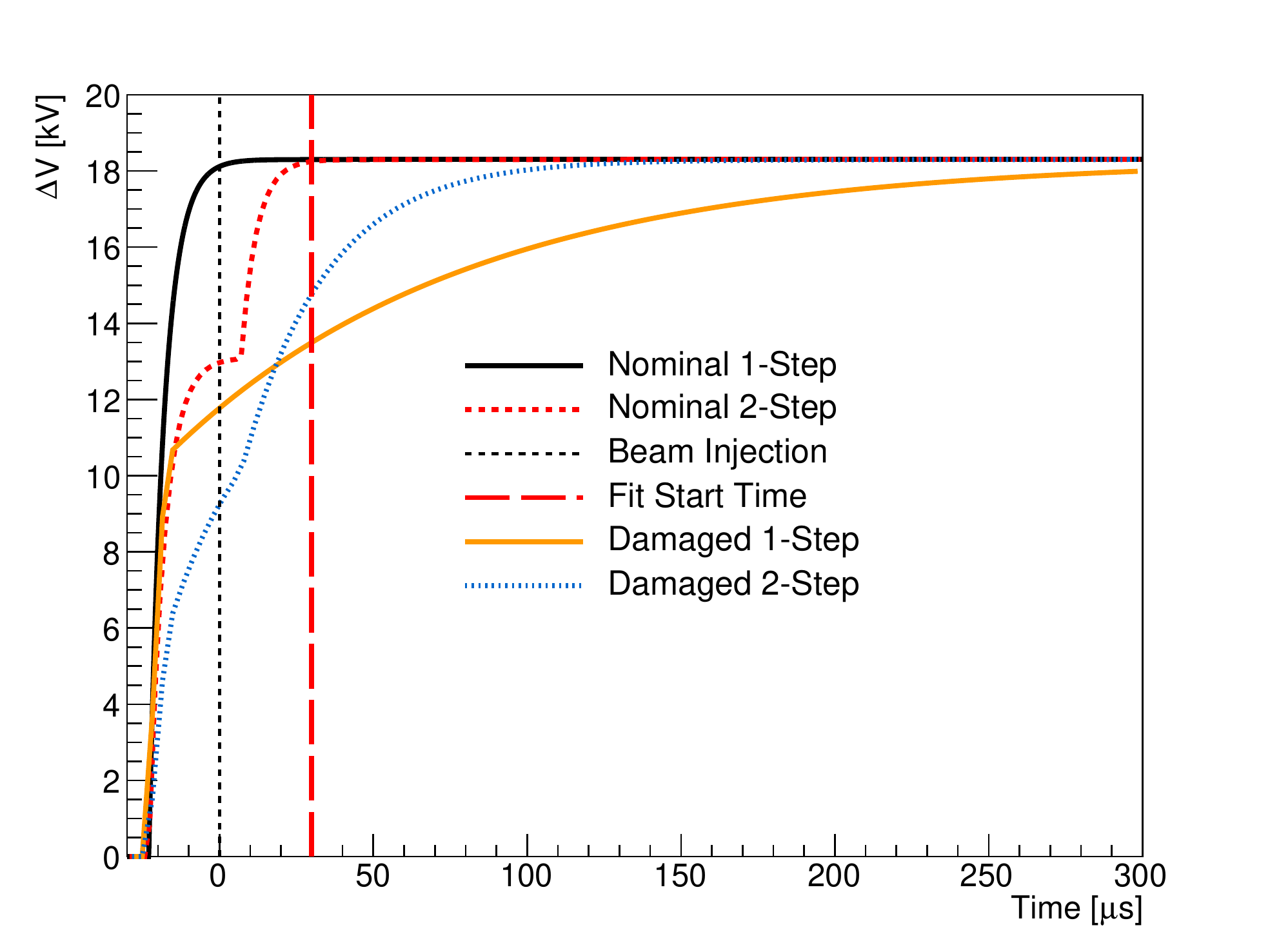}
\caption{Charging profiles for the 30 nominal ESQ plates driven by either one-step (black line) or two-step (dotted red line) power supplies.  The two damaged resistors (solid orange and dotted blue lines), connected to the same one- and two-step power supplies, exhibit markedly different charging profiles during the data-fitting period.  The vertical dotted black line at time $t = 0$ represents the arrival time of the muons in the storage ring.  The vertical dashed red line at \SI{30}{\micro\second} indicates the time at which the precession data fits begin.  The resistors deteriorated slightly for \runoned and the precession fit start time was delayed to \SI{50}{\micro\second} to compensate for the longer time charging profiles.}\label{fig:goodbadresistors}
\end{figure}

The charging traces of the ESQ plates for the \runone\ configuration are shown in Fig.~\ref{fig:goodbadresistors}.  As noted, 30 of 32 charging profiles follow the nominal and ideal pattern as shown in the solid black and dotted red traces for the one-step and two-step supplies, respectively.  However, following the completion of \runone\ data taking, it was discovered that two resistors had dynamically changed their resistances when high voltage was applied.  The resulting traces, measured after the data-taking period, are shown in dotted blue and solid orange in Fig.~\ref{fig:goodbadresistors}.  These resistors were connected to the upper and lower plates of the Q1L system. Because they are asymmetric and did not rise to their proper voltage prior to the fit start time, they introduced a perturbation to the stored muon spatial distribution versus time in fill. As will be evident in later sections, this was likely responsible for the larger-than-expected rate of lost muons, and it created a time-dependent phase shift from the correlations of decay position to average phase owing to the detector acceptance.  The changing voltages within the measurement time also led to a change in the storage ring tune values, the consequence of which is a dependence of the CBO frequency versus time in fill.  The $\omega_\text{CBO}(t)$ variation is well measured and included in the fits that determine the anomalous precession frequency. These damaged resistors were replaced prior to \runtwo.

The temporal bunch structure of the injected muon beam does not fill the storage ring uniformly.  Observing how this bunch spreads out owing to the finite momentum distribution is central to determining the stored muon momentum distribution (see Sec.~\ref{sec:fastrotation}).  For Runs-1a, 1b, and 1c, the measuring period to determine \wam\ begins \SI{30}{\micro\second} after injection, an optimization of minimizing systematic uncertainties and maximizing statistical significance. During \runoned, the damaged resistors deteriorated further, causing greater perturbations to the stability of the muon storage distribution at early times in the fill.   A delayed fit start time of \SI{50}{\micro\second} is used for this dataset, allowing the ESQ plate voltages to more closely reach to their target values.

\subsection{Instrumentation employed to study the muon beam and decay positrons}
\label{sec:experiment}

The detectors employed to measure the incoming and stored muon bunches are the T0, Injected Beam Monitoring System (IBMS), Calorimeter, and Tracker systems.  They are orientated as shown in Fig.~\ref{fig:ring}. Their signals are recorded using dead-time-less digitizers and saved fill by fill for offline processing.
The injected muon bunch first passes through the T0 and IBMS detectors, located at the entrance to the storage ring. They are used to measure the intensity of the incoming beam and its temporal and spatial profile and to establish the average entrance time of the bunch on a fill by fill basis, which is required for the analysis of the momentum distribution as discussed in Sec.~\ref{sec:fastrotation}. The T0 detector is a 1.0-mm-thick plastic scintillator with dual photomultiplier readout.  The two  IBMS detectors each consist of a horizontal and vertical array of 16 0.5-mm-diameter scintillating fibers.

Twenty-four electromagnetic calorimeters are positioned symmetrically around the inside radius of the storage ring, adjacent to the storage volume, but outside of the vacuum chamber; see Fig.~\ref{fig:ring}.
The scallop profile in the chamber allows decay positrons that curl toward smaller radii to exit through a thin, nearly perpendicular aluminum window before striking a detector.
Each calorimeter station consists of 54 lead-fluoride Cherenkov crystals read out individually on the downstream side by large-area silicon photomultipliers (SiPMs)~\cite{Khaw:2019yzq,Kaspar:2016ofv}.  The signals are continuously digitized at 800\,megasamples per second. The precise time alignment of the 1296 crystals and the system gain stability are enabled using a laser system as described in Ref.~\cite{Anastasi:2019lxf}.  Reconstruction of positron showers in the calorimeter crystals yields energy, time of hit, and impact position.
The coincidence of signals in three consecutive calorimeters, each depositing an energy typical of a minimum ionizing muon of about 170\,MeV, is used to identify muon losses and measure the loss rate versus time in fill.

The straw-tracker systems are located at approximately $180^\circ$ and $270^\circ$ with respect to beam injection.
They reside within the vacuum chamber in the scallop region just upstream of a calorimeter but outside of the muon storage volume.
 Both stations consist of eight modules,  each made of 128 5-mm-diameter straws oriented at $\pm 7.5^{\circ}$ with respect to the vertical.  They are used to track the decay positrons with the intention of tracing the decay trajectory back to its point of tangency inside the storage volume, a good proxy for the decay position of the parent muon.  High-quality tracks are selected to construct distributions of the beam position at the tracker locations.  These measurements are corrected for momentum-dependent detector resolution and the nonuniform acceptance of the detector.  The magnitudes of the corrections are estimated using the \ringsim package described below.  These trackers are critical to all beam dynamics topics discussed in this paper.  They provide the muon profile versus time, which is used to determine key storage properties, such as the betatron horizontal and vertical frequencies, their time dependence owing to the damaged resistor influence, and the vertical distribution for the pitch correction.

\subsection{Magnetic field considerations}
\label{sec:magfield}
The highly uniform magnetic field amplitude $B=|\vec{B}|$ within the storage volume allows for the extraction of the spatial field structure with NMR probe measurements~\cite{\field}.
The magnet iron poles and iron shims are precisely aligned in order to minimize the rms variations of the multipole terms as a function of the azimuth. The reduction of the field nonuniformities improves several systematic uncertainties in the measurement of the magnetic field, including the extraction of the frequency from the NMR probe measurements and the uncertainty due to limited knowledge of the probe positions.  Priority is given to the reduction of the lowest-order multipoles that couple to the moments of the muon beam.
Surface correction coils on the surfaces of the pole pieces are utilized to reduce the azimuthally averaged field multipole strengths.
Typically, the azimuthally averaged multipoles are reduced to below ${\sim}1\,\mathrm{ppm}$, as shown in Table~\ref{tbl:azi_multipoles_DT}.

The main measurement uncertainties in the mapping of the field, detailed in Ref.~\cite{\field}, stem from motional effects of the mapping device and the probe positions. Because of the overall field uniformity, despite the significant accuracy demands, the requirements for positioning of the probes are not detrimentally restrictive and can be met in practice. Specifically, a laser alignment survey of the mapper transiting around the ring and a sensitivity analysis determine the impact of the probe position uncertainties on the multipole strengths. This effect generates uncertainties for the azimuthally averaged dipole (12\,ppb), normal quadrupole (27\,ppb), and skew quadrupole (4\,ppb).  The azimuthal variations in the magnetic field and muon distributions are incorporated in the muon weighting analysis ($<20\,\mathrm{ppb}$) as well as the beam dynamics simulations described in Sec.~\ref{sec:siumulationtools}.  The multipole strengths, tracked between field maps by stationary NMR probes outside the storage volume, are highly stable over time; the ranges observed in the \runone dataset are summarized in Table \ref{tbl:azi_multipoles_DT}. Additional experimental uncertainties from magnetic field tracking (56\,ppb) and the presence of fast transients (99\,ppb) are carefully quantified, resulting in total systematic uncertainties on the determination of the muon-weighted magnetic field of approximately 114\,ppb for the \runone dataset.

\begin{table}[htpb!]
\begin{ruledtabular}
\begin{tabular}{lcccc}
		    &  Mean (ppm) & SD (ppm) & Pk-Pk (ppm) \\ \hline
		Dipole & 1\,000\,000 & 1.11 & $-2.10 - 1.56$ \\
		Normal quadrupole & 0.22 & 0.34 & $-0.72 - 0.84$ \\
		Skew quadrupole & 0.75 & 0.22 & $0.17 - 1.01$ \\
		Normal sextupole & $-1.38$ & 0.12 & $-1.56 - -1.14$ \\
		Skew sextupole & 0.57 & 0.12 & $0.34 - 0.75$ \\
		Normal octupole & 0.02 & 0.01 & $-0.01 - 0.03$ \\
		Skew octupole & 0.31 & 0.02 & $0.27 - 0.34$ \\
		\end{tabular}
\end{ruledtabular}
\caption{Statistical characterization of the azimuthally averaged multipole strengths (normalized to $r=\SI{45}{\milli\meter}$) relative to the dipole (1.45\,T) tracked during the \runone dataset. The mean, standard deviation (SD), and peak-to-peak (Pk-Pk) range correspond to the variations of these azimuthally averaged multipole strengths over time and are significantly smaller than the variation in azimuth. The Pk-Pk of the dipole denote lower and upper deviations relative to its mean. The uncertainty on the muon-weighted dipole term is approximately 114\,ppb for the \runone dataset~\cite{\field}. \label{tbl:azi_multipoles_DT} }
\end{table}

\subsection{Beam dynamics simulation tools}
\label{sec:siumulationtools}
Many of the results discussed in this paper incorporate comparison with, or results from, beam dynamics simulations.  Several compact simulation packages were used to rapidly estimate effects, but it is the work of three sophisticated and complementary methods that drive the results.  To determine critical information, such as momentum-time beam correlations, all three are typically used.
In all cases, it is critical that the simulation programs are cross-checked against a set of benchmarks showing that they can evaluate analytically calculated effects with high precision~\cite{Metodiev:2015twa}. It is also imperative that they are first tuned to match measurements of the incoming beam properties, the storage distribution within the ring, the CBO amplitudes, frequencies, and their time dependence, and the stored momentum distribution. We describe their essential features below.

\subsubsection{\GEANT-based \ringsim}
\label{sec:gm2ringsim}
The \ringsim program~\cite{Arvanitis:2014lza} is a \GEANT~\cite{Agostinelli:2002hh} based model of the storage ring and the final focus beam line used to steer the beam into the ring. The model includes all of the active detectors and most of the passive components installed in the storage ring. The geometry is constructed from a mixture of \GEANT native solids and \CADmesh objects~\cite{Poole:2012cad}. Calorimeters and straw-tracker modules are fully described, as are complex solids, such as the vacuum chambers and their inner structures, and the ESQ and kicker plates.  The geometry for the straw-tracker modules includes coordinates as determined in alignment surveys.

Four sets of realistic time- and space-dependent electric and magnetic fields are implemented. These include a pure dipole magnetic field in the storage region and a radially dependent fringe field that extends toward the center of the ring.  Radial magnetic field maps and additional multipole perturbations in the storage region, as described in Sec.~\ref{sec:magfield}, can optionally be included.  The inflector magnetic field is implemented as a map.  The time dependence of the kicker magnetic field is taken from direct magnetometer measurements made at the center of the plates (see Fig.~\ref{fig:kicker-T0}). The spatial field map within the kicker region is obtained using the finite-element magnetics modeling package \OPERA~\cite{Opera3D}. The strength and timing of the kicker with respect to the injected beam can be adjusted independently for each of the three kicker plates.  The fields associated with the ESQ plates are implemented as a multipole expansion.  They are dynamically evolved through the scraping periods at the beginning of each muon fill, and they can accommodate the perturbations and independent time constants caused by the damaged resistors during \runone.

Simulated datasets are typically generated using two types of ``particle guns." The {\it beam gun} imports muon distributions at the end of the final focus beam line as determined from \GBeamline simulations of the Muon Campus beam and injects them into the storage ring.
The beam gun allows for the injection of a mixture of particles, facilitating studies of proton and positron contamination.
The {\it gas gun} omits the injection process completely and instead fills the storage region phase space of the ring with muons that then decay at that location. This distribution of muons is matched to reflect the measured vertical and radial offsets of the beam within the storage ring as well as the measured coherent betatron oscillation amplitudes. The gas gun is particularly useful for positron acceptance and reconstruction studies.  In both cases, muon spin is appropriately evolved during time in fill, and proper spin-dependent muon decays are employed.

\subsubsection{\COSYINFINITY}

The \COSY-based model~\cite{Tarazona:2019ivq} is a data-driven computational representation of the storage ring. Dedicated packages in \COSYINFINITY~\cite{Makino:2006sx} for the design and analysis of particle optical systems provide the framework for beam physics studies and symplectic tracking simulations in the storage ring. The beam dynamics of the injected muon beam is recreated with high fidelity by representing the magnetic and electric guide fields in the storage region based on measurements of the beam.

The magnetic field inhomogeneities of the storage ring are determined from the magnitude of the magnetic field as measured by the NMR probes (see Sec.~\ref{sec:magfield}). Because of the high uniformity of the magnetic field along the vertical direction in the storage region, a reliable extraction of magnetic multipole strengths from the experimental data is performed and implemented in the model as a series of magnetic multipole lattice elements.
Each ESQ station is modeled as an optical element superimposed on the magnetic field. The nonlinear action of the ESQ on the beam's motion is captured by accounting for the high-order coefficients of the electrostatic potential's transverse Taylor expansion. These coefficients are calculated by recursively iterating the horizontal midplane coefficients---modeled with conformal mapping methods~\cite{TUPAG22}---to satisfy Laplace's equation in curvilinear optical coordinates. The effective field boundary and fringe fields of the ESQ are calculated~\cite{TUPAG22} using \COULOMB's~\cite{COULOMB} boundary element method field solver.

With the aforementioned electric and magnetic guide fields implemented in the \COSY-based model, the orbital and spin equations of motion are well defined and integrated to produce high-order transfer maps via differential algebra methods~\cite{Berz_1999aa}. Transfer maps are computed and combined to recreate either azimuthal segments or the entire storage ring.

Beam-tracking simulations are performed by preparing transfer maps of the storage ring. The muon beam, represented as an array of orbital and spin coordinates around the closed orbit, is transformed turn by turn with transfer maps that encapsulate the time-evolving guide fields as they vary throughout the beam fill. Symplecticity is enforced during beam tracking with high-order transfer maps to account for the truncation of components beyond the map order, in this way controlling energy conservation and error propagation. To account for beam collimation, special routines  were developed to efficiently remove muons beyond the collimator apertures and only at the azimuthal locations where they are inserted. This tool, together with the symplectic enforcement during tracking, allows for reliably studying muon loss rates in the storage ring.

The beam conditions after the action of the injection kickers are obtained by preparing high-order transfer maps of the azimuthal segments where the three kickers are placed. Using these maps combined with the transfer maps of the other components of the storage ring, a beam distribution obtained from simulations of the Muon Campus beam lines and the inflector is transferred into the ring's storage region. The structure and time-dependent strength of the kicker magnetic fields  are constructed from magnetometer measurements (see Fig.~\ref{fig:kicker-T0}). Alternatively, the initial beam distribution after injection is also prepared by calculating with a non-negative least-squares solver the probability density functions of the beam, based on data from the straw-tracking detectors and the fast-rotation analysis (see Sec.~\ref{sec:fastrotation}).

The \COSY-based model has been extensively used to calculate lattice configurations (e.g., periodic Twiss parameters, betatron tunes, closed orbits, and dispersion functions) of the storage ring with and without the damaged ESQ resistors. It provided a reconstruction of the special electric field behavior of the ESQ electric fields during \runone for the assessment of beam dynamics systematic uncertainties. Further, it has provided a numerical model to link tracker measurements with direct high-voltage (HV) probe measurements of the damaged ESQ resistors, and it has been used to choose optimal configurations of the storage ring to minimize muon losses.

\subsubsection{\label{sec:bmad} \BMAD}

\BMAD refers to a subroutine library~\cite{Sagan:2006sy} for simulating the dynamics of relativistic beams of charged particles
and an associated format for defining beam line elements. So defined, the full complement of the analysis tools of the library can be used to investigate the particle dynamics.
Particle-tracking methods include Taylor maps, symplectic integration, and Runge-Kutta integration through field maps.
Taylor maps can be predefined or constructed by tracking.

The \BMAD formatted representation of the \gmtwo experiment is comprised of three distinct branches:
i) the M5 beam line; ii) the injection channel and inflector; and iii) the storage ring, including a static magnetic field, time-dependent quadrupole electric fields and kicker magnetic fields, and collimator apertures.
The beam lines are assembled as a sequence of elements with fixed length.
The electromagnetic fields in each element are defined by field maps,  multipole expansions, or analytic expressions.
Time dependence for pulsed kickers and ramped ESQ plates requires custom code.
Specialized custom routines are used to incorporate arbitrary kicker pulse shapes or ESQ voltage time dependence.
The curvilinear coordinate system can be represented by beginning with a full three-dimensional map and then extracting an azimuthal
slice of the map (in $x, y$ at fixed $\phi$) or with a fitted multipole expansion of McMillan functions.
The curvature necessarily introduces nonlinearities that are not faithfully included in a two-dimensional Cartesian expansion.
The \BMAD code allows specification of the quadrupole electric field in terms of field maps or multipoles.
The main magnet is represented as a map or analytic function with uniform field.
A uniform radial component can be specified.
Measured field errors are incorporated analytically.
The azimuthal dependence of the error field is expanded as a solution to Laplace's equation in cylindrical coordinates in order to ensure consistency with Maxwell equations.
The magnetic field through the hole in the back-leg iron and cryostat and main magnet fringe field is based on a three-dimensional \OPERA~\cite{Opera3D} map.
A distinct map is computed for the field in the inflector. The fringe and inflector maps are superimposed as appropriate.

\section {Determination of the Stored Muon Momentum Distribution}
\label{sec:fastrotation}
The contribution to \wam from the electric field depends on the momentum distribution of the stored muon beam or, equivalently,
the equilibrium radial distribution. Since the azimuthal speed of the stored muons is nearly uniform over the 9~cm aperture, to a good approximation a muon's rotation frequency is
inversely proportional to its equilibrium radius $x_e$. Because the muons are stored over a range of $x_e$, a beam bunched at $t=0$ will steadily
debunch, as the higher-frequency muons at smaller radii advance with respect to the lower-frequency muons at larger radii.

\begin{figure}[htbp]
\includegraphics[width=\columnwidth]{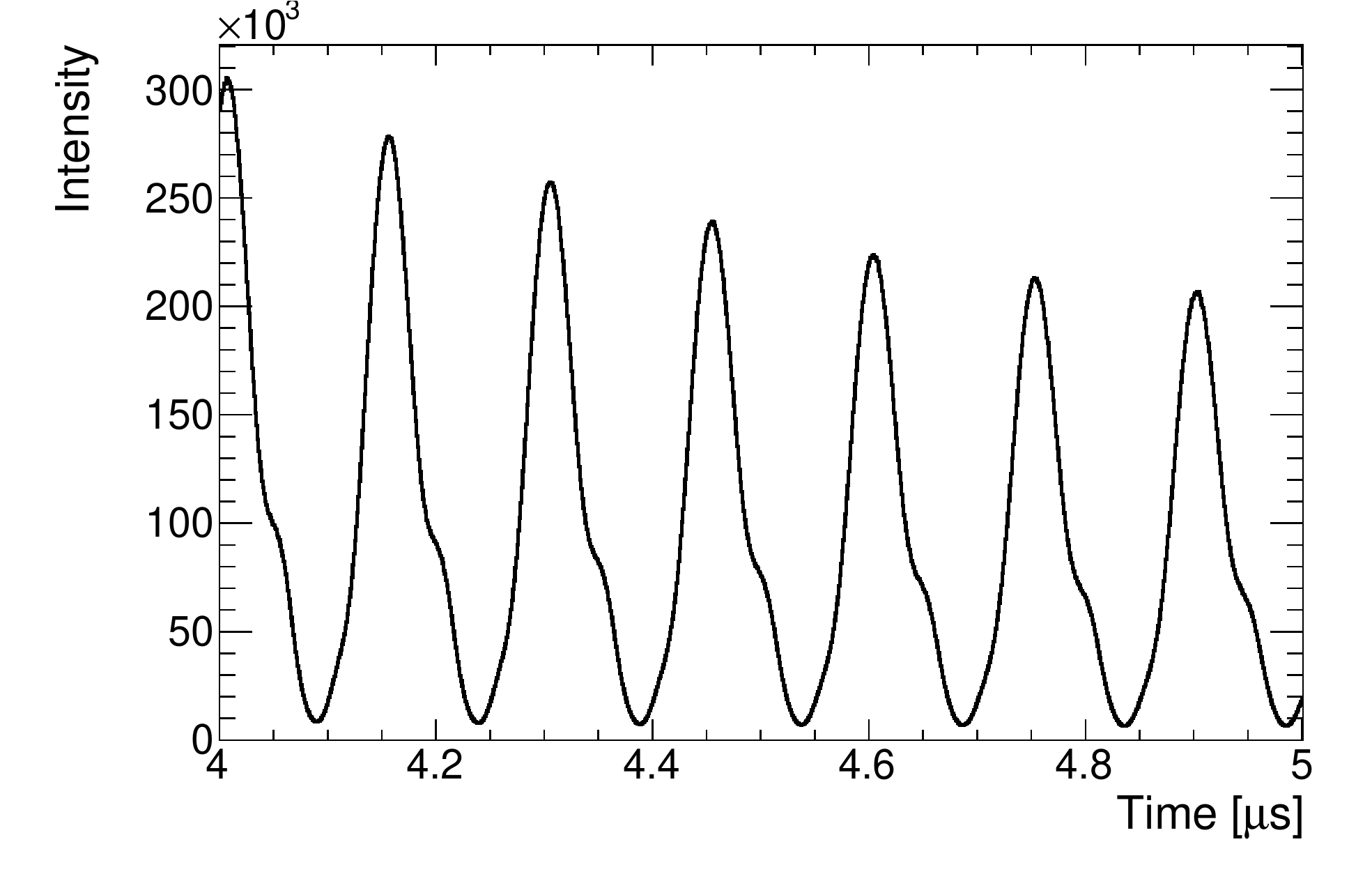}
\includegraphics[width=\columnwidth]{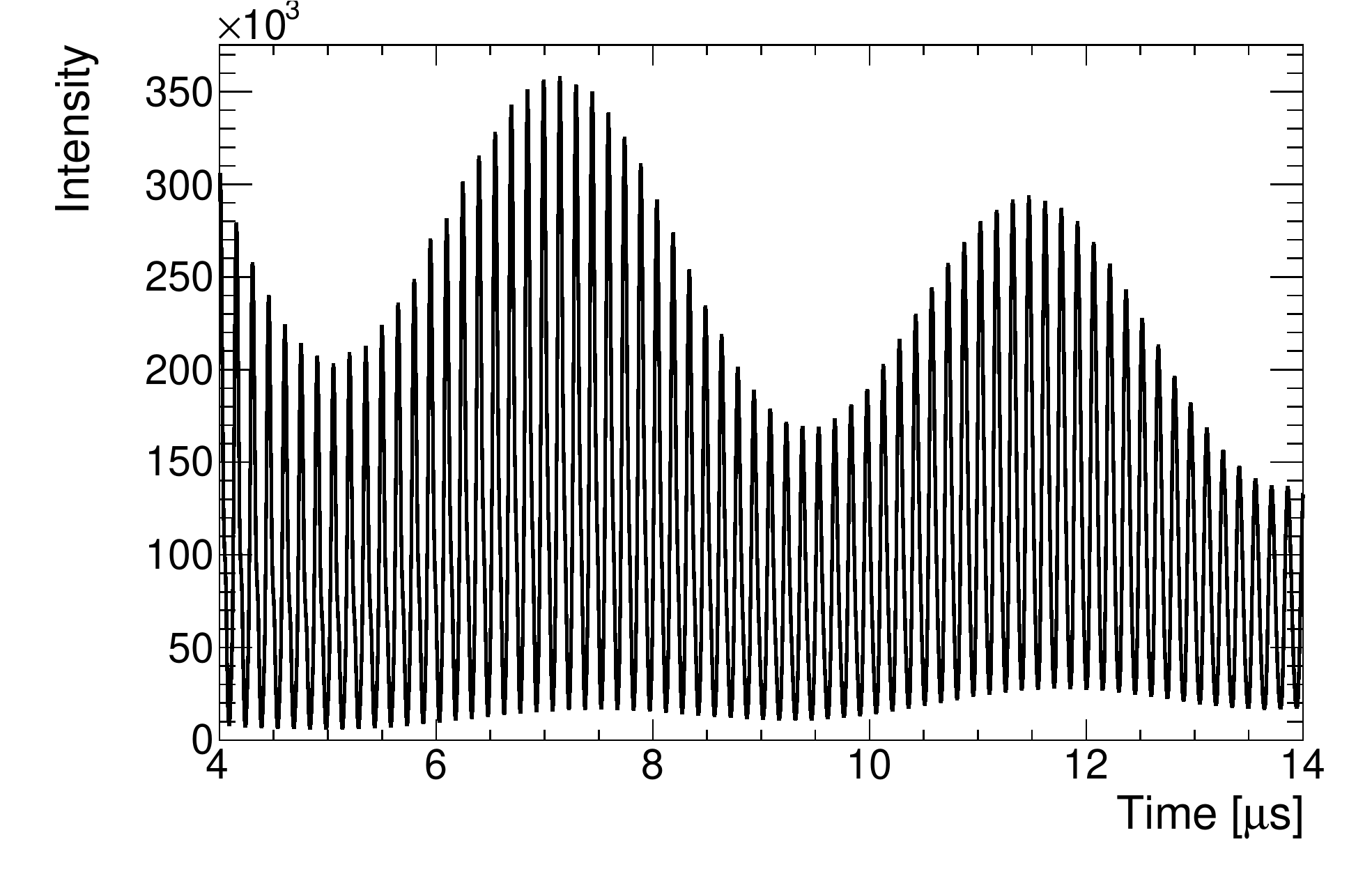}
\caption{Positron counts as a function of time as seen by all the calorimeters combined for the \runoned dataset for the time ranges: (top) 4--5 and (bottom) 4--14\,\SI{}{\micro\second} with respect to the beam injection. The time binning period is 1\,ns. The amplitude modulation in the bottom panel is from the muon spin rotation frequency \wam.}
\label{fig:allcalo_intensity}
\end{figure}

A technique~\cite{Orlov:2002ag} based on Fourier transformation yields a frequency spectrum that can be converted to radius and momentum.
An alternative method extracts the radial distribution by a direct fit to the debunching signal of the muon beam~\cite{Combley:1974tw,Bailey:1972eu}.
In both cases, the input data are provided by the calorimeters,
which measure the time dependence of the intensity of the decay positron distribution.
The positron counts from the 24 calorimeters are merged together with a time offset of $T/24$ per calorimeter where $T$ is the cyclotron period, approximately 149\,ns.
The upper panel of Fig.~\ref{fig:allcalo_intensity} shows the intensity 4--5\,\SI{}{\micro\second} after injection; the lower panel expands the time range to 4--14\,\SI{}{\micro\second}. The individual turns around the ring, referred to as ``fast rotation,'' are distinct. The slower modulations in the upper envelope are caused by muon decay and the \wam precession frequency. By the nominal precession fit start time of 30\,\SI{}{\micro\second}, the rapid cyclotron frequency modulation has largely dephased and eventually disappears.
To isolate this fast rotation signal, the calorimeter time spectra are divided by a fit\footnote{In the full fits of the precession data, the function has additional terms, but the simple form of Eq.~\ref{eq:wiggle_func} is perfectly sufficient for this Fourier analysis.} to the envelope modulation using Eq.~\ref{eq:wiggle_func}, which effectively removes all the significant physics
signals {\it except} the
fast rotation itself.

\subsection{Momentum-time correlation}
\label{sec:momtimecorr}
The Fourier and debunching methods, discussed below in Secs.~\ref{sec:fouriermeth} and \ref{sec:debunchingmethod}, both assume that the momentum distribution in the captured muon pulse is uncorrelated with the longitudinal position in the pulse train (see Fig.~\ref{fig:kicker-T0}).
If the momentum and time are uncorrelated, the distribution is maximally bunched when it enters the ring.
The peak intensity of the fast rotation signal only decreases with time.
The behavior is symmetric with respect to time reversal. One can imagine that the peak intensity likewise decreases moving backward in time.

In reality, a momentum-time correlation is introduced by the injection kicker. The efficiency with which muons are captured in the ring
depends on momentum and the amplitude of the kicker.
If the kicker field is low, acceptance is higher for high-momentum muons; if the kicker field is high, low-momentum muons are favored.
Since the magnetic field of the kicker varies over the time duration of the incoming muon pulse,
so do the momenta of muons that are stored in the ring.
Simulations are used to characterize the momentum-time correlation of the captured muons and to estimate the systematic
uncertainty in the measurement of the momentum distribution from the fast rotation analysis.

\subsection{Frequency domain fast rotation analysis: Fourier method}
\label{sec:fouriermeth}
For a stored muon beam in which the cyclotron frequencies and injection times are independent, the frequency distribution of the ensemble can be extracted by the cosine transform of the fast rotation signal $S(t)$ ~\cite{Orlov:2002ag}:
\begin{equation}
\hat S(\omega) = 2\int_0^{\infty} S(t + t_0) \cos(\omega t) \, dt,
\label{eq:costransform}
\end{equation}
where $t_0$ is an effective time of symmetry for the ensemble. The optimal $t_0$ is determined by imposing that the transform $\hat{S}(\omega)$ must vanish in the unphysical frequency region outside the range that can be stored.

\begin{figure}[bt]
\includegraphics[width=\columnwidth]{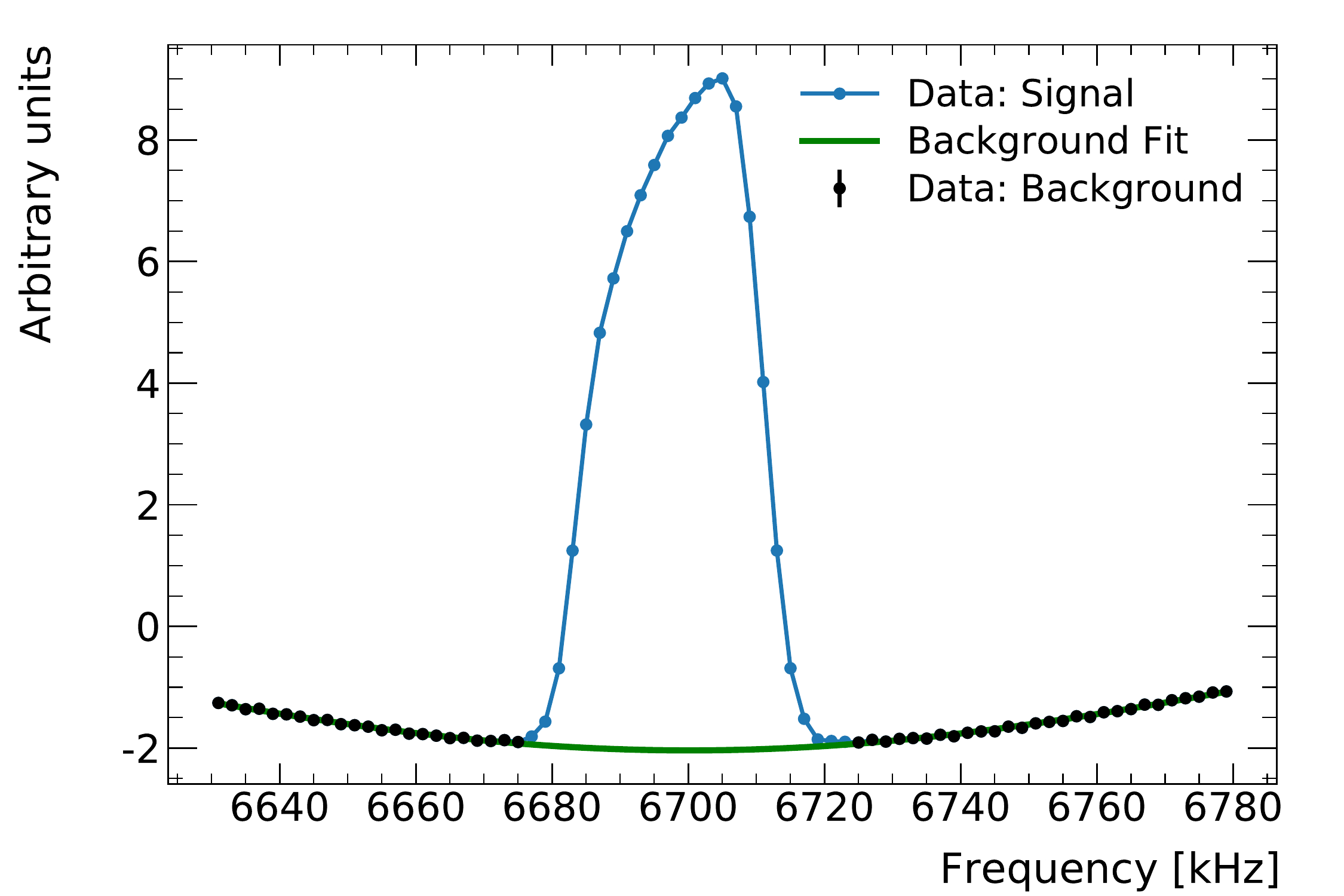}
\caption{The frequency distribution obtained with a cosine transform of the fast rotation signal from the \runoned dataset. The background is a consequence of the missing data before the start time of \SI{4}{\micro\second}. A cardinal sine (sinc) background fit has been applied to the black points and is subtracted from the whole frequency range as a correction.}
\label{fig:bkgfit}
\end{figure}

The problem is complicated by the fact that the first few microseconds of the fast rotation signal are contaminated by beam positrons and by
muons that will not be stored.  The analysis is based on the intensity signal that begins at about 25 turns, or $t_s=4.1$\,\SI{}{\micro\second} after injection, by which time the
beam positrons have been lost
and the performance of the calorimeter SiPMs has largely recovered from the intense flash of particles at injection. The available cosine transform is, therefore, missing the contribution before the start time $t_s$, which introduces a background modulation $B(\omega)$ to the frequency spectrum.  As shown in~\cite{Orlov:2002ag}, this background can be estimated using an inverse cosine transform and a model for the frequency distribution $\hat S_0$ with the result
\begin{equation}
\begin{aligned}
B(\omega)
&\approx{\frac{1}{\pi}} \int_{\omega_-}^{\omega_+} \hat S_0(\omega^\prime) \, \frac{\sin[(\omega^\prime-\omega)(t_s-t_0)]}{(\omega^\prime - \omega)} \, d\omega^\prime ,
\end{aligned}
\label{eq:frbkg}
\end{equation}
where the limits of integration correspond to the physical range of frequencies that can be stored.
An ansatz for $\hat S_0(\omega)$ is hypothesized and the parameters determined by a fit to the background.
If $\hat{S}_0(\omega) = A\delta(\omega~-~\omega_B)$, where $A$ and $\omega_B$ are fit parameters for amplitude and frequency, respectively, then
\begin{equation}
B(\omega) \approx \frac{A}{\pi}\frac{\sin[(\omega_B - \omega)(t_s - t_0)]}{(\omega_B - \omega)}.
\label{eq:sinc_background}
\end{equation}
Empirically, this model is found to give a good fit to the background for start times before \SI{5}{\micro\second}.
More sophisticated functional forms  give good background fits for start times up to $t_s\sim 25$\,\SI{}{\micro\second} in Monte Carlo simulation, and, in practice,
the fitted spectrum is nearly independent of start time for $\SI{4}{\micro\second}<t_s<\SI{25}{\micro\second}$.
An example of the frequency transform $\hat{S}$ of Eq.~\ref{eq:costransform} is shown in Fig.~\ref{fig:bkgfit}, with the optimal background fit using Eq.~\ref{eq:sinc_background}.
The background fit is subtracted, and the corrected transform is taken as the distribution of cyclotron frequencies.

\subsection{Time domain fast rotation analysis: debunching method}
\label{sec:debunchingmethod}
An alternative approach, pioneered by the CERN~II collaboration, is based on a simple model of the beam's debunching at early times in the measurement
period~\cite{Combley:1974tw,Bailey:1972eu}. Consider first the contribution to the fast rotation signal from a narrow bin in time and momentum space. The signal is initially rectangular but grows increasingly trapezoidal as it revolves around the ring, due to the uniform momentum spread within the bin.
Mathematically, all the essential physics is
captured in the propagator function $ \beta_{ijk}$, which describes this segment's contribution to the overall signal in the detector at time $t_j$.
Indices $i$ and $k$
identify the segment's equilibrium radius and position in time within the injected muon bunch, respectively. Using superposition, an ensemble of segments with
joint distribution $f_i I_k$ is given by
\begin{equation}
  S_j = \sum_{ik} \beta_{ijk} f_i I_k.
\end{equation}
Here, $S_j$ is the calorimeter signal in time bin $j$, $f_i$ describes the radial distribution, and $I_k$ describes the time profile of the injected beam.
A least-squares fit to
the signal $S_j$ determines $f_i$ and $I_k$.

In the BNL experiment~\cite{Bennett:2006fi}, many of the calorimeters were live on the first turn following injection. It was, therefore, simple to make a very good
guess of the stored beam's initial time profile $I_k$. As noted above in Sec.~\ref{sec:fouriermeth}, detector signals in the Fermilab experiment cannot be used before \SI{4}{\micro\second} after injection.
Therefore, the original CERN method was replaced with a pair of fits, for $f_i$ and $I_k$, which are iterated until the results are stable. In the first
pair of fits, the time profile is taken from the calorimeter signal at \SI{4}{\micro\second}, and the momentum distribution is determined by the fit. Then, that momentum distribution is used to update the injected time profile in a second fit. The determination of the momentum and time distributions are computationally identical. Between 50 and 100 iterations of this double fit are generally required for convergence.

\subsection{The radial distributions for \runone}
\begin{figure}[htbp]
\includegraphics[width=\columnwidth]{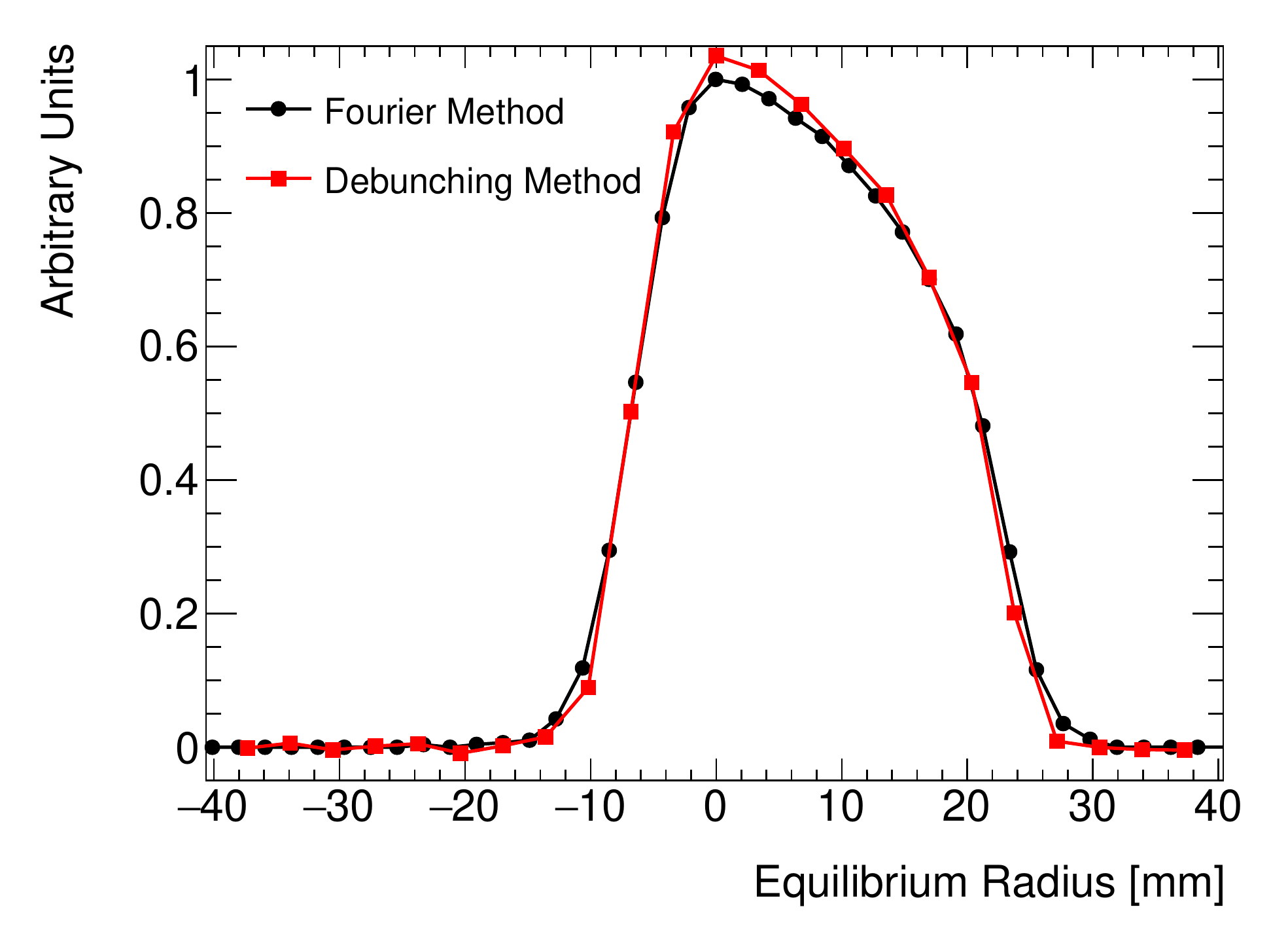}
\includegraphics[width=\columnwidth]{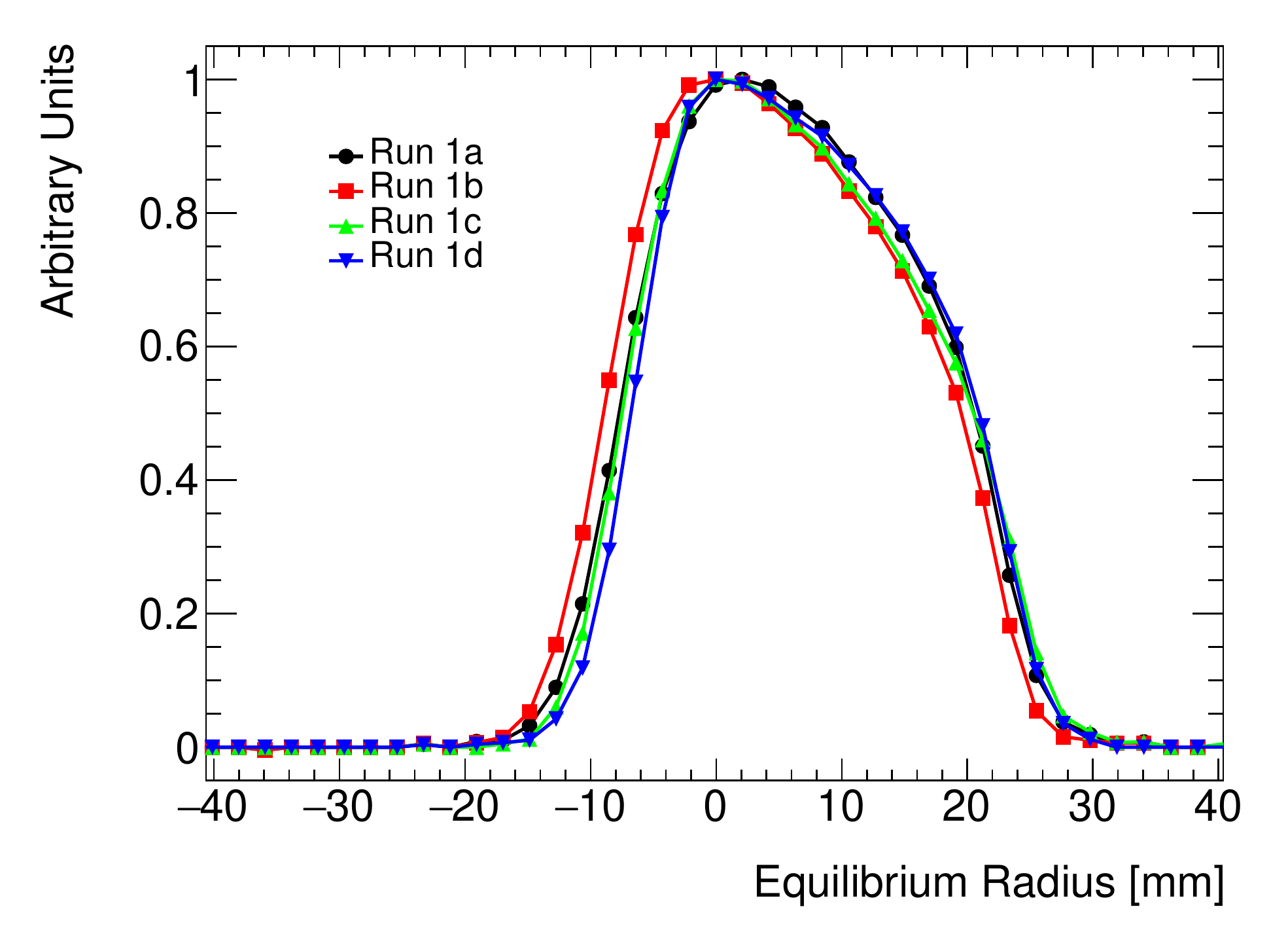}
\caption{Top: the radial (closed-orbit) distribution extracted by both the Fourier method and the debunching analysis for the \runoned dataset. Bottom: the radial distribution for the four \runone datasets as determined by the Fourier method.  In both plots, the equilibrium radius is defined such that a magic-momentum muon is at 0\,mm.}\label{fig:fourierdebunching}
\end{figure}

An example of the radial distribution extracted by both the Fourier method
and the debunching analysis is shown in the top panel in Fig.~\ref{fig:fourierdebunching}. The agreement is sufficiently good that either can be used to extract the electric field correction.
The radial distributions for the \runonea, 1b, 1c, and 1d periods as determined by the Fourier method are shown in the bottom panel in Fig.~\ref{fig:fourierdebunching}.
The slightly different distributions are attributed primarily to the different kicker strengths, as indicated in Table~\ref{tb:mastertable}.
The means of the radial distributions---in all cases---do not fall on the magic radius.  This fact will enter in the calculation of the electric field correction in Sec.~\ref{sec:efield}.
The radial offsets and widths as determined by the Fourier method are given in Table~\ref{tb:fastrotation}.

\begin{table}[h!]
\begin{ruledtabular}
\begin{tabular}{ccc}
Dataset & $x_e$ (mm) & $\sigma$ (mm) \\
\hline
\runonea & $6.1 \pm 1.2$ & $9.2 \pm 0.2$ \\
\runoneb & $4.9 \pm 1.2$ & $9.2 \pm 0.2$ \\
\runonec & $6.3 \pm 1.2$ & $9.2 \pm 0.2$ \\
\runoned & $6.7 \pm 1.2$ & $8.9 \pm 0.2$ \\
\end{tabular}
\end{ruledtabular}
\caption{The radial offsets $x_e$ and widths $\sigma$ for the four \runone\ datasets obtained by the Fourier method.}
\label{tb:fastrotation}
\end{table}

 \section{\label{sec:efield} Electric Field Correction $C_e$}

The electric field term in Eq.~\ref{eq:trueomega} produces a rest frame magnetic field that affects the measured anomalous precession frequency \wam.
For the simple case where we neglect the vertical betatron motion and $\vecwa = \vec \omega_s - \vec \omega_c$, Eq.~\ref{eq:trueomega} can be simplified to
\begin{equation}
  \wam = \frac{|q|}{ m}a_\mu B_y \left[ 1 - \beta \frac{E_r}{ c B_y} \left(1-\frac{m^2 c^2}{a_\mu p^2}\right) \right]\,
  \label{eq:Efield_omega}
\end{equation}
where \wam is a scalar frequency and the subscripts $r$ and $y$ denote the radial and vertical components respectively.
The electric field term vanishes at the magic momentum $p_0 = (mc)/\sqrt{\amu}$, or when $E_r = 0$
which is the case at $R_0$ as a result of the design of the ESQ system.
In practice, the stored muon distribution has a finite momentum spread and is not centered, as 
discussed in Sec.~\ref{sec:fastrotation} and shown in Fig.~\ref{fig:fourierdebunching}.
The mean radial electric field experienced by a muon oscillating about an equilibrium radius $x_e$ in an ideal electric quadrupole is
\begin{equation}
\langle E_r\rangle = \kappa x_e = \frac{ n \beta c B_y}{ R_0} x_e \, ,
\end{equation}
where $\kappa$ is the electric field gradient. Defining $p = (p_0 + \Delta p)$ and using
\begin{equation}
\frac{\Delta p}{p_0} = (1-n)\frac{x_e}{R_0}\, ,
\label{eq:DispRelation}
\end{equation}
where the approximation that the dispersion is a constant value of $R_0/(1-n)$ is sufficient, we average over the entire muon distribution to obtain the electric field correction
\begin{equation}
C_e \approx 2n(1-n)\beta_0^2  \frac{ \langle x^2_e\rangle}{ R^2_0}  \, ,
\label{eq:Ce-equation}
\end{equation}
where $\beta^2_0 = p^2_0/[m^2 c^2 + p^2_0]$.  The correction must be applied to the measured \wam to obtain the anomalous precession frequency \wa used to determine \amu. The use of an analytic expression for the electric field correction has been previously considered~\cite{Semertzidis:2003zs,Farley:2004hp,Miller:2012opa}.
The precision goals of our experiment require careful consideration of this correction.
An extensive discussion of electric quadrupole nonlinearity factors is given in Appendix~\ref{ap:Efield} and numerical tests to justify the validity of the use of Eq.~\ref{eq:Ce-equation} are given in Appendix~\ref{ap:EpitchNumeric}.

\subsection{Measuring the muon radial distribution and calculating $C_e$}

The fast rotation analysis using the Fourier method (see Sec.~\ref{sec:fouriermeth}) yields the distribution of cyclotron frequencies $f$, which are converted to equilibrium radii $R$ by the relation $R(2\pi f) = v$, assuming fixed muon velocity $v$. The radial offsets $x_e$ relative to the magic radius $R_0 = 7112~\text{mm}$ are, therefore, $x_e = v/(2\pi f) - R_0$. This conversion yields the distribution of equilibrium radial offsets, as in Fig.~\ref{fig:fourierdebunching}. The electric field correction depends on the mean and width of this distribution, via $\langle x_e^2 \rangle = \sigma_{x_e}^2 + \langle x_e \rangle^2$.

The recovered radial mean and width exhibit some variation when the fast rotation analysis is repeated over a range of positron energy bins. This is a consequence of variations in calorimeter acceptance with positron energy and radial decay position, supported by simulation with \ringsim. The final $C_e$ is, therefore, weighted over positron energy bins according to the statistical power of \wam in each bin. For an asymmetry-weighted analysis,\footnote{The positron spectrum constructed by weighting each $e^{+}$ contribution to the time series by its energy-dependent asymmetry, shown to be the optimal approach~\cite{Bennett:2007zzb}.} this is the average of $C_e$ over positron energy bins between 1 and 3.1~GeV, weighted by $N(E)A(E)^2$, where $N(E)$ is the number of positron counts in energy bin $E$ and $A(E)$ is the fitted asymmetry of the \wam modulation in that bin. The resulting corrections $C_e$ are tabulated with their dominant uncertainties in Table~\ref{Tab:EFieldCorrection}, as discussed below.

The statistical uncertainties in $\langle x_e \rangle$, $\sigma_{x_e}$, and $C_e$ are estimated by repeating the fast rotation analysis over an ensemble of pseudodata. Each positron count $N_i$ in the measured time spectrum is shifted by $\pm \sqrt{N_i}$ at random, assuming uncorrelated Poisson statistics. The analysis is repeated over about 1000 randomly altered signals, and the ensemble standard deviation of each recovered quantity is taken as its statistical uncertainty. For subsets of the data with different total positron counts $N$, the statistical uncertainty was confirmed to scale as $\sqrt{N}$.

The fast rotation Fourier analysis method relies on several chosen parameters, including the start and end times of the cosine transform, the frequency bin spacing, the frequency distribution model used in the background fit, and the set of frequency bins included in that fit. In each case, a scan is performed over the range of appropriate choices, and the standard deviation in the result is taken as the corresponding systematic uncertainty. The total uncertainty attributed to these parameters is the average of their linear sum and quadrature sum, accounting for probable correlations.

Furthermore, Eq.~\ref{eq:costransform} does not extract the frequency distribution perfectly as intended when there is any systematic relationship between cyclotron frequency and injection time.
Referred to here as momentum-time correlation, this relationship is expected as discussed in Sec.~\ref{sec:momtimecorr}. The corresponding uncertainty in $C_e$ has been estimated as 52~ppb, which is the average discrepancy between truth and reconstruction using \ringsim and \BMAD simulations.

The expression for $C_e$ in Eq.~\ref{eq:Ce-equation} is derived under the assumption of continuous ESQ plates and ideal alignment of the quadrupole field relative to the target muon orbit. The effects of discrete ESQ plates, position misalignments, and voltage errors have been studied using \BMAD simulations with surveyed ESQ positions and their measurement uncertainties. The resulting uncertainty in $C_e$ is estimated to be about 6.4~ppb.

The field index $n$ has been measured using the value of $\omega_\text{CBO}$ fitted during the production of the fast rotation signal and the relationship $\omega_\text{CBO} = \omega_c - \omega_x = (1 - \sqrt{1 - n})\omega_c$ from Sec.~\ref{sc:experiment}.  However, the CBO frequency is not constant throughout the fill, rather approaching a stable value with two exponential time constants based on the beam scraping procedure and damaged ESQ resistors. This time dependence has been measured using the tracking detectors (see Sec.~\ref{sec:time_changing_cbo}), and the uncertainty in $n$ has been estimated using the rms spread in $\omega_\text{CBO}$ over the \wam measurement period.

\begin{table}
\begin{ruledtabular}
\begin{tabular}{rrrrr}
Dataset & \runonea & \runoneb & \runonec & \runoned \\
\hline
$C_e$ & 471 & 464 & 534 & 475 \\
\hline
Stat. uncertainty  & $<1$ & $<1$ & $<1$ & $<1$ \\
\hline
Fourier method  & 8 & 13 & 14 & 4 \\
Momentum-time  & 52 & 52 & 52 & 52 \\
ESQ calibration & 6 & 6 & 6 & 6 \\
Field index & 2 & 2 & 2 & 4 \\
\hline
Syst. uncertainty  & 53 & 54 & 54 & 53 \\
\end{tabular}
\end{ruledtabular}
\caption{Electric field corrections $C_e$ (ppb) and uncertainties combined in quadrature (ppb) for the four \runone groups.}
\label{Tab:EFieldCorrection}
\end{table}

 \section{\label{sec:pitch} Pitch Correction $C_p$}
The ESQ system used to vertically confine the muon beam creates vertical betatron oscillations, that is, periodic
 up-down pitching of the vector $\vec \beta$.
The $\amu \hat \beta \times \vec B$ term in Eq.~\ref{eq:trueomega} affects the value of \wam. The magnitude of the term is reduced when $\hat\beta$ and $\vec B$ are not perpendicular, as is the case here.
The vertical betatron frequencies $\omega_y$ listed in Table~\ref{tab:beamfreq} are an order of magnitude larger than the muon spin rotation frequency \wa, which avoids depolarizing spin resonances.
These topics were first recognized and discussed in Refs.~\cite{Farley:1972zy,Farley:1990gm}, where the pitch correction $C_p$ is derived as 
\begin{equation}
  C_p \approx \frac{n}{2} \frac{\langle y^2\rangle} {R_0^2} = \frac{n}{4} \frac{\langle A^2\rangle} {R_0^2}\, .
  \label{eq:pitch_cor}
\end{equation}
The vertical oscillation amplitude $A$ can be extracted from measurements by the tracker detector system. 
The validity of Eq.~\ref{eq:pitch_cor} has subsequently been confirmed and explored further~\cite{Field:1974pe,Silenko_2006,Metodiev:2015twa}.
Appendix~\ref{ap:EpitchNumeric} describes our numerical simulations that also reaffirm Eq.~\ref{eq:pitch_cor} and furthermore permit modeling of the uncertainty on $C_p$ owing to the use of flat electrodes, ESQ plate misalignment, and voltage errors.

\subsection{Measuring the muon vertical distribution and calculating $C_p$}

\begin{figure}[h!]
 \begin{center}
\includegraphics[width=\columnwidth]{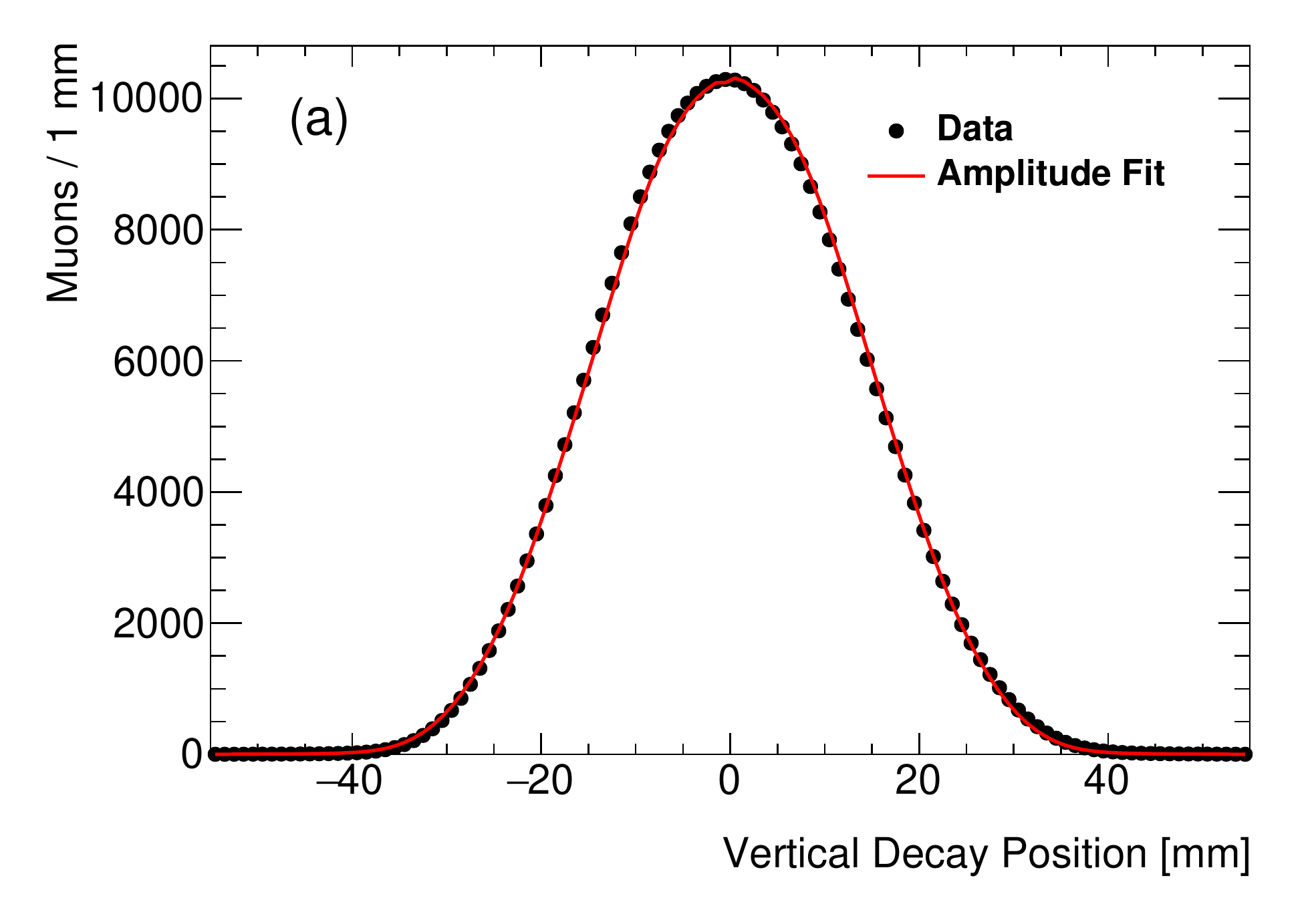}
\includegraphics[width=\columnwidth]{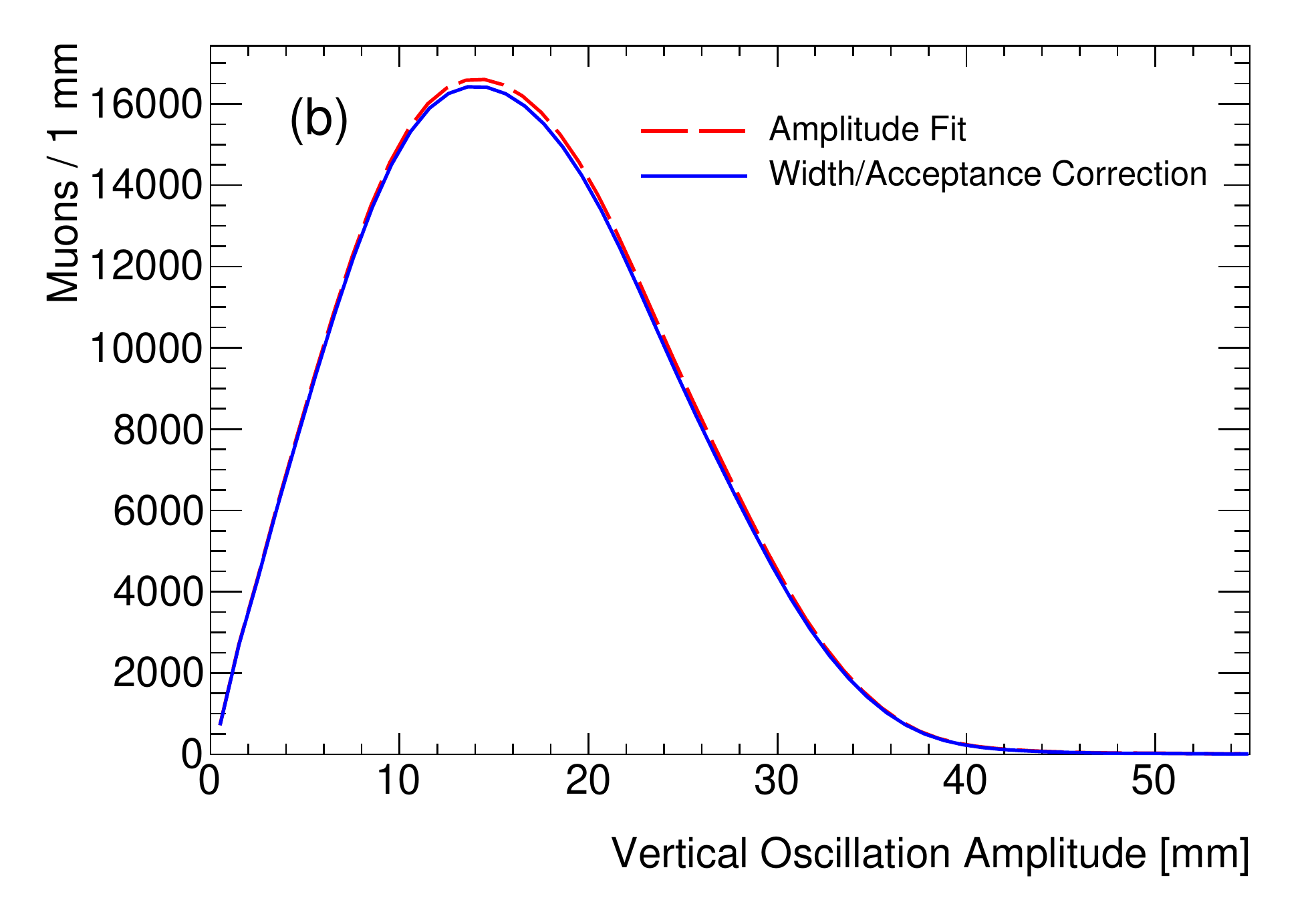}
\caption {(a) The vertical distribution of muons from a subset
  of \runonea after applying tracker resolution and acceptance
  corrections. (b) The fitted distribution of vertical oscillation
  amplitudes, before (dashed red line) and after (solid blue line) the azimuthal averaging and calorimeter
  acceptance corrections described in the text.}
\label{fig:Pitch-a-y}
 \end{center}
\end{figure}

The vertical distribution of decay positrons is measured by the straw
trackers, as shown for a subset of the data in
Fig.~\ref{fig:Pitch-a-y}a.
Throughout \runone, the temperature in the experimental hall slowly increased from $24.5^\circ$\,C to $28.5^\circ$\,C and exhibited typical diurnal fluctuations of roughly $1^\circ$\,C.
These changes produced a slowly varying radial component to the magnetic field, which
 caused the vertical mean of the beam to change with time.
To account for this effect, these data were subdivided into
shorter running periods, and a weighted average of their corresponding pitch
corrections was computed. When determining the vertical muon distribution,
time cuts are applied, which restricted the tracker data
to the same time interval used for the measurement of the anomalous
spin-precession frequency.

The tracker measurements yield a good estimate of the
true vertical distribution of the muon beam
at the location of the tracker stations.
However, it is necessary to take into account
azimuthal variations around the storage ring from
the discrete ESQ sections. The effective
vertical distribution seen by the calorimeters that measure \wam must
also be determined.

In order to address azimuthal variations in the vertical distribution of the
muon beam, the vertical beta functions $\beta_y$ are evaluated as a function
of azimuthal coordinate $\phi$ using \ringsim and
\COSY. By taking the ratio of $\sqrt{\beta_y(\phi)}$ with the value
at each tracker station, the scale factor which relates the vertical width at
each tracker station to any other azimuthal coordinate is obtained. The
vertical distribution from a single tracker station
is then averaged over azimuth by
stretching or shrinking the width ratio
in each azimuthal slice using $\sqrt {\beta_y(\phi)}$.

Not all decay positrons hit a calorimeter and enter into the determination of
\wam.  To account for this, each calorimeter's acceptance has been estimated as
a function of transverse and azimuthal decay position using \ringsim (see Fig.~\ref{fig:acceptance-maps}).
In each azimuthal slice of the storage ring, the transverse
acceptance function is integrated over the radial dimension to produce an effective
vertical acceptance function. Then, during the azimuthal averaging procedure,
the vertical distribution in each azimuthal slice is masked by the
corresponding calorimeter acceptance function after stretching by the
vertical width ratio. The nominal results use the acceptance functions for
all calorimeters combined, treating the acceptance per calorimeter as a
cross-check. Furthermore, the calorimeter acceptance functions are subdivided
by positron energy bin. Therefore, the pitch correction is evaluated using
calorimeter acceptance from each positron energy bin, and the results are
averaged according to the statistical power of \wam in each energy bin.

When considering calorimeter acceptance, the pitch correction can no longer be
evaluated using $\langle y^2 \rangle$ as in Eq.~\ref{eq:pitch_cor}. This is because
the simple relation between the pitch angle and $\langle y^2 \rangle$
along an oscillation breaks down when the vertical positions are not evenly weighted.
Instead,  the right-hand side of Eq.~\ref{eq:pitch_cor} is used with $\langle A^2\rangle$
from the distribution of oscillation amplitudes, which accurately describes the distribution
of measured pitch angles when vertical acceptance is present. The amplitude distribution
may be recovered from the trackers' vertical decay distribution by defining
the fit function
\begin{equation}
\langle N_j^y \rangle = \sum_i N_i^A P(y_j | A_i)\, ,
\end{equation}
where $\langle N_j^y \rangle$ is the expected number of decays in the
$j$th position bin, $N_i^A$ is a fit parameter for the
number of muons in the $i$th amplitude bin, and $P(y_j | A_i)$ is the calculable constant probability that a muon from the
$i$th amplitude bin decays to the $j$th
position bin. The amplitude distribution is extracted from a fit to tracker
measurements of the vertical counts in each bin, $N_j^y$, after correcting for the intrinsic tracker acceptance and resolution. The fit result and
corresponding amplitude distribution are shown in Figs.~\ref{fig:Pitch-a-y}a and \ref{fig:Pitch-a-y}b, respectively.
The amplitude distribution is stretched and averaged over azimuth as described above,
and each amplitude bin $A_i$ is weighted by the average of the calorimeter
acceptance over all position bins $y_j$ using $P(y_j|A_i)$. The result of these corrections is shown by the solid blue line in Fig.~\ref{fig:Pitch-a-y}b.
Finally, the pitch correction is calculated using Eq.~\ref{eq:pitch_cor} with $\langle A^2 \rangle$.

Since the corrections have a small effect, the statistical uncertainty of $C_p$ can be estimated using the statistical uncertainty of
the measured vertical width $\sigma_y$, propagated to $C_p \propto \sigma_y^2$. The dominant systematic uncertainty is from the
straw-tracker
alignment and reconstruction.  There are
smaller contributions from the acceptance
and resolution corrections.
Other possible sources of uncertainty come from the estimation procedure for $C_p$ described above, as well as possible errors in alignment and calibration of the ESQ system that are described in detail in Appendix~\ref{ap:EpitchNumeric}.

\begin{table}[h]
\begin{ruledtabular}
\begin{tabular}{rrrrr}
Dataset& \runonea & \runoneb & \runonec & \runoned \\
\hline
$C_p$ (ppb) & 176 & 199 & 191 & 166 \\
\hline
Stat. uncertainty  & $<1$ & $<1$ & $<1$ & $<1$ \\
\hline
Tracker reco. & 11 & 12 & 12 & 11 \\
Tracker res. and acc.  & 3 & 4 & 4 & 3 \\
$\beta_y(\phi)$ and calo. acc. & 1 & 1 & 2 & 1 \\
Amplitude fit & 1 & $<1$ & 1 & 3 \\
ESQ calibration & 4 & 4 & 4 & 4 \\
\hline
Syst. uncertainty & 12 & 14 & 14 & 12 \\
\end{tabular}
\end{ruledtabular}
\caption{Pitch corrections $C_p$ (ppb) and uncertainties (ppb) for each run group in \runone.  The total systematic uncertainty is the quadrature sum of the individual contributions.}
\label{Tab:PitchCorrection}
\end{table}

The resulting values for $C_p$ are summarized in
Table~\ref{Tab:PitchCorrection}.  The corrections vary from 166 to 199\,ppb
with the main driver behind the range being the different ESQ settings
for the different datasets. Differences with the same $n$ value arise due to
different radial magnetic fields.  The statistical uncertainty is negligible,
and the systematic uncertainty is well under control at the 12--14\,ppb level.
 
\section{\label{sec:time_changing_fields}Dynamic Effects Owing To Time-changing Fields}

\subsection{\label{sec:time_changing_cbo}Changes to the betatron frequencies}

\begin{figure}[htp]
  \begin{center}
  \includegraphics[width=\columnwidth]{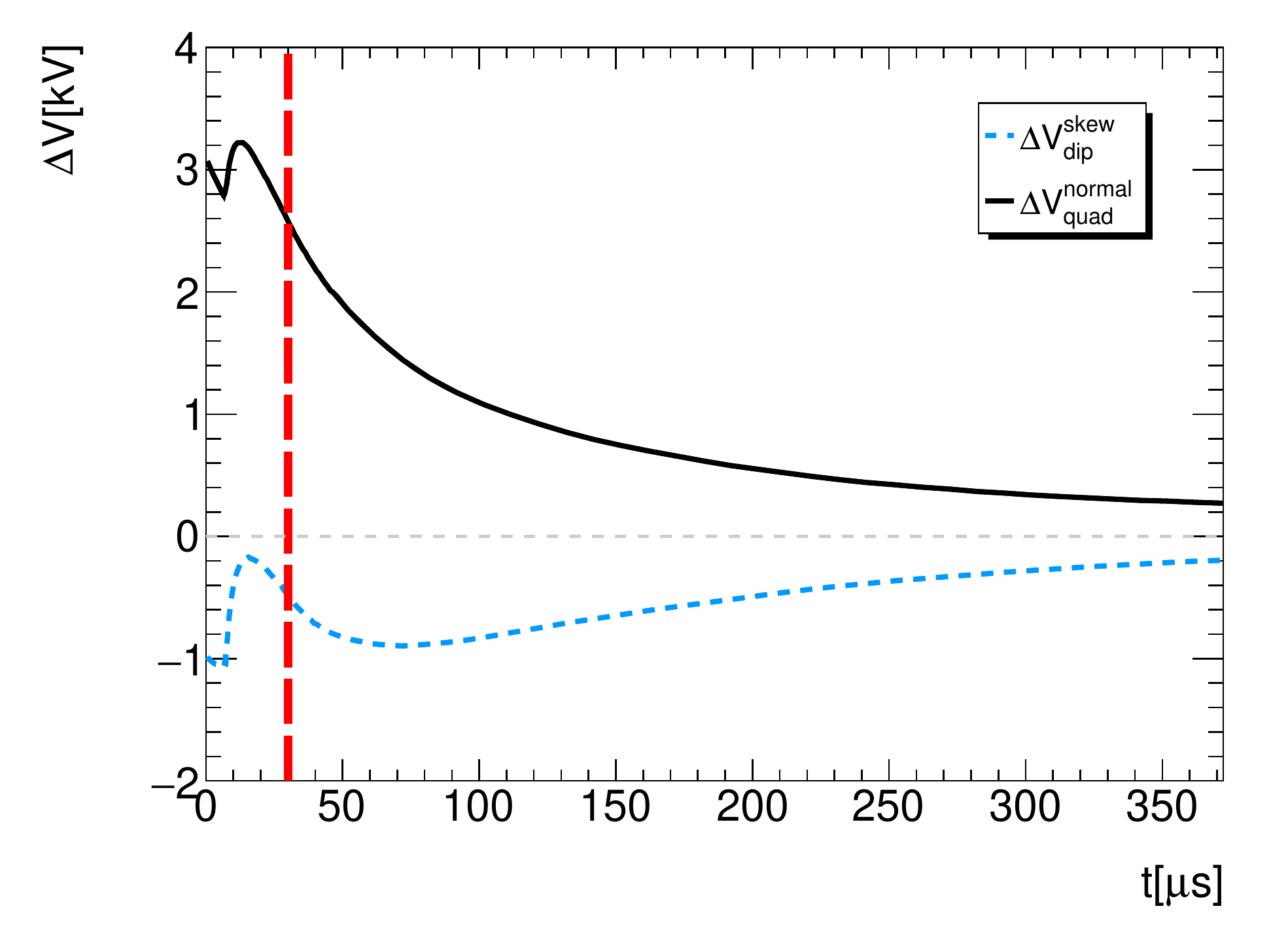}
  \end{center}
  \caption{Additional electrostatic potential at $r=5$\,cm versus time for the ESQ powered through damaged resistors. The skew dipole (dashed blue line) and normal quadrupole (solid black line) terms are shown relative to the nominal case and correspond to a 2D Taylor expansion based on data collected after \runone using an HV probe. The vertical dashed red line at \SI{30}{\micro\second} after injection indicates the nominal start of the precession fits. Scraping ends at \SI{7}{\micro\second} after injection, visible at the abrupt kinks in the curves. Nominal ESQ plates rise with a nominal $RC$ time constant of $\tau\approx 5$\,\SI{}{\micro\second}, but these ESQ plates show prolonged and asymmetric time constants which lead to focusing and vertical steering errors during the data-taking period.}
  \label{fig:Q1L-br}
\end{figure}

The slower voltage increase on the Q1L upper and lower ESQ plates as a result of the damaged resistors in \runone\ introduces time dependencies of the storage ring lattice parameters (see the dotted blue and solid orange traces in Fig.~\ref{fig:goodbadresistors}). Electric normal quadrupole and skew dipole terms are largely proportional to the sum and difference in high voltage between the top and bottom electrodes, respectively, making the beta functions, radial dispersion function, and the closed orbits time-dependent during the measurement period. Figure~\ref{fig:Q1L-br} illustrates the difference in the electrostatic potential in the Q1L ESQ section versus time in fill with respect to the nominal case for all other ESQ sections, which had normal resistors  and stable voltages by \SI{30}{\micro\second}.
The beta functions are a consequence of the focusing gradient configuration in the ring, which depends on the normal quadrupole terms at each ESQ section. The solid black line in Fig.~\ref{fig:Q1L-br} illustrates the added quadrupole field at Q1L that introduces a time dependence to the beta functions. The vertical closed orbit is distorted by skew dipole terms from the guide fields, the time dependence of which corresponds to the value of the dashed blue line in Fig.~\ref{fig:Q1L-br}.

\begin{figure}[t]
  \begin{center}
  \includegraphics[width=\columnwidth]{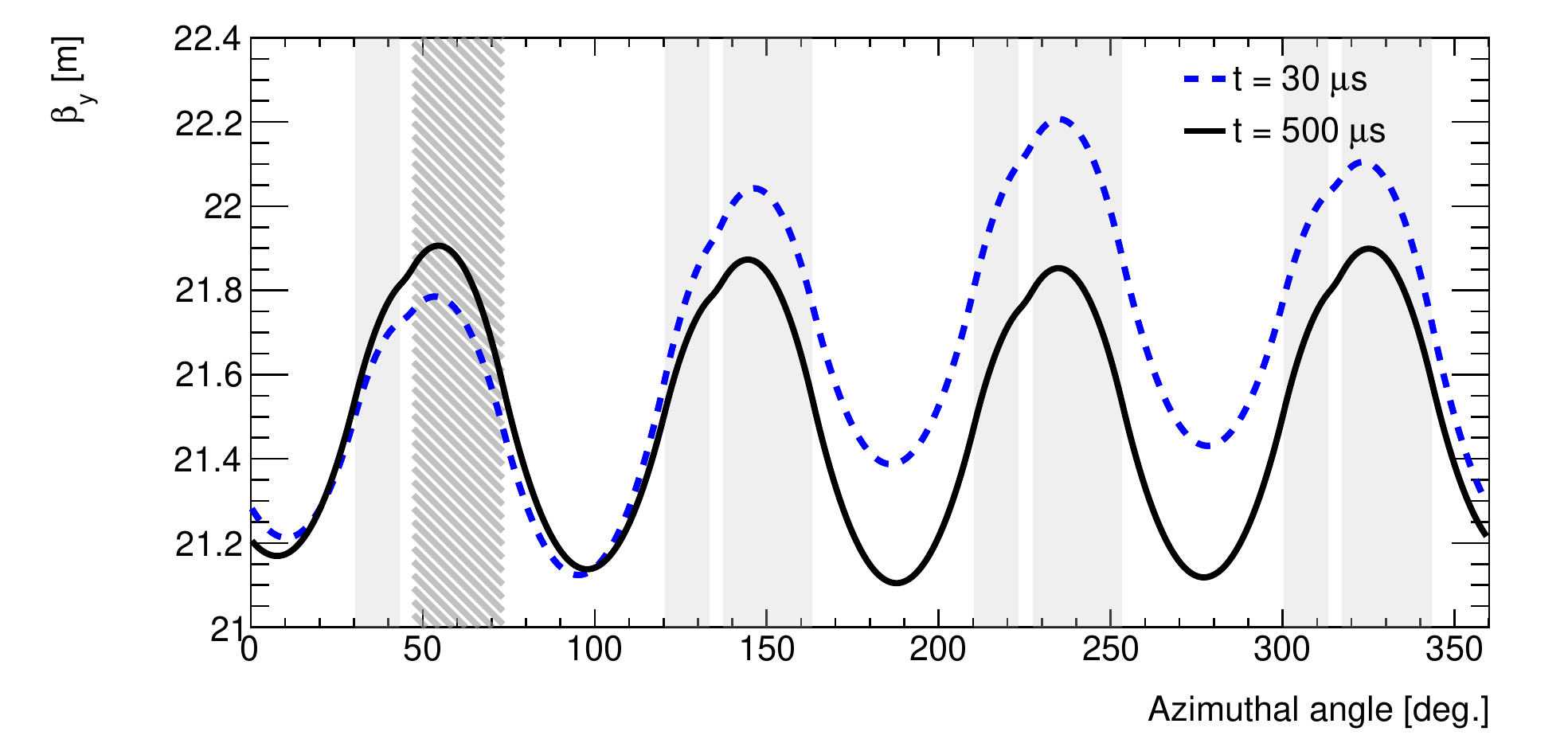}
  \includegraphics[width=\columnwidth]{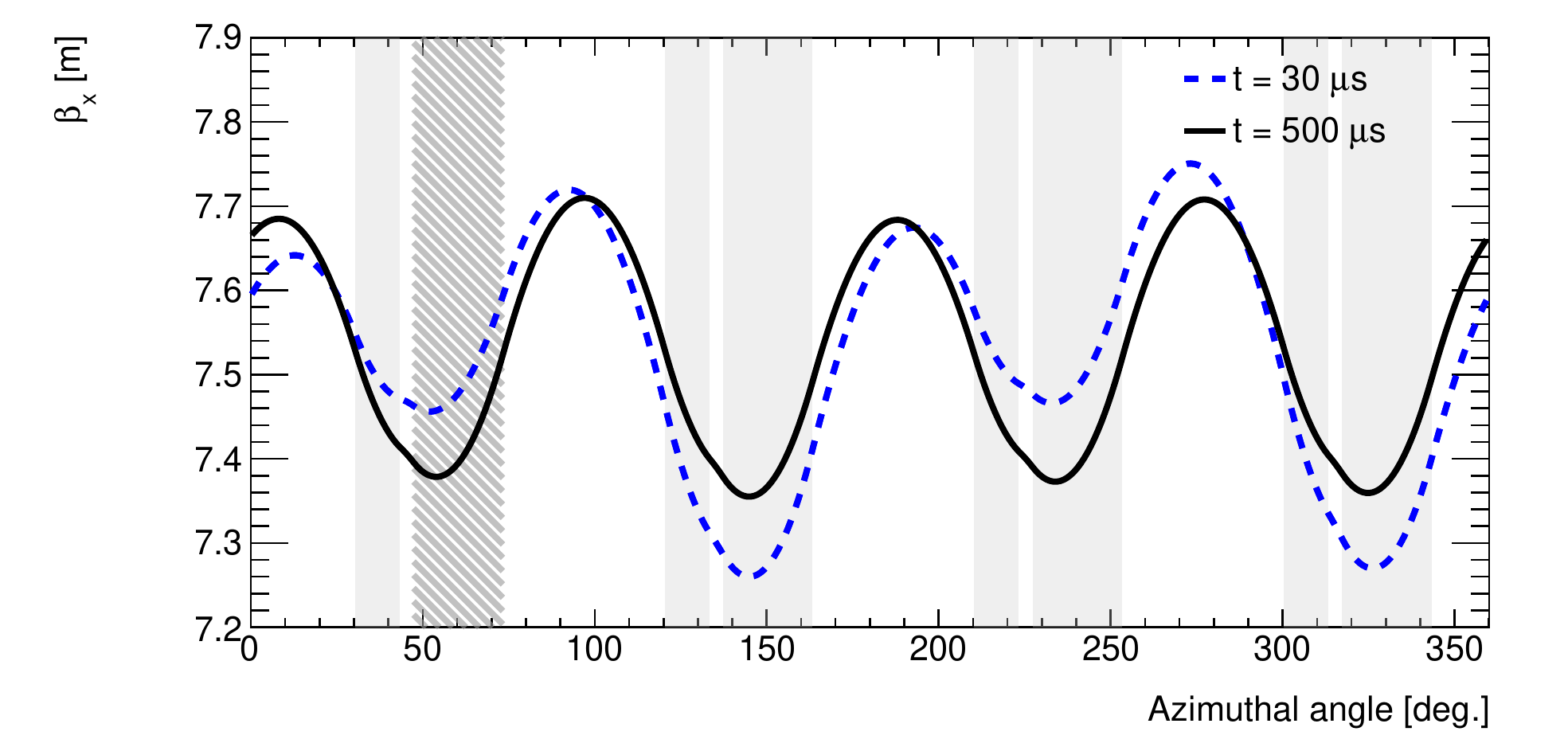}
 \includegraphics[width=\columnwidth]{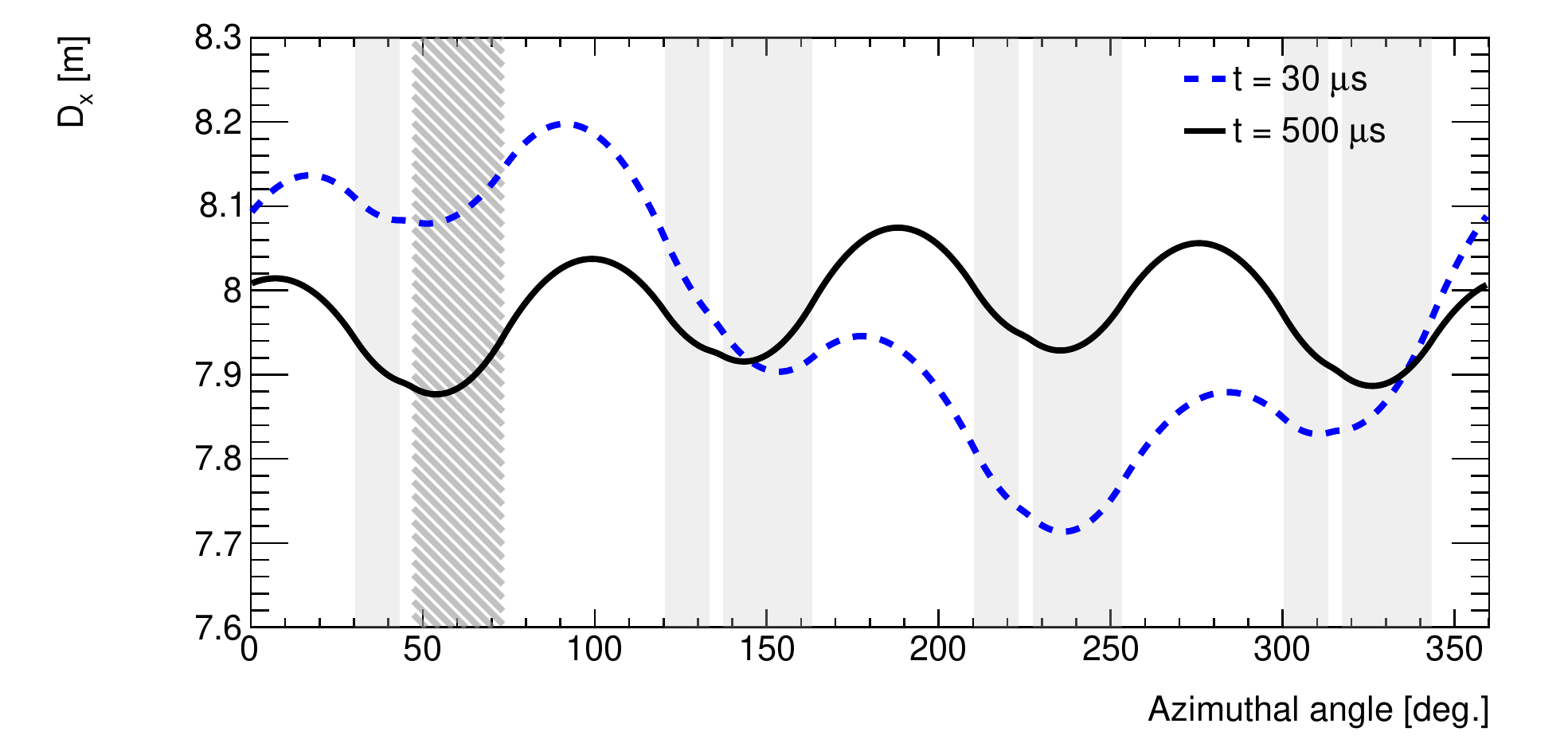}
  \end{center}
  \caption{The calculated $\beta_y$ (top), $\beta_x$ (middle), and dispersion $D_x$ (bottom) functions versus azimuth in the ring at times 30 and \SI{500}{\micro\second} after injection for the \runonea dataset. The calculation includes the effect of the damaged resistors used in the Q1L ESQ section (hatched region) and the measured magnetic field distortions versus azimuth.
The vertical shaded regions  correspond to the locations in azimuth of the short and long ESQ sections.}
  \label{fig:betay}
\end{figure}

Figure~\ref{fig:betay}  shows the calculated beta functions $\beta_y$ and $\beta_x$ and the dispersive function $D_x$  for \runonea early and late in the fill; the latter is well after the ESQ voltages are at their intended values. The vertical shaded regions in this figure correspond to the locations in azimuth of the short and long ESQ sections, with Q1S at $\phi=30^\circ$. The dynamic functions match the time dependence of $\omega_\text{CBO}$ and the vertical width (VW) frequency associated with the vertical breathing of the muon beam, and the slow changes of vertical beam mean $y$ and width $\sqrt{\langle y^2 \rangle}$.

The tracker detectors are capable of reconstructing the stored muon distribution at different times within the fill and are, therefore, used to measure the betatron oscillation parameters as well as any slow drift of the beam position or width.  The betatron frequencies can be determined to high precision: In fact, it was analysis of tracker data that first drew attention to a possible time dependence of the electric quadrupole field.  The measured betatron frequencies are necessary to verify the tune, and the measured betatron amplitudes are used in the optimization of kick strength and inflector deflection angle.

After the end of scraping, with stable voltages during the measuring period, $\omega_{\text{CBO}}$ and $\omega_{\text{VW}}$ should not change.  However, in \runone, they continued to evolve after 30\,\si{\micro\second} by 1.5\% in Runs-1a, 1b, and 1c and by 3.0\% in \runoned.

\begin{figure}[htp]
\includegraphics[width=\columnwidth]{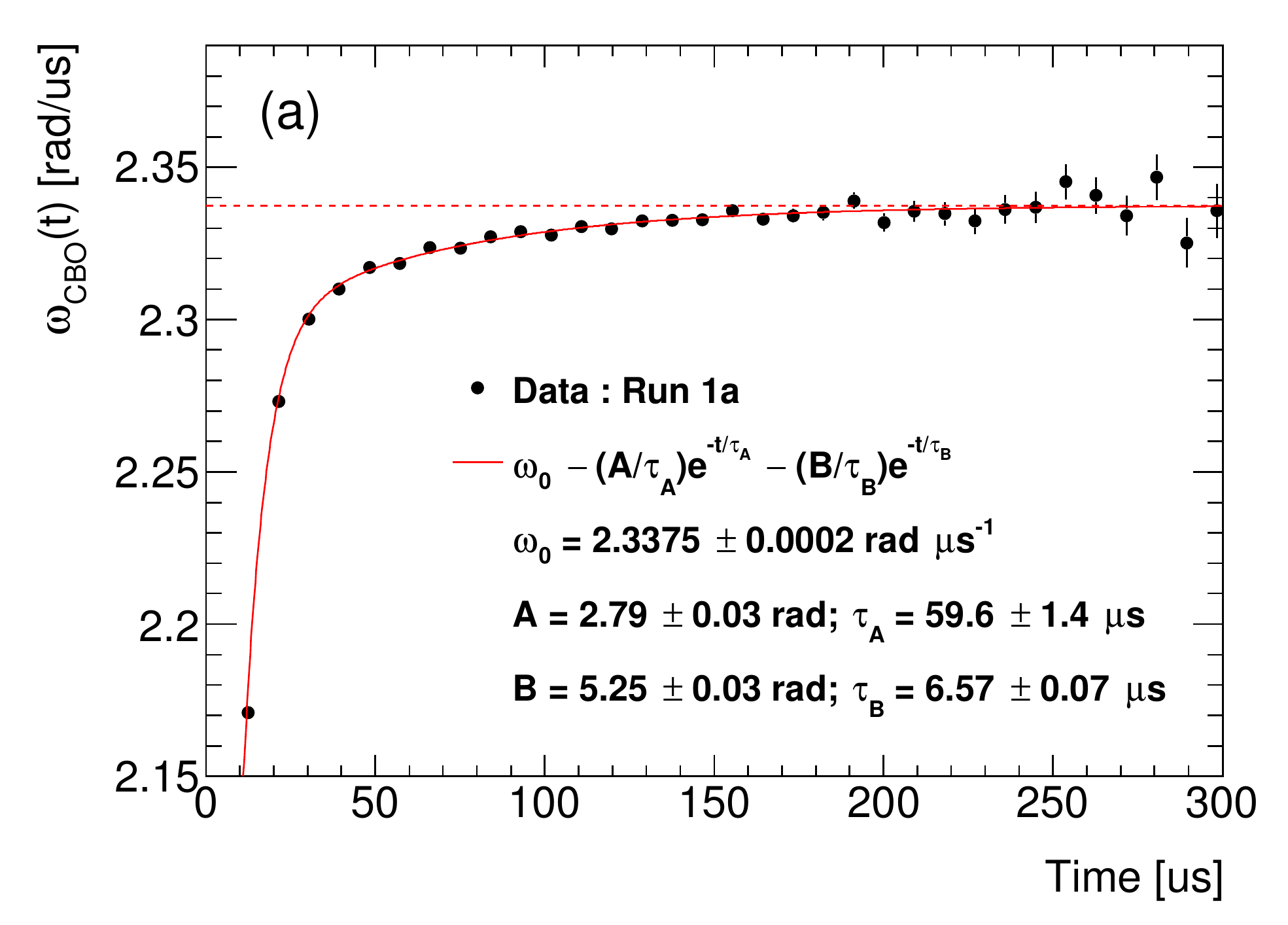}
\includegraphics[width=\columnwidth]{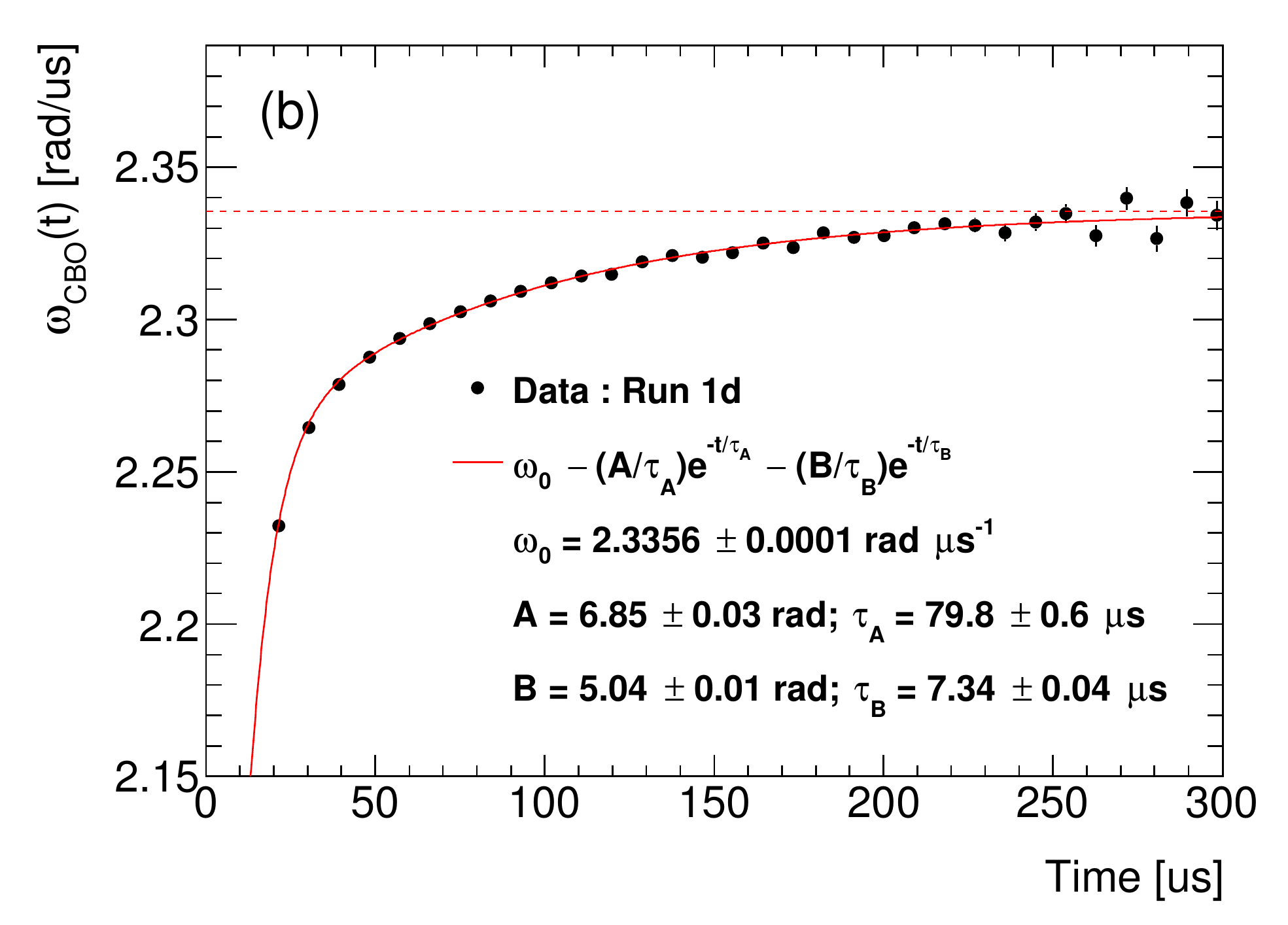}
  \caption{Tracker measurements of $\omega_{\text{CBO}}$ in different time slices during the muon fill from the 270\si{\degree} tracker station.  (a) and (b) show the \runonea and \runoned datasets, respectively.  The fit function and parameters are noted in the figure, together with their uncertainties. The difference between the two datasets is attributed to worsening performance of the damaged ESQ resistors.  The time dependence of the frequency is included in the fits of the positron data from the calorimeters to accurately incorporate the CBO acceptance dependence.}
  \label{fig:cbovstime}
\end{figure}

Figure~\ref{fig:cbovstime} shows tracker measurements of $\omega_{\text{CBO}}$ for Runs-1a and 1d. The fit function contains two exponential terms that describe $\omega_{\text{CBO}}(t)$ well.  The first of these is a fast (${\sim}7$\,\si{\micro\second}) term, which relates to the changing field during scraping.  The second term has a longer time constant of ${\sim}60$\,\si{\micro\second} in Runs-1a, 1b, and 1c and ${\sim}80$\,\si{\micro\second} in \runoned.  This term arises from the damaged resistors and the change in their average value during \runoned due to their further deterioration.

The fits to the precession data require an accurate model of $\omega_{\text{CBO}}(t)$ to obtain a good $\chi^2$ and stable fit results when changing the fit start time.  We note that the correlation between \wam and $\omega_{\text{CBO}}$ in the fit is small (${\sim}2.4\%$).  Varying the time dependence of the CBO frequency within allowable bands determined from the tracker measurements yields a less than 10\,ppb shift to the fitted \wam frequency.  The effect from the VW is even smaller than that from the CBO, and it has a negligible effect on the precession fits.

\subsection{Changes to the muon spatial distribution}
\begin{figure}[htbp]
\includegraphics[width=\columnwidth]{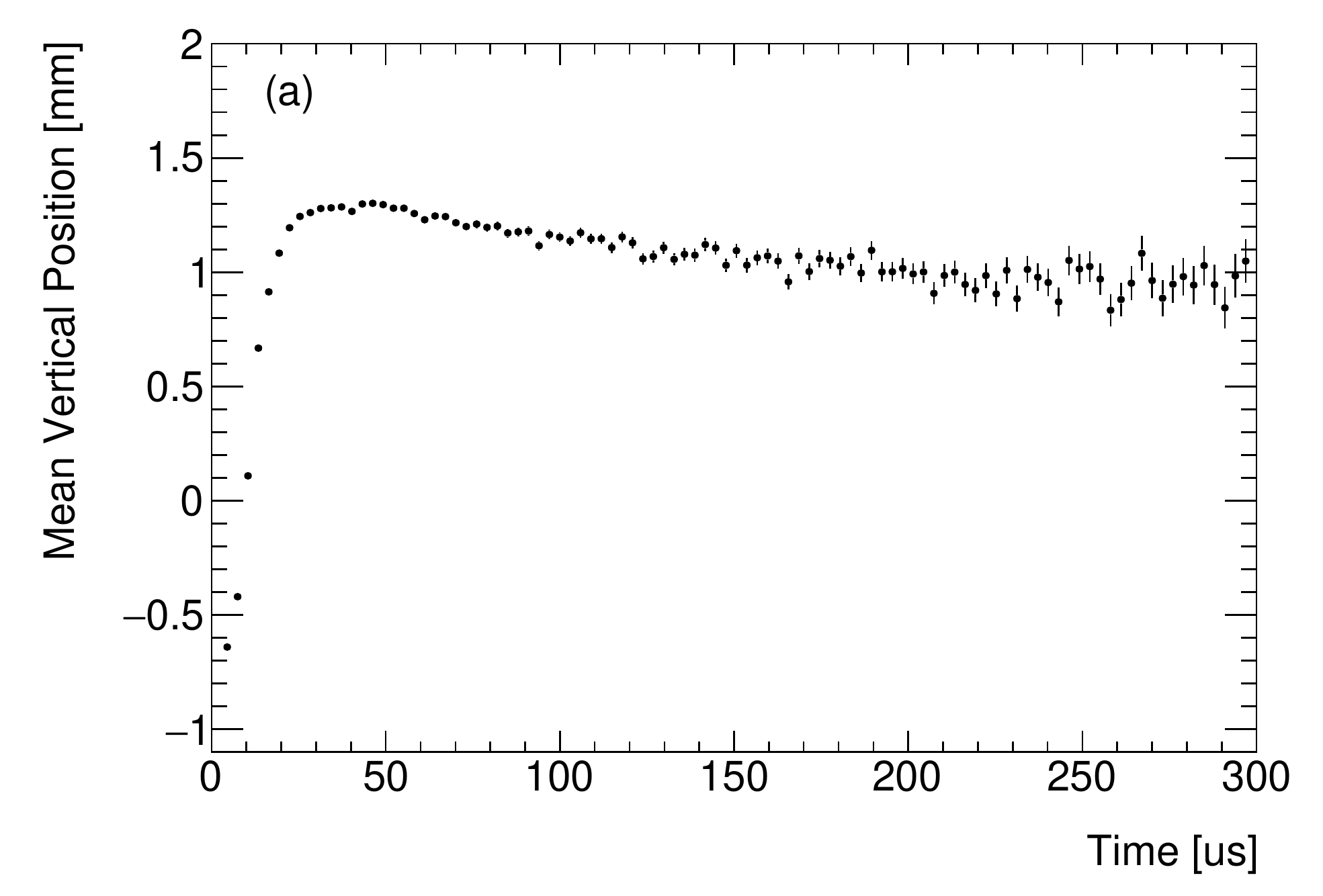}
\includegraphics[width=\columnwidth]{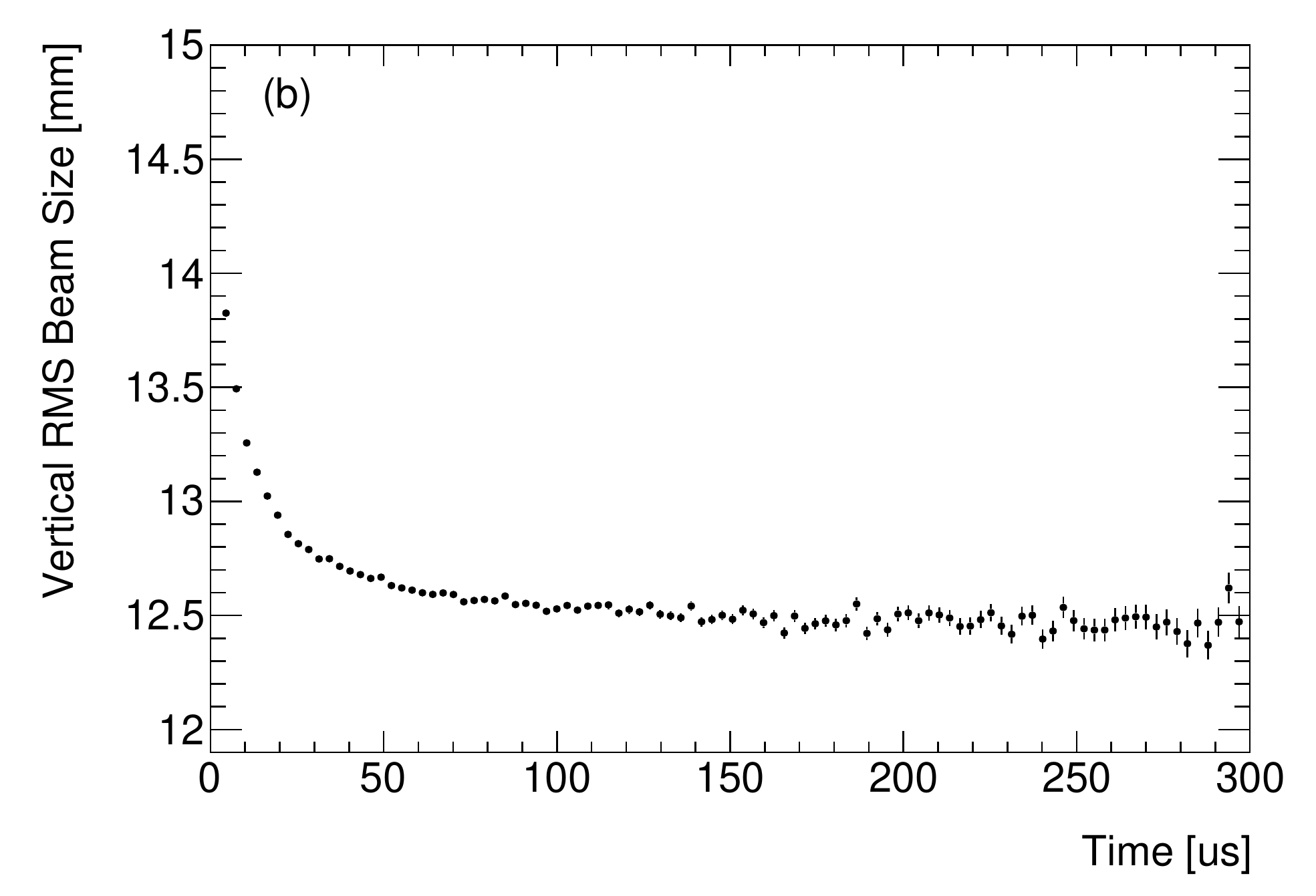}
\caption{The 180\si{\degree} tracker station determination of the vertical mean (a) and width (b) versus time in fill for \runoned. The rapid oscillations owing to vertical and horizontal betatron motion have been randomized out to reveal the underlying time dependence of the mean and width (see also Sec.~\ref{sec:randomization}).}
\label{fig:Tracker-Ymean-Yrms}
\end{figure}

Because the damaged resistors were connected to the same Q1L plates, and because the circuitry at station Q1L had $RC$ characteristics that differed from the nominal charging conditions (see Fig.~\ref{fig:goodbadresistors}), we observe a physical effect on the mean and width of the muon distribution.
The asymmetrical charging of the top and bottom plates in Q1L created an unbalanced quadrupole component of the ESQ field.
This led to a time-dependent focusing gradient, resulting in submillimeter drifts in the beam widths and radial closed-orbit distortions from the time-dependent optical lattice. The vertical closed orbit also manifested an in-fill temporal evolution owing to an introduced skew dipole field at Q1L.
Figure~\ref{fig:Tracker-Ymean-Yrms} displays the time-varying vertical mean and rms measured by the 180\si{\degree} tracker for the \runoned dataset, which is by far the worst case.
In the range $30 - 300$\,\SI{}{\micro\second}, the vertical mean shifts by approximately 0.5\,mm. Note that the absolute vertical scale is not known to better than ${\sim}1$\,mm  owing to alignment uncertainty and the local radial magnetic field.
The vertical rms versus time in fill is unstable at the nominal fit start time of \SI{30}{\micro\second} but by \SI{50}{\micro\second} has flattened out, which is when the measurement period for this dataset starts.
As noted previously, $\beta_y$, which is proportional to $\langle y^2 \rangle$, varies around the ring, so one does not expect the magnitude of the width change to be constant versus azimuth.

\begin{figure}[htp]
  \begin{center}
  \includegraphics[width=\columnwidth]{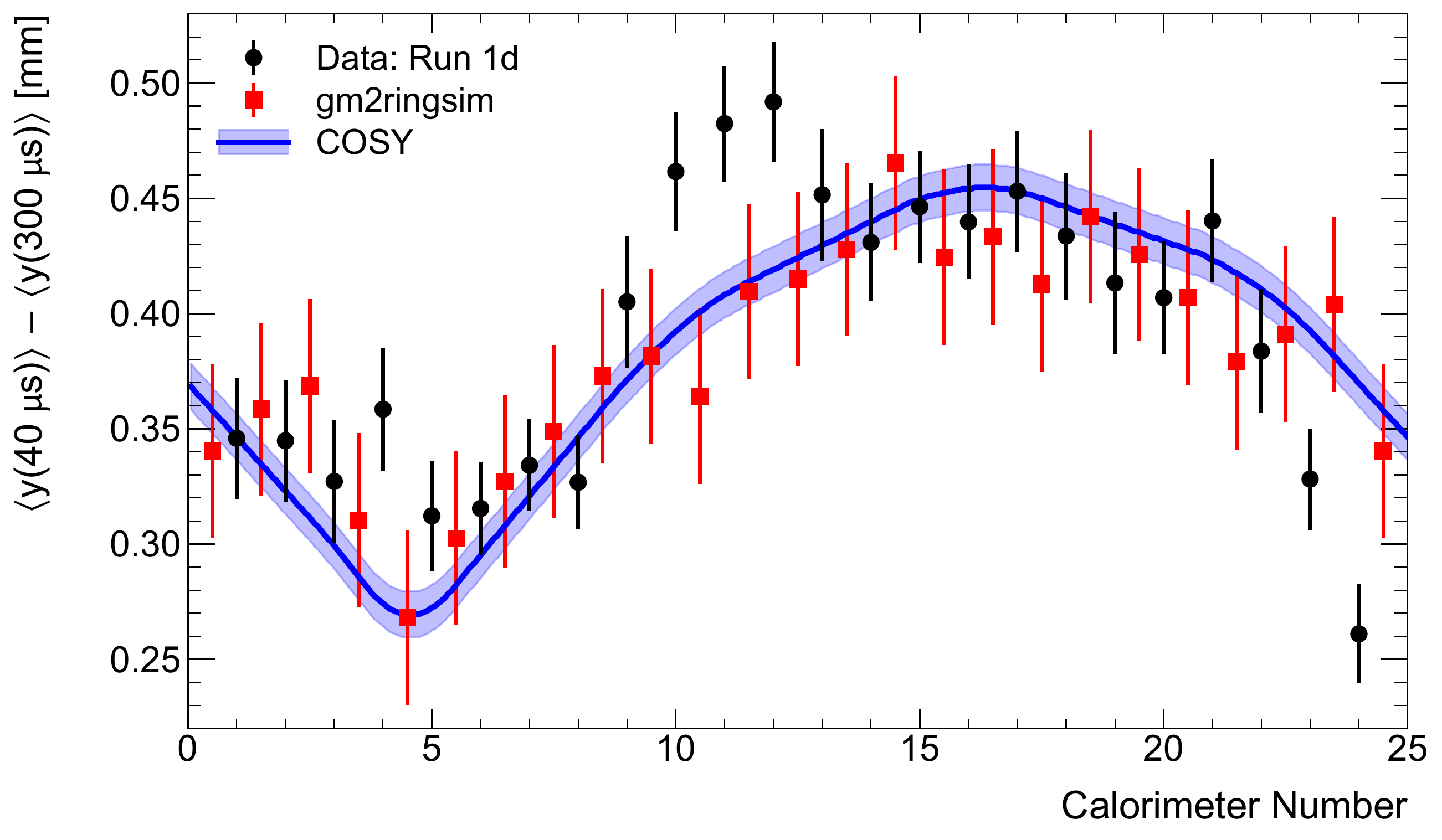}
  \end{center}
\caption{The change in the vertical mean over the time period from 40 to 300\,\SI{}{\micro\second} in a fill during \runoned as measured by the 24 calorimeter stations.  The calorimeter data are corrected for acceptance. The blue line corresponds to the \COSY model of the vertical closed-orbit evolution during the fill and its prediction of the vertical mean change around the ring. The red squares are from the \ringsim muon tracking simulation. This implementation of \COSY does not utilize beam tracking, so the error band is purely systematic, while the \ringsim errors are statistical.
Both simulations have been anchored to data from the tracking station at $270^\circ$ and have been shifted azimuthally to align with the maximum acceptance for each calorimeter.  The azimuthal structure predicted by the simulations due to the damaged resistors is clearly visible in the data.}
\label{fig:DeltaYvsTime}
\end{figure}

The tracker station measurements are limited to the 180\si{\degree} and 270\si{\degree} locations in the storage ring. However, the 24 calorimeter stations are positioned at regular intervals covering the complete azimuth.  The segmented calorimeters provide a measure of vertical mean versus time in fill at each location. The \COSY model of the vertical closed-orbit evolution during the fill predicts the vertical mean change around the ring.  Similarly, \ringsim can determine the vertical mean change by using internal virtual tracking planes at different azimuthal locations.
Figure~\ref{fig:DeltaYvsTime} shows the calorimeter measurements of the vertical mean change from 40 to 300\,\SI{}{\micro\second} and the corresponding predictions of the two simulation programs. The azimuthal variation in the data supports the implementation of the damaged resistors in the simulation and the projection of the dynamics at all azimuthal locations needed for the  phase-acceptance correction discussed in Sec.~\ref{sec:phase-momentum}.

 \section{\label{sec:muon_loss_phase}Muon Loss Correction $C_{ml}$}
Several driving mechanisms can lead to loss of muons during storage. For example, the scattering of particles off of the residual gas in the vacuum chamber, noise from residual high-frequency electromagnetic fields in the system, the sampling of nonlinear fields near the aperture, and nonlinear resonances are potential mechanisms.  In \runone, the muon loss rate was higher than expected owing to a combination of factors including the damaged resistors and the nonoptimized kick.  However, measurements still show integrated loss rates of less than a percent during the fill.

\begin{figure}[tb]
\includegraphics[width=\columnwidth]{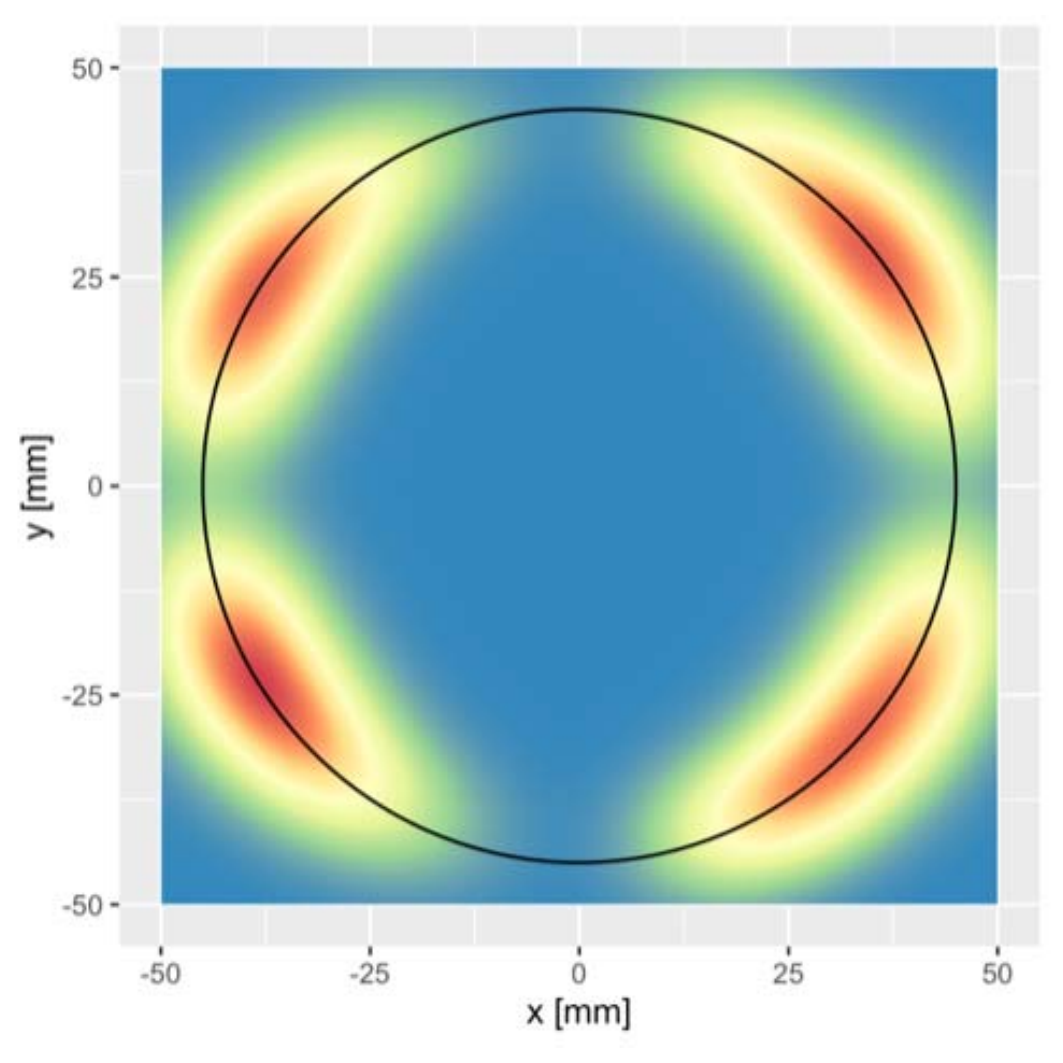}
\caption{The intensity distribution of where muons first strike the $r_0$ = 45\,mm radius collimators (black circle). For any particular muon, this will occur when its horizontal and vertical betatron oscillations conspire such that the transverse displacement is at $r_0$ and, at the same time, it is at the azimuthal location in the storage ring of a collimator.   Some muons, which will eventually be lost, can survive hundreds to thousands of turns before this condition is met.}
\label{fig:MLcollimatorhits}
\end{figure}

In general, a muon will be scattered out of the storage region after it strikes one of a set of circular collimated apertures that limit the transverse phase space admittance and its momentum dependence. These collimators have an aperture of radius $r_0 = 45\,$mm and are centered on the ideal orbit.  Figure~\ref{fig:MLcollimatorhits} shows where on the collimators these muons strike first, a consequence of circular apertures and normal amplitude distributions.

Monte Carlo beam line simulations and simple analytical calculations both predict that a correlation exists between the injected muon average spin phase and the particle momentum.
If such a spin-momentum correlation exists, muons that permanently escape from the storage volume during data taking
can potentially bias \wam by inducing slow drifts in the phase.
The $\varphi_0$ fit parameter in Eq.~\ref{eq:wiggle_func} depends on the average initial spin orientation of the muons that produce the detected decay positrons. In this context, an ensemble of muons is said to have a spin-momentum correlation if $d \varphi_{0} / d\langle p\rangle \neq 0$, where $\langle p\rangle$ is the mean value of the muon momentum for the ensemble of decaying muons that produces the positron spectra being used to measure \wam. The population of stored muons is depleted only by decays, while the population of muons that will be lost is depleted at a faster rate due to decays and losses.
The stored and lost muon populations have different momentum distributions, and so the different rate of depletion creates a time-dependent average of the muon momenta: $d \langle p\rangle / d t \neq 0$. The spin-momentum correlation will combine with the time-dependent muon losses to produce a time-dependent phase:
\begin{equation}
        \frac{d\varphi_{0}}{d t} = \frac{d\varphi_{0}}{d\langle p\rangle} \frac{ d\langle p\rangle}{d t}  \neq 0.
	\label{eq:spin_corr_def}
\end{equation}

The three subsections that follow address, in turn, the following topics: (1) the data-driven determination of the absolute rate of muon losses during a fill, (2) the data- and simulation-driven determination of $d\varphi_{0} / d\langle p\rangle$ at the fit start time, and (3) the data- and simulation-driven determination of the $d\langle p\rangle / d t$ during a fill.  With these rates and correlations in hand, one can evaluate the impact on the muon loss correction factor $C_{ml}$.

\subsection{Muon loss rate determination}
\label{sec:muonlossrate}

Muons that exit the storage ring during the $30 - 650$\,\SI{}{\micro\second} measuring period deplete the population faster than the expected time-dilated decay $e^{-t/\gamma\tau_\mu}$.
The shape of the muon loss function $L(t)$ is accurately measured from the rate of coincident signals in three consecutive calorimeter stations, where each station records an energy deposit of $\approx 170$\,MeV and the time between stations is $\approx 6.4$\,ns, corresponding to the energy deposit of a minimum ionizing muon and its time of flight between stations, respectively.
The muon precession fit includes a term that multiplies the overall normalization $N_0$ such that
\begin{equation}
  N_0 \rightarrow N_0 \Lambda(t) = N_0 \left (1 - K_\text{loss}\int_{0}^{t} e^{t'/\gamma\tau_\mu}L(t')dt' \right ).
\label{eq:kloss}
\end{equation}
The scale parameter $K_\text{loss}$ can be accurately extracted from the precession analysis to provide the absolute scale of the muon loss function~\cite{precession-run1}; this is necessary to estimate the phase-related systematic error.
We note that the loss function is a property of the beam and should, therefore, be the same for all calorimeters and energy bins in the precession analyses.
An important measure of the rigor of this method is that $K_\text{loss}$ is independent of which calorimeters are used or which energy bins are selected in the precession fits.

\begin{figure}[tb]
\includegraphics[width=\columnwidth]{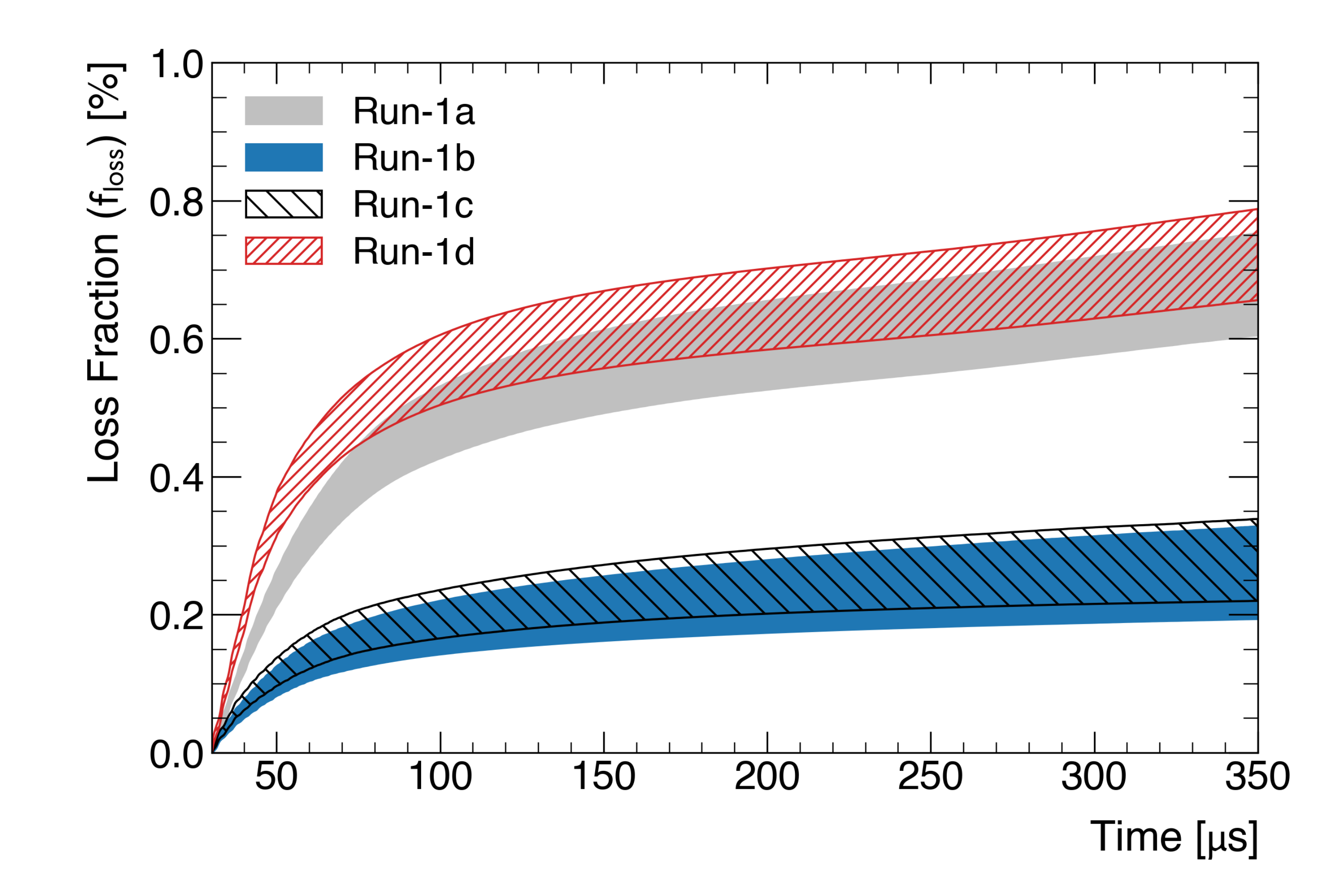}
	\caption{The integrated fractional muon loss versus time in fill from \SI{30}{\micro\second} following muon injection.  The four curves are from the different run groups.  The smaller loss fraction curves are from the $n = 0.120$ datasets (1b and 1c) and the greater loss fractions are from the $n = 0.108$ datasets (1a and 1d). The absolute scale is determined from the $K_\text{loss}$ parameter following final precession frequency fits.  The uncertainty bands on the curves come from two different precession frequency analyses and whether a small empirical slow correction term to ensure stability of $K_\text{loss}$ versus energy is included.}
\label{fig:muonlossprob}
\end{figure}

Figure~\ref{fig:muonlossprob} shows the accumulated loss fraction ($f_\text{loss}$) for the four datasets in \runone, defined for $t>t_s$ as
\begin{equation}
  f_\text{loss}(t) = \frac{K_\text{loss}}{\Lambda(t_s)}\int_{t_s}^{t} e^{t'/\gamma\tau_\mu}L(t')dt'.
\end{equation}
This gives the fraction of muons that have been lost from the storage ring with respect to the number present at the fit start time.  Although all curves rise steeply at early times and gradually at later times,  it is clear there are two distinct groups, which are associated with the two different tune $n$ values (see Table~\ref{tb:mastertable}). We note that the loss fraction in this figure is approximately 8 times larger than the loss fraction measured for \runtwo, when the damaged resistors had been replaced.  This is strong evidence that the dynamic beam motion during storage led to a high degree of scraping on the collimators at early times, until the beam relaxed to its nominal central value when the voltages had stabilized.

\subsection{\label{sec:phase-momentum}Phase-momentum correlation determination}

Nonzero spin-momentum correlations $d\varphi_{0} / d\langle p\rangle$ are generated in the Muon Campus beam line and muon storage ring.
These arise primarily during the circulation of the muons around the Delivery Ring (DR), which is composed of FODO cells and bending dipole magnets.
In particular, a dipole bending magnet will change the angle between the muon spin and momentum by $\Delta\varphi \approx a_{\mu}\gamma\eta_{b}$,
where $\eta_{b}$ is the angle at which the muon bends through the dipole field and the momentum dependence is embedded in the $\gamma$ factor.
For $k$ full revolutions of the DR, the angle between the spin and momentum will advance by $\Delta\varphi \approx 2\pi a_{\mu}\gamma k$, where the sign of phase is defined in the sense that the spin angle precesses with the functional form $\cos(\omega t + \varphi)$.
For a hypothetical muon distribution entering the DR with no phase-momentum correlation, the four turns around the DR will imprint a change in $\varphi$ of 8.6\,mrad per 1\% of $\Delta p/p_0$ on the overall distribution.

\begin{figure}[tb]
\includegraphics[width=\columnwidth]{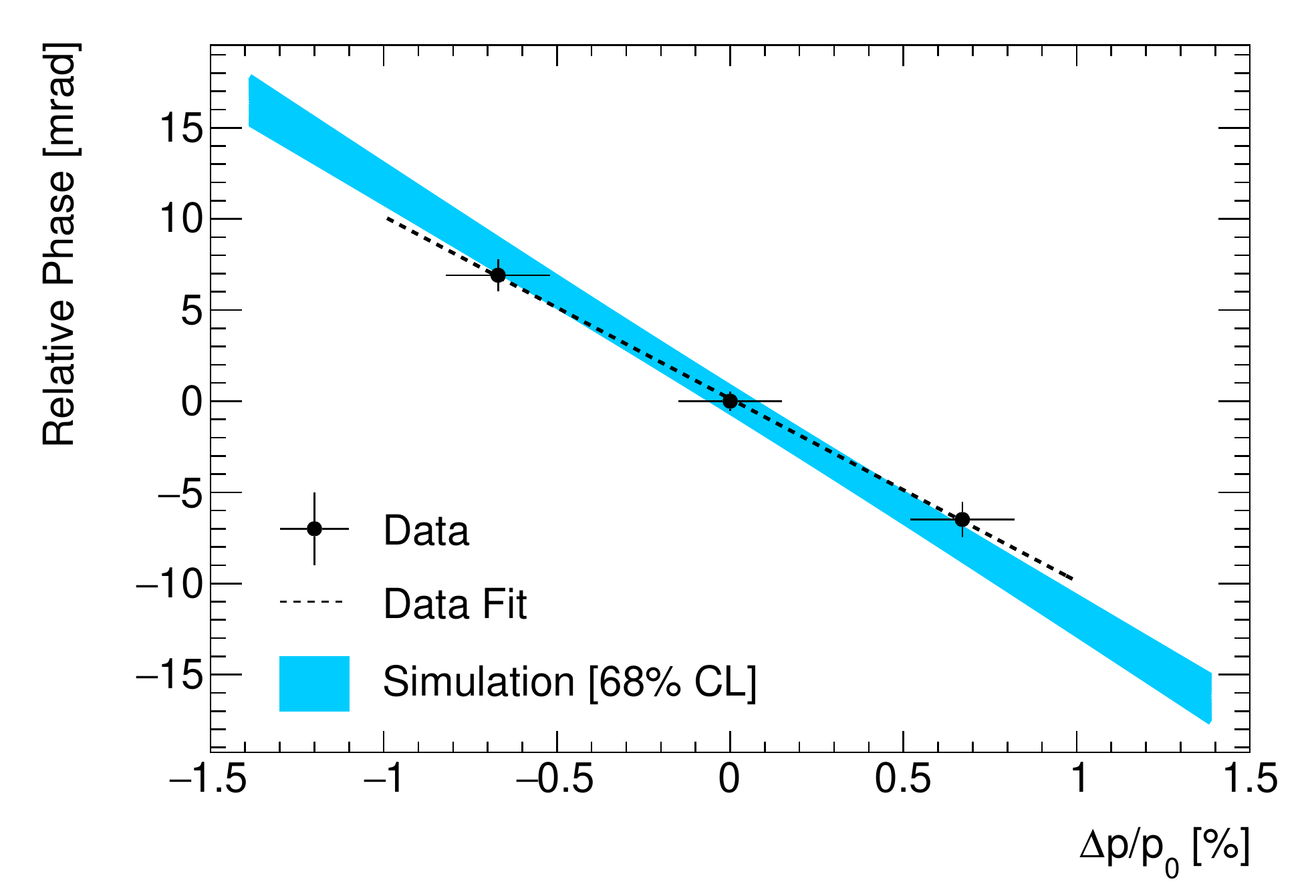}
	\caption{Phase-momentum correlation from an end-to-end simulation (blue band) and from a data-driven approach (black). The simulation gives the result at the entrance to the storage ring.  The three data points are obtained by fits to muon precession frequency data at nominal, reduced, and increased central magnetic field values.  The phase reported for these data represented muons stored and fit during the measuring period.  The phase dependence on momentum from the data is
$-10.0 \pm 1.6$~mrad/\%$\Delta p/p_0$.}
\label{fig:phase_momentum}
\end{figure}

A complete end-to-end simulation has been performed to determine the muon distribution phase space at the exit of the inflector from muons born in all distinct target and beam line regions.  The simulation tracks spin and kinematic variables from the production in the target to the delivered beam.  A plot of the average phase-momentum correlation from this simulation is shown as the blue band in Fig.~\ref{fig:phase_momentum}.

It is possible that this spin-momentum distribution is further perturbed during the storage ring kick, because, as illustrated in Fig.~\ref{fig:kicker-T0}, the kick does not apply an equal impulse to muons that are distributed longitudinally throughout the incoming bunch. Although this effect appears to introduce a negligible spin-momentum correlation in the simulation, it was possible to perform a direct measurement of the correlation that exists during the measuring period and compare it to the simulation as shown in Fig.~\ref{fig:phase_momentum}.

Three special runs were made with the magnetic field of the storage ring set at nominal (1.45\,T), reduced (-0.68\%), and increased (+0.67\%) values.    At each setting, an approximately $\pm0.15\%$ momentum slice of the broad incoming beam is stored, with its central value corresponding to the momentum of the adjusted field settings.  The statistical precision on a few hours of beam is sufficient to determine the average spin phase at injection, the precession frequency, and the time-dilated muon lifetime.   The values are obtained from fitting the positron versus time plots to Eq.~\ref{eq:simpleomega}. The change to the precession frequency is proportional to the magnetic field values and readily serves to determine the actual field (and by proxy, momentum) setting.
The black points in Fig.~\ref{fig:phase_momentum} show the results of these  direct measurements.  The fitted slope of $(-10.0 \pm 1.6)$\,mrad$/(\%\Delta p/p_0)$ agrees well with the simulation.  The error quoted is from the fit alone.

\subsection{Lost-muon momentum correlation determination}

A set of special measurements was made to determine the behavior of the muon losses as a function of time in fill and momentum.
The two damaged resistors were reinserted into their \runone\ locations during a short period of systematic tests at the beginning of \runthree. One 8-h period of data collection was acquired in otherwise nominal conditions, to reestablish the \runone\ time dependence of the CBO frequency (see Sec.~\ref{sec:time_changing_cbo}) and to provide data that could be used in a \COSY simulation of the storage ring behavior under these conditions; see below.
The Delivery Ring momentum collimators were used to bias the incoming muon momentum distribution.
The collimators can be driven separately on both the high- and low-momentum sides and can traverse the entire horizontal width of the beam.

\begin{figure}[tb]
\includegraphics[width=\columnwidth]{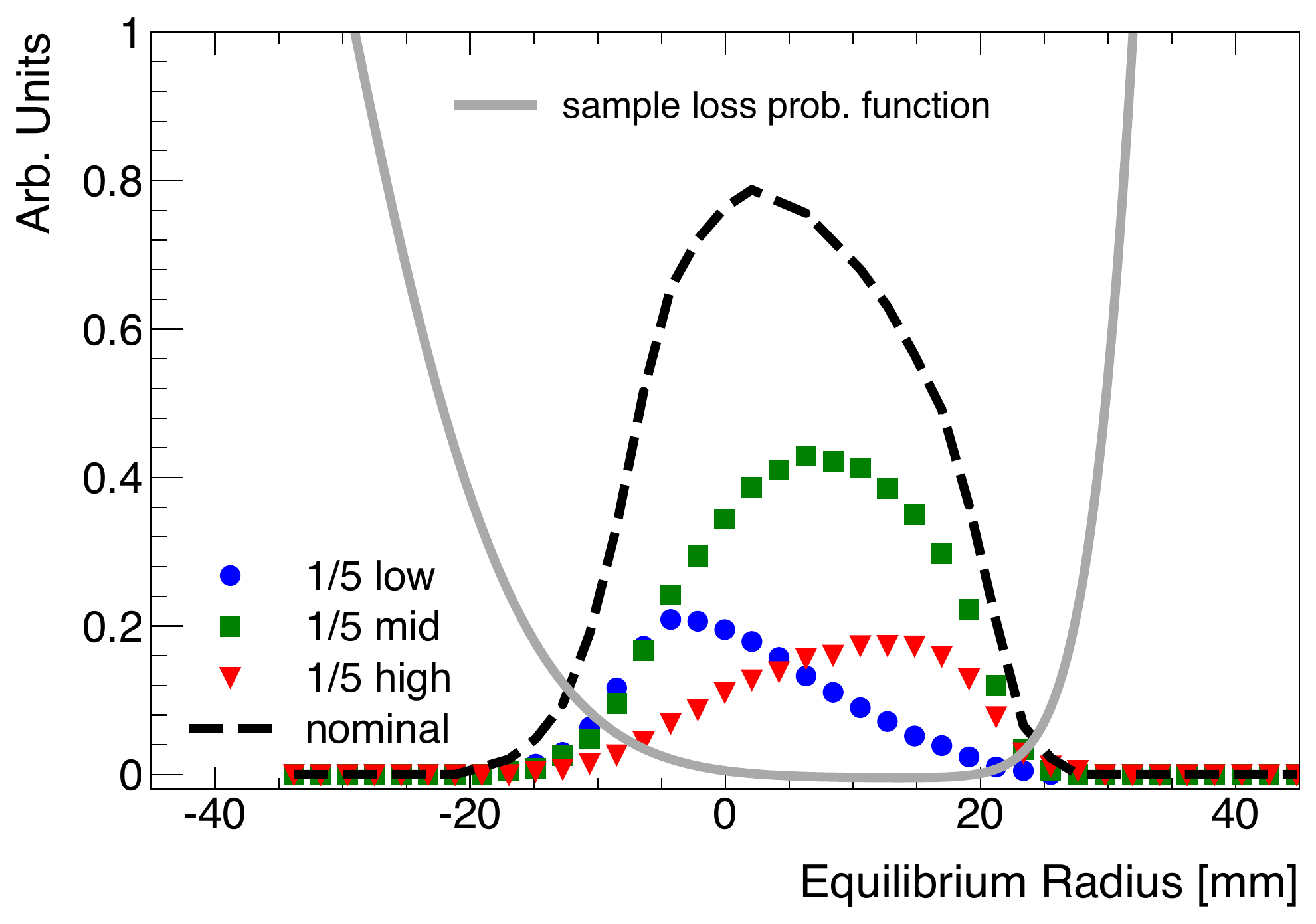}
	\caption{Four of the eight stored momentum distributions $F_i(x)$ obtained from adjusting the DR high- and low-momentum collimators. The area under each curve is normalized for each DR setting to show the fractional intensity of stored muons with respect to the nominal distribution.  The solid gray line is an illustrative muon loss probability function $l(x,t)$ from a model fit to all eight distributions for a time early in the fill.  In this time window, the loss probability is greater at low momentum than at high momentum.}
\label{fig:perturbedFRs}
\end{figure}

Figure~\ref{fig:perturbedFRs} shows horizontal stored-muon distributions $F_i(x)$ from a subset of these special runs. The dashed line corresponds to the nominal, full-acceptance distribution used for normal data taking.  The three colored distributions are from runs where the low, high, or both momentum collimators were used to bias the stored momentum distribution.  The fractions in the legend indicate the intensity with respect to the nominal case.  For example, a collimator would be moved until the muon storage rate was reduced to that fraction.
The muon loss rate versus time in fill was measured for each collimator setting, providing eight distributions (not all shown in the figure) that were used in the analysis below.

\begin{figure}[tb]
\includegraphics[width=\columnwidth]{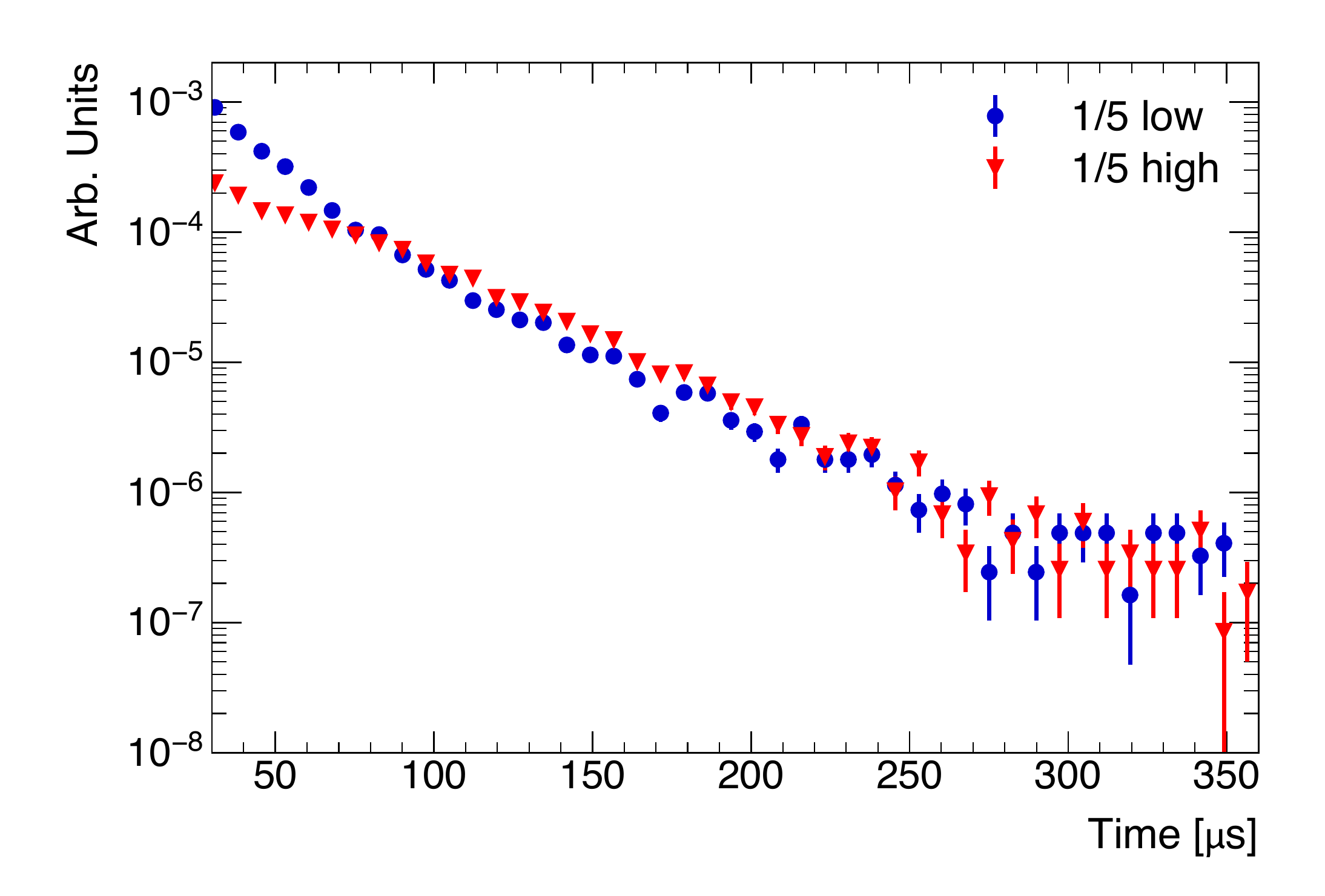}
	\caption{The muon loss function $L(t)$, normalized to beam intensity, for the lowest (blue) and highest (red) momentum stored muon distributions; see corresponding momentum distributions in Fig.~\ref{fig:perturbedFRs}.  Asymmetry in the loss rates is seen early in the fill.}
\label{fig:high-low-muonlossrates}
\end{figure}

An example of the two extremes---1/5-low (blue) and 1/5-high (red)---correspond to the relative muon loss rates shown in Fig.~\ref{fig:high-low-muonlossrates}.
The low-momentum muons are lost disproportionately early in the fill and the high-momentum muons are lost more often at later times; the other distributions provide intermediate cases.

The eight distributions were used to parameterize an analytical loss-rate function, whose form was motivated by simple simulation phase space studies. Various models were used  to assess the reliability of the conclusions from input assumptions.
The solid gray curve in Fig.~\ref{fig:perturbedFRs} represents a loss rate probability function determined from the integrals of the muon loss versus time distributions from 30 to 70\,\SI{}{\micro\second}. This is simply meant to be illustrative, as the function evolves in shape throughout the fill.

What is needed is a time-dependent muon loss probability function $l(x,t)$, which can be applied to the nominal momentum distribution to yield the time dependence of the average momentum of the stored muons.  That information is readily translated into a time-dependent spin phase of the stored muons, $\varphi(t)$.
The parameters of the function $l(x,t)$ are determined by fitting the eight distributions over increasingly long time ranges from the fit start time $t_{s} = 30$\,\SI{}{\micro\second}.  At each time $t$, the fit is performed using eight equations of the form
\begin{equation}
\int_{R_\text{min}}^{R_\text{max}} F_i(x) l(x, t) dx = \frac{1}{\mathrm{H_{i}}}\int_{t_{s}}^{t}e^{t'/\gamma\tau_{\mu}} L_i(t') dt',
\label{eq:loss_function}
\end{equation}
where $i$ runs from 1 to 8 for each of the special runs and $R_\text{min}$ and $R_\text{max}$  represent the minimum and maximum, respectively, of the possible radii of the stored muons; $F_i(x)$ is the measured intensity of the fast rotation distribution for that run as a function of the equilibrium radius, normalized to 1; and  $\gamma\tau_{\mu} \approx 64.4$\,\SI{}{\micro\second} is the time-dilated muon lifetime.
The empirical loss function $l(x, t)$, which depends on time and radius, is determined in the fit.
The measured integrated triples spectrum $L_i(t')$  is integrated from the fit start time to time $t$.  The extra term $e^{t'/\gamma\tau_{\mu}}$ is included in order to follow the convention of the muon loss term in the decay positron fit function, as expressed in Eq.~\ref{eq:kloss}.  The normalization by $\mathrm{H_{i}}$, the total number of positrons measured in that dataset, ensures that the eight special runs can be correctly compared.

An analytic form of $l(x, t)$ is assumed in order to perform the fit.  From simulations, $l(x, t)$ is expected to peak near the edges of the storage distribution, as muons that have a high or low equilibrium radius are more likely to be lost.  Several forms of $l(x, t)$ with this qualitative behavior were compared, including a sum of two Gaussians and a piecewise sum of two parabolas.  The same analytic form was used throughout the fill, and the fit parameters were allowed to vary with time to account for the changing behavior of the lost muons.

At each time, the remaining stored distribution $F_\text{curr}(x, t)$ is calculated using the equation
\begin{equation}
F_\text{curr}(x, t) = F_{0}(x)-f_\text{loss}(t)\frac{F_{0}(x)l(x, t)}{\int_{R_\text{min}}^{R_\text{max}}F_{0}(x)l(x, t)dx},
\end{equation}
where $F_{0}(x)$ is the fast rotation distribution (see Sec.~\ref{sec:fastrotation}) of the full physics dataset, normalized to 1, and represents the radial distribution of the stored muons at the fit start.  The second term represents the total number of muons that have been lost up to that time, scaled by the fractional loss correction $f_\text{loss}(t)$, which emerges from the decay positron fit, as seen in Fig.~\ref{fig:muonlossprob}.

\begin{figure}[tb]
\centering
\includegraphics[width=\columnwidth]{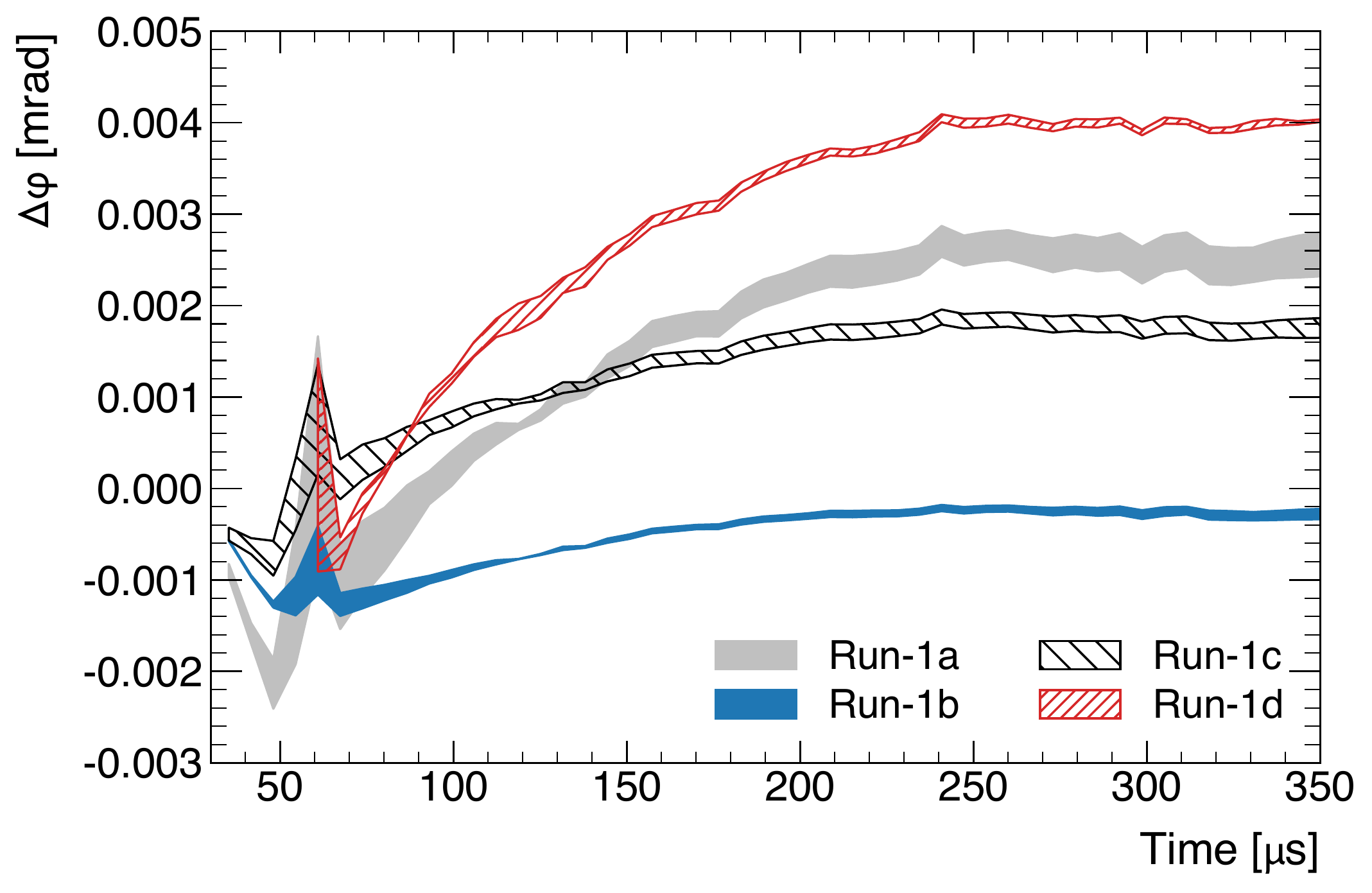}
\caption{The expected phase shift versus time in fill of the remaining, stored muon population.  The uncertainty bands arise from the use of three different $l(x)$ functions.}\label{fig:phases_vs_time}
\end{figure}

The average radius of the stored distribution is then extracted at each time from $F_\text{curr}(x, t)$.  This average radius can be converted into momentum units using Eq.~\ref{eq:DispRelation}.
The average $\Delta p/p_0$ of the stored distribution is converted to $\varphi$ using the measured phase-momentum correlation of $(-10 \pm 1.6)$\,mrad$/(\%\Delta p/p_0)$.
The $\varphi(t)$ determined by this method for the four \runone\ datasets, using three different analytical forms of $l(x)$, is shown in Fig.~\ref{fig:phases_vs_time};  $\varphi(t)$ behaves very similarly, regardless of the form of $l(x)$ used.

\subsection{Value of the muon loss correction}

The phase angle $\varphi(t)$ is parameterized using a high-degree polynomial function, which is used to generate a set of points based on the five-parameter decay positron fit function Eq.~\ref{eq:wiggle_func} with the empirical $\varphi(t)$ inserted.  A fit is then performed on the generated points using $\varphi(t)=\varphi_{0}$, as is assumed in the physics analysis, with all parameters allowed to float.  The bias to \wam is extracted by comparing the input \wa value to the value extracted from fitting with a constant $\varphi$.  This analysis is repeated for all four datasets.

The correction to \wam from this effect, $C_{ml}$, is given in Table \ref{tab:muon_loss_table}.  The muon loss-induced phase change artificially increases the measured value of \wam, so $C_{ml}$ is negative.  It varies from $-17$ to $-3$\,ppb depending on the dataset, with \runonea and \runoned having larger shifts because of their higher muon loss rates.  Three sources of uncertainty  are common to all run groups, and they are added linearly.  The measured phase-momentum correlation of $(-10 \pm 1.6)$\,mrad$/(\%\Delta p/p_0)$ contributes a 16\% uncertainty ($1-3$\,ppb).  The choice of analytical form of $l(x,t)$ contributes a $0-2$\,ppb uncertainty. The use of different $f_\text{loss}(t)$ functions from different analyses contributes $1-2$\,ppb.  To quantify the latter two sources of uncertainty, the full analysis described in the previous section was repeated for every combination of $l(x,t)$ and $f_\text{loss}$ function.  This procedure yields $\sigma(C_{ml}) = 2-6$\,ppb, which is small on the scale of other \runone uncertainties.
\begin{table}
\begin{ruledtabular}
\begin{tabular}{rrrrr}
Dataset & \runonea & \runoneb & \runonec & \runoned \\
\hline
$C_{ml}$ & -14 & -3 & -7 & -17  \\
\hline
Phase-momentum  & 2 & 0 & 1 & 3  \\
Form of $l(t)$ & 2 & 0 & 1 & 1  \\
$f_\text{loss}$ function & 2 & 1 & 2 & 2  \\
\hline
Linear sum ($\sigma_{C_{ml}}$) & 6 & 2 & 4 & 6  \\
\end{tabular}
\end{ruledtabular}
\caption{Muon loss correction $C_{ml}$ (ppb) with three sources of uncertainty contributing to $\sigma_{C_{ml}}$ (ppb). \label{tab:muon_loss_table}}
\end{table}
 \section{\label{sec:phase_acc}Correction for Muon Distribution Time Dependence $C_{pa}$}

The phase of the \wa oscillation at the moment of a muon's decay is related to the orientation of the muon spin vector relative to its momentum at injection into the storage ring.  The sign of the phase is defined by the convention in Eq.~\ref{eq:wiggle_func} that the positron intensity spectrum modulation is described by a term
proportional to $A \cos(\wa t + \varphi)$.
As discussed in Sec.~\ref{sec:phase-momentum}, the average phase of the incoming polarized muon beam is determined by upstream beam line components and the number of turns in the DR; its value does not affect the extraction of \wam.
Ultimately, the observed phase is what a calorimeter detects, an integration of decays from all locations that produce a positron signal in an energy bin $E$.  The phase of the fitted distribution corresponds to the injection phase $\varphi_0$ plus any average orientation of the muon spin with respect to its momentum that maximizes the anomalous precession signal.

\begin{figure}[htp]
  \begin{center}
  \includegraphics[width=\columnwidth]{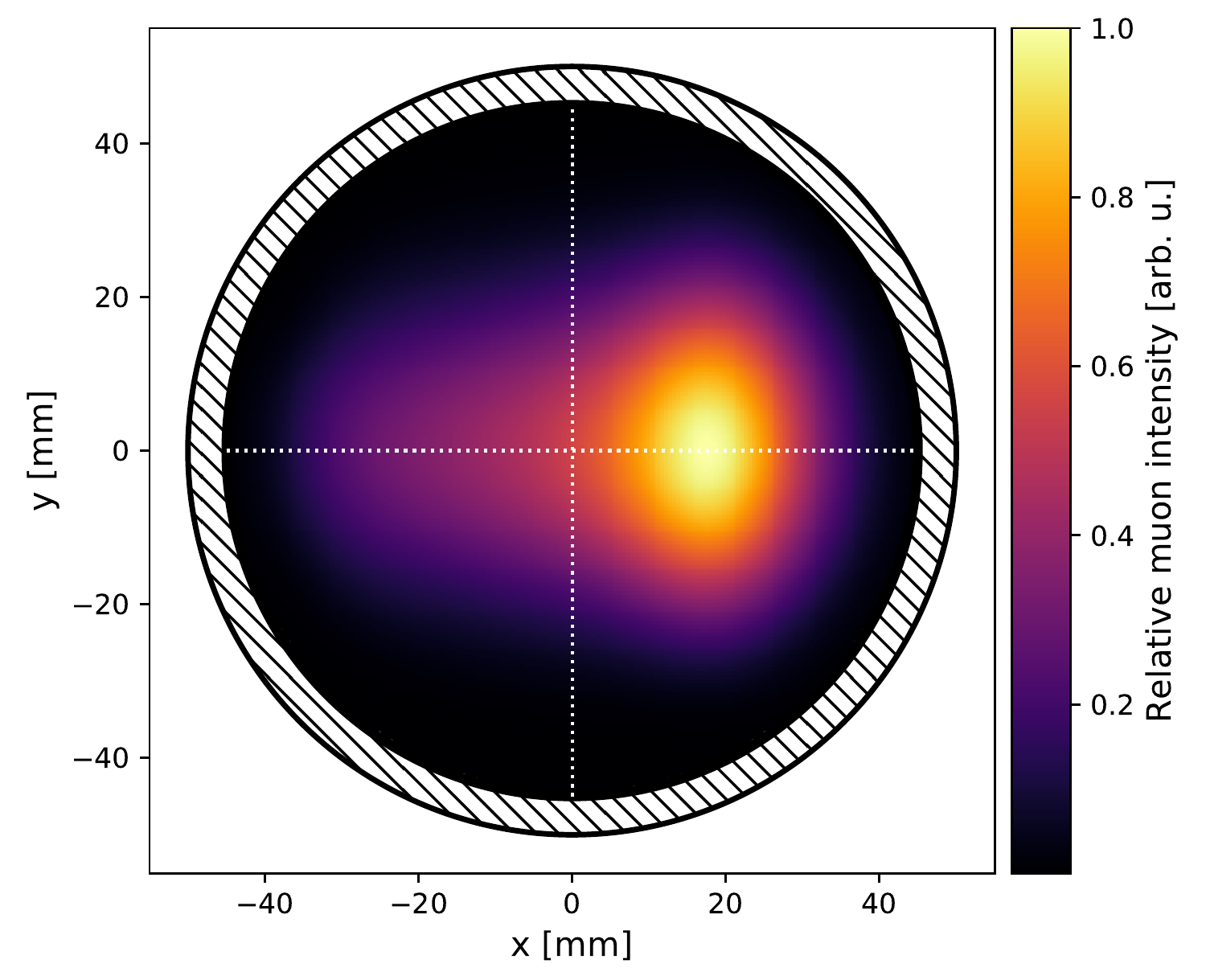}
  \end{center}
  \caption{An azimuthally averaged muon spatial distribution for \runonea\ as measured by the trackers.   The (hatched) collimator defines the 45-mm-radius transverse storage aperture.}
  \label{fig:muondistribution}
\end{figure}

In Sec.~\ref{sec:time_changing_cbo}, we described how the calorimeter acceptance is dependent on the transverse decay coordinate.
An azimuthally averaged transverse distribution is shown in Fig.~\ref{fig:muondistribution}.  This static image does not convey the radial and vertical oscillations of the mean and width at frequencies associated with the betatron oscillations.  They require modification to the normalization, asymmetry, and phase terms in Eq.~\ref{eq:wiggle_func} to fit the spectrum to obtain \wam.  The form of these modifications is discussed in Refs. \cite{\precession,Bennett:2006fi}, and an example form used in our \runone \wa analysis is shown in Appendix~\ref{ap:fitfunction}.

\begin{figure}
\centering
\includegraphics[width=\columnwidth]{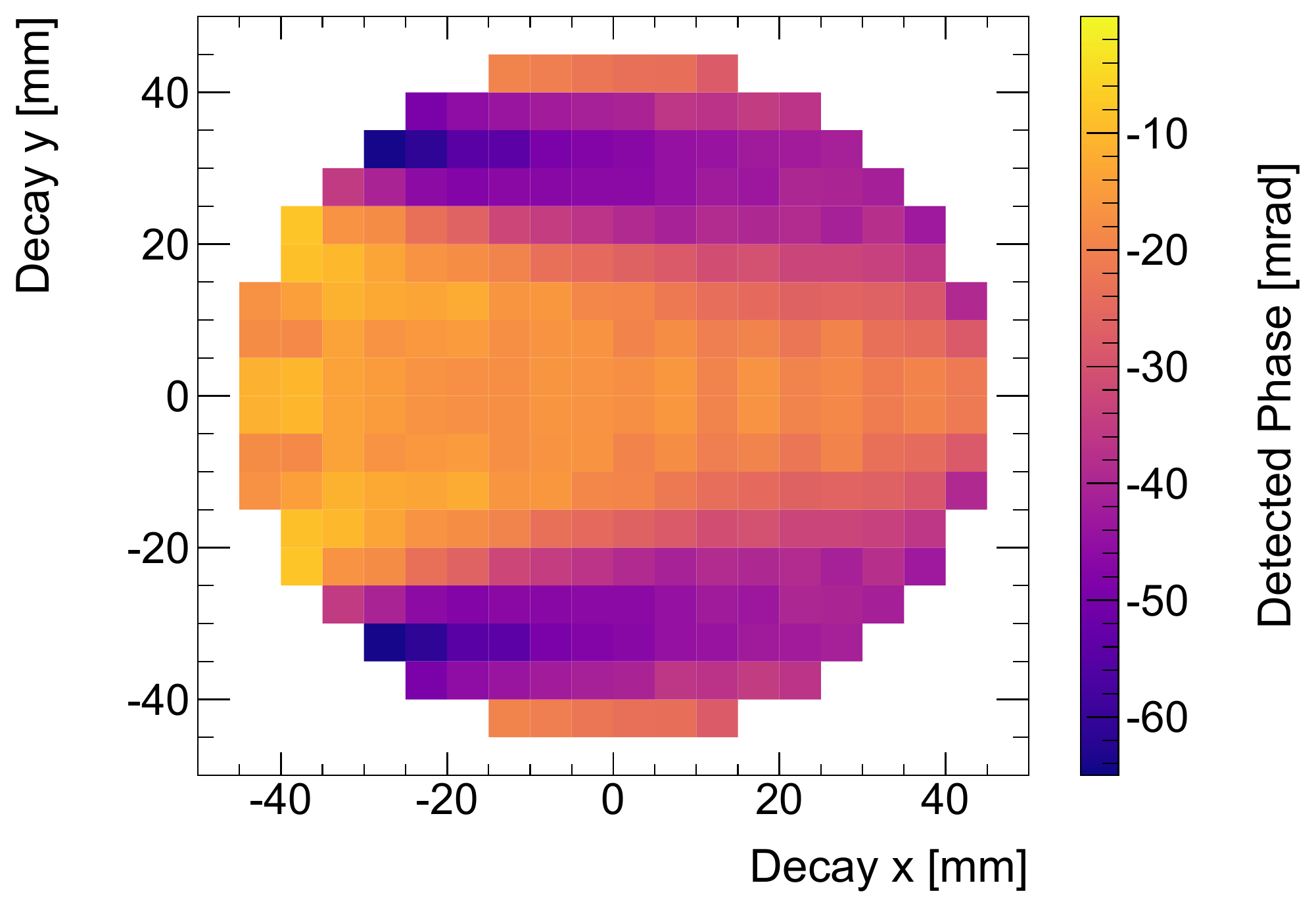}
\caption{The azimuthally averaged phase maps for the asymmetry-weighted analysis.}\label{fig:phasemap}
\end{figure}

The phase for a given $(x,y)$ decay coordinate depends on the orientation of the muon's spin that maximizes the acceptance.
The 24 calorimeters are finite sized, placed to the inside of the muon trajectory, and positioned at uniformly spaced azimuthal locations.
Therefore, the spin orientation of a muon that maximizes acceptance into this system is not parallel to its momentum but rotated slightly radially inward.
This rotation, captured by an effective phase shift $\varphi_{pa}$, is a function of transverse decay coordinate $(x,y)$ because of acceptance effects.
Figure~\ref{fig:phasemap} is a ``phase map'' averaged over azimuth and weighted by the asymmetry method used to extract \wam from the positron intensity time spectra.  The map varies more strongly in the vertical direction and less so radially.  The procedure to create phase maps using \ringsim is described in Appendix~\ref{ap:phase-acceptance-maps}.  Briefly, muon decays are generated over all $(x,y,\phi)$ coordinates with full muon spin precession versus time in fill properly included.  Individual positron intensity spectra from decays originating in a matrix of $(x,y)$ transverse bins are fit to determine \wa and $\varphi_{xy}$.  The procedure naturally includes the effects of acceptance, and, indeed, acceptance and asymmetry maps are also obtained for each calorimeter.

The spectrum summed across all detectors and all decay coordinates has a modulation frequency \wa and a net,  spatially averaged phase $(\varphi_0 + \bar{\varphi}_{pa})$.   As long as  the  muon distribution remains constant throughout the measurement period, the net phase is constant.
The dependence of the decay-coordinate phase $\varphi_{pa}(x,y)$ on detector acceptance was understood well enough from the E821 experiment to help guide our voltage stability specifications in the development of the new ESQ power-supply network.
However, as explained in Sec.~\ref{sec:time_changing_fields}, the two damaged resistors in \runone spoiled the voltage stability requirement on the Q1L upper and lower electrodes, which, in turn, led to a time dependence of the mean and width of the muon distribution.
In this section, we evaluate how this effect results in a time dependence of the average coordinate-dependent phase contribution that is not included in the fits to extract \wam from the positron spectrum.  A correction factor $C_{pa}$ is needed to correct \wam to the true \wa needed to determine \amu.

\begin{figure}[htp]
  \begin{center}
  \includegraphics[width=\columnwidth]{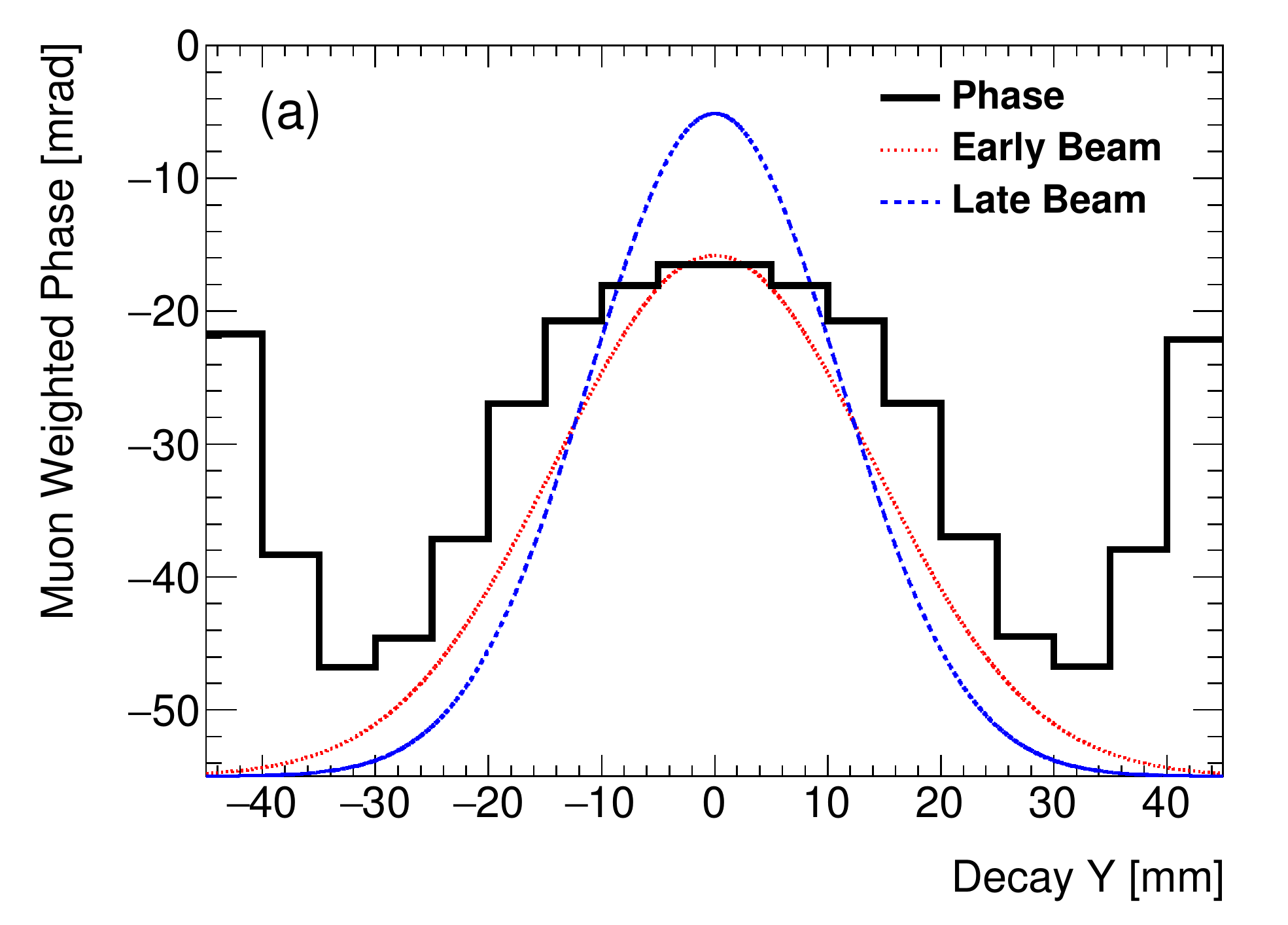}
  \includegraphics[width=\columnwidth]{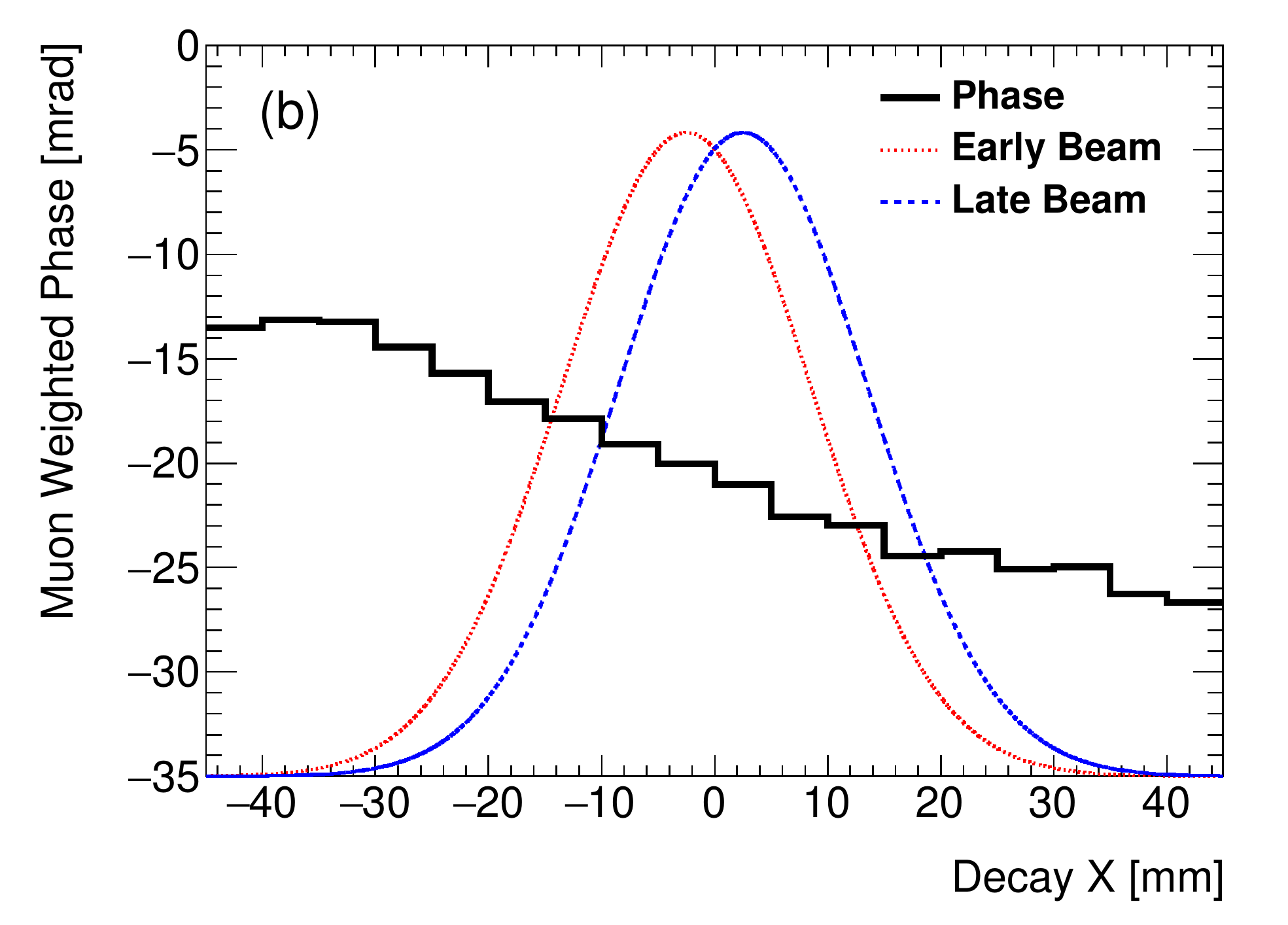}
  \end{center}
  \caption{Phase versus vertical (a) and horizontal (b) decay coordinate.  The Gaussian profiles show (at an exaggerated scale) how muon distributions might evolve from early (dotted red line) to late (dashed blue line) times in the fill.  The largest effect is from the reduction in the vertical width (a), where the phase changes to the distribution add coherently.  The mean shift (not shown) is smaller, because the increase on one side is balanced on the opposite side.  Conversely, the larger phase shift in the radial projection is from the mean motion (b), while the width change (not shown) is largely cancelled. }
  \label{fig:phaseprojections}
\end{figure}

Figure~\ref{fig:phaseprojections} illustrates how the time dependence of the muon distribution vertical and horizontal means and widths can lead to an average phase shift $\bar{\varphi}_{pa} \rightarrow \varphi_{pa}(t)$.
The exaggerated Gaussian profiles are representative of how such a muon distribution would evolve from early (dotted red line) to late (dashed blue line) times in a fill.  The average phase $\bar{\varphi}_{pa}$ at any time is calculated by taking a weighted average of the phase values in the map projection, with the weights from the beam intensity (assuming uniform asymmetry and acceptance).  The largest effect is from the reduction in the vertical width, where $\bar{\varphi}_{pa}$ is evidently different early compared to late in the fill.   In contrast, a small shift of the vertical mean both gains and loses phase nearly symmetrically.  The radial phase projection is nearly linear, and, therefore, a mean shift will cause a variation in the phase, whereas a width change is relatively balanced.  The radial effects are relatively small, and it is the coherent reduction in the vertical width (see Fig.~\ref{fig:Tracker-Ymean-Yrms}b)  that dominates the correction to \wam  and must be evaluated.

An additional effect that contributes to the phase-acceptance correction has its origin in CBO decoherence.
Early in a fill, the betatron oscillations are largely in phase and the beam moves coherently back and
forth in the radial direction.
When the beam has fully decohered, all calorimeters sample
the full radial distribution at once, and so the relative acceptance of the full aperture determines the average phase.
Conversely, early in the fill when the oscillations are still coherent, the calorimeters sample muon decays from only a subset of the radial
distribution at any particular time. In this scenario, the relative acceptance between different radial
positions is not important.
As a result of this difference between the early and later
stages of CBO decoherence, the average phase for an individual calorimeter drifts from early-to-late times in the fill with an impact
as large as $\mathcal{O}(100)$~ppb.
However, detectors on opposite sides of the ring see the CBO oscillations out of phase with one another, so when considering the sum of the calorimeter data there is a strong cancellation with no net contribution to the $C_{pa}$ correction at a statistically significant level.   We do assign an uncertainty owing to the imperfect cancellation of detector pairs that are $\pi$ radians out of CBO phase.
We note that a similar reduction is realized when comparing individual calorimeter positron decay spectra fits to those of the sum.
The terms that describe the CBO oscillation amplitudes are 6$-$7 times smaller for the summed spectra compared to fits to
individual calorimeters,  and the reduction would be more complete if the detector acceptances were identical.

The strategy to evaluate the net phase shift to each of the four datasets in \runone\ involves the following steps: (1) Create high-fidelity maps of acceptance, asymmetry, and phase for each calorimeter (see Appendix~\ref{ap:phase-acceptance-maps}). (2) Determine the time-dependent muon spatial distributions $M^T(x,y,t)$ for each tracker station $T$ and for each dataset. (3) Evolve those distributions using beam dynamics models and simulations to produce $M^c(x,y,t)$ for all azimuthal locations where calorimeters ($c$) are placed. (4) Fold $M^c(x,y,t)$ distributions with  acceptance, asymmetry, and phase maps to obtain phase shift versus time in fill for each calorimeter, $\varphi^c_{pa}(t)$. (5) Generate Monte Carlo data using the full fit function, including the predicted behavior of the phase, $\varphi^c_{pa}(t)$.  Fit these pseudodata using the normal fit function to determine the difference in the extracted \wam compared to the input.  This difference yields the correction factor $C_{pa}$ for each calorimeter.

\subsection{Measurement of the time-changing muon distribution}
\label{sec:randomization}

The two tracker stations produce the time evolution of the muon distribution maps over the course of a fill, $M^T(x,y,t)$.  These distributions are corrected for resolution and acceptance as described in Sec.~\ref{sec:experiment}.  Betatron oscillations cause the distribution to change rapidly, which makes extraction of the slower terms due to the damaged resistors more difficult.  These oscillations are removed from the data by randomizing the time information for each track according to a uniform distribution $U(-T/2,T/2)$, where $T$ is the time period of the oscillation.  The modulations at $\omega_{\text{CBO}}$,  $\omega_{\text{2CBO}}$, $\omega_{\text{VW}}$, $\omega_{y}$, and \wam are all removed.  This randomization procedure also has the effect of removing the CBO decoherence phase-acceptance effect, which mostly cancels in the sum of all calorimeter data as described above.  Figure~\ref{fig:Tracker-Ymean-Yrms} shows the smooth nature of the vertical projections of the mean $\langle y \rangle$ and width $\sqrt{\langle y^2 \rangle}$ versus time after the randomization procedure.
Naively, one might assume that an average of the two tracker maps will provide the most accurate representation of the muon distribution versus time and decay coordinates.  However, as indicated in Fig.~\ref{fig:betay}, the early-to-late behavior of the storage ring optical lattice functions varies significantly versus azimuth, because the damaged resistors are located in only one section of the storage ring.  The correction is, therefore, evaluated independently starting from each tracker station, and the final result is taken as their average.

\subsection{Estimation of beam distribution around the storage ring}
The trackers measure $M^T(x,y,t)$ at two locations around the ring, but the extraction of \wam is performed using calorimeters at 24 azimuthal locations.  The damaged resistors were located on the opposite side of the ring with respect to the two tracker positions.  This asymmetry leads to differences in the muon transverse distribution measured by the trackers and in the proximity of calorimeters elsewhere in the ring.
Both \COSY and \ringsim predict similar relative behavior of $\langle y \rangle(t)$ by azimuthal location, as is shown with a comparison to calorimeter data in Fig.~\ref{fig:DeltaYvsTime}.
But the reduction of the vertical width (Fig.~\ref{fig:Tracker-Ymean-Yrms}b)---which cannot be measured by the calorimeters---is more important to the $C_{pa}$ calculation. The $M^T(x,y,t)$ distributions are evolved separately to all azimuthal locations using the predictions from either \COSY or \ringsim. This provides a complete spatial and time distribution of the muons $M^c(x,y,t)$ in the vicinity of a calorimeter based on a measurement at a single point in the ring.
The time dependent muon profiles were calculated at an azimuthal angle approximately \SI{22}{\degree} upstream of each calorimeter's front face, where acceptance is maximal.

Similarly to the treatment for the pitch correction (Sec.~\ref{sec:pitch}), the vertical distribution is anchored to the tracker data at its azimuthal location, and the distribution at other parts of the storage ring is obtained by scaling the width as a function of azimuth relative to that at the tracker station:
\begin{equation}
  y_\text{pred}(\phi, t) = y_\text{tkr}(t) \frac{y_\text{pred}^\text{rms}(\phi,t)}{y_\text{pred}^\text{rms}(\phi_\text{tkr},t)}.
\end{equation}
The width ratio is evaluated directly from virtual tracking planes in \ringsim and via the ratio of beta functions in \COSY ($\sqrt{ \beta_y(\phi, t) / \beta_y(\phi_\text{tkr}, t)}$). From such transformations, vertical beam drifts are properly projected around the ring, and further effects from permanent vertical closed-orbit distortions are accounted for in the systematic errors of the correction.

The contribution to $C_{pa}$ from the radial motion of the beam is subdominant, but for a complete treatment the radial beam changes are also included.  A similar procedure to that for the vertical changes is followed, although there are added complications due to the interplay of muon momentum and radial position as well as closed-orbit distortions due to azimuthal variation of the main dipole field~\cite{\field}. Each entry from the measured distribution is scaled as
\begin{equation}
    x_\text{pred}(\phi, t) = \frac{x_\text{pred}^\text{rms}(\phi, t)}{x_\text{pred}^\text{rms}(\phi_\text{tkr}, t)}\cdot (x_\text{tkr}(t) - \bar{x}_\text{tkr}(t)) + \bar{x}_\text{pred}(\phi, t) ,
\end{equation}
where the predicted values for $x_\text{rms}(\phi,t)$ and $\bar{x}(\phi,t)$ are taken from tracking planes in \ringsim and calculated in \COSY using $\beta_x(\phi,t)$, $D_x(\phi,t)$, and the reconstructed distribution of muon momenta.  Measurements of the main dipole storage field are used to calculate the closed-orbit distortion included in $\bar{x}_\text{pred}(\phi,t)$.  After including both vertical and radial manipulations, the procedure has been shown to provide excellent agreement between calculations of $C_{pa}$ using either extrapolated $M^c(x,y,t)$ or beam-tracking planes in a closed-loop simulation test.

\subsection{Phase-acceptance correction: Results and uncertainty evaluation}
\begin{figure}[htp]
  \begin{center}
  \includegraphics[width=\columnwidth]{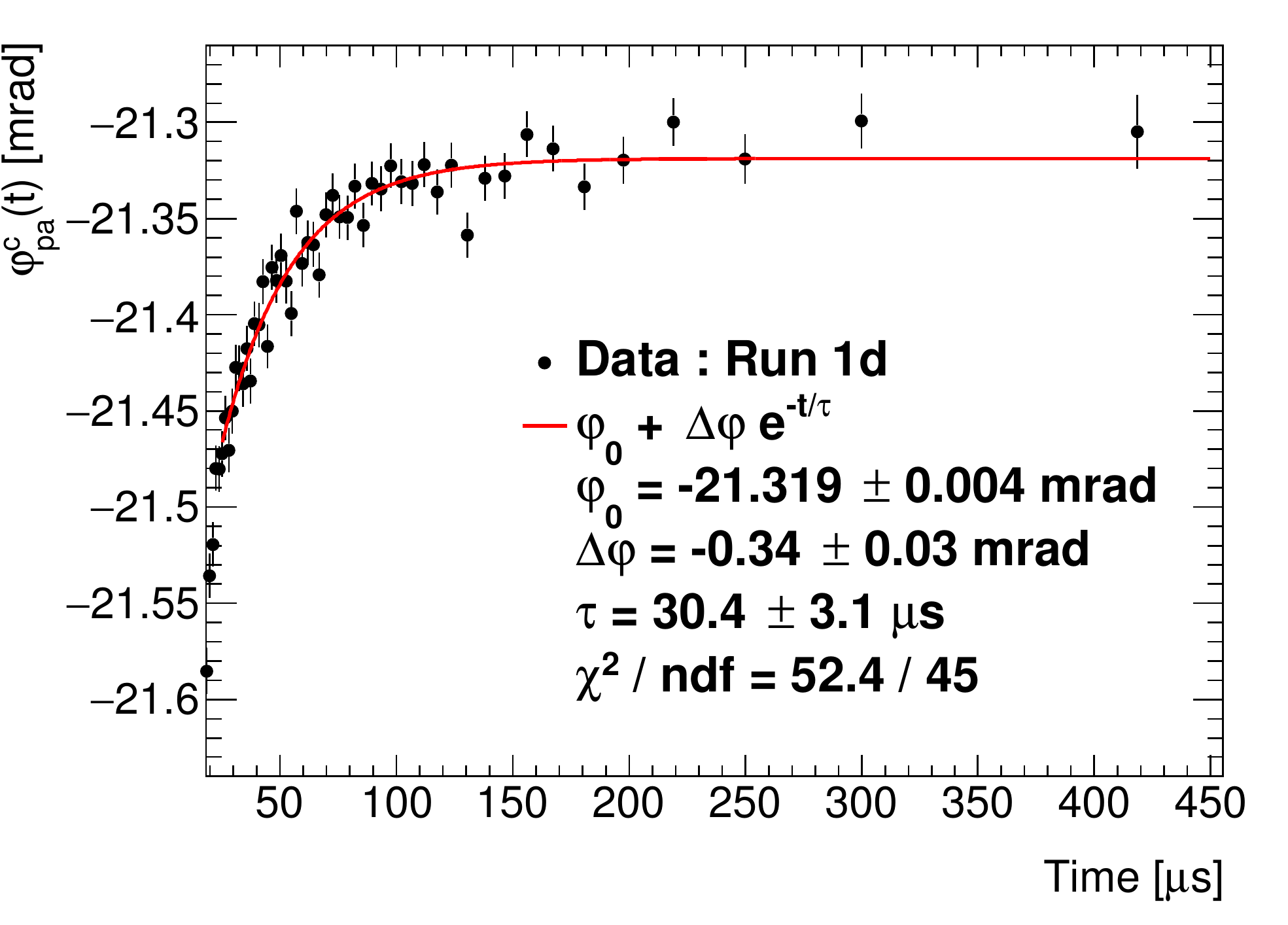}
  \end{center}
\caption{Calculation of $\varphi_{pa}^c(t)$ for calorimeter 19 in \runoned\ using data from the tracker station at \SI{180}{\degree} and Eq.~\ref{eq:master}. The fit start time for the extraction of \wam for this dataset is \SI{50}{\micro\second} after injection in order to mitigate the effect of $\varphi_{pa}^c(t)$.}
\label{fig:phaseVStime}
\end{figure}

The muon distribution is combined with the simulated maps to extract the time-dependent phase for each calorimeter $\varphi_{pa}^c(t)$ using the following weighted sum:
\begin{widetext}
\begin{equation}
  \varphi^{c}_{pa}(t) = \arctan\cfrac{\sum_{ij} M^c(x_i,y_j,t) \cdot \varepsilon^c(x_i,y_j) \cdot A^c(x_i,y_j) \cdot \sin(\varphi^c_{pa}(x_i,y_j))}
                                     {\sum_{ij} M^c(x_i,y_j,t) \cdot \varepsilon^c(x_i,y_j) \cdot A^c(x_i,y_j) \cdot \cos(\varphi^c_{pa}(x_i,y_j))}\, .
\label{eq:master}
\end{equation}
\end{widetext}
The sum is over all spatial bins.
The time dependence enters via the muon distribution $M^c$. 
The acceptance, asymmetry, and phase maps for calorimeter $c$ are represented by $\varepsilon^c$, $A^c$, and $\varphi_{pa}^c$, respectively.
An example evaluation of $\varphi^c_{pa}$ for the \runoned\ dataset is shown in Fig.~\ref{fig:phaseVStime}.  As outlined earlier, the phase increases with time primarily due to the decreasing vertical width of the beam.

The size of the mismeasurement of \wam as a result of the time-dependent phase $\varphi_{\text{pa}}^{\text{c}}(t)$ is estimated using simulated data. A histogram is generated for each calorimeter according to the full fit function used to extract \wam, an example of which is described in Appendix~\ref{ap:fitfunction}.   The spectrum includes betatron oscillations of the normalization, asymmetry, and phase and a modified \gm phase term that has a parameterization of $\varphi_{\text{pa}}^{\text{c}}(t)$. The function parameters for each calorimeter are set according to their data-derived values.  The resulting simulated data are then fit with the full fit function but excluding the modified phase term.  The difference between the values of \wam that were input and extracted from the fit give the value of the correction $C_\text{pa}$ for that calorimeter.

The final values correspond to the average of the corrections for all calorimeters, obtained starting from measurements from each of the tracker stations and from performing the procedure using both \COSY and \ringsim.  Given that the phase increases with time as shown in Fig.~\ref{fig:phaseVStime}, the measured \wam\ is larger than the true value.  The correction $C_\text{pa}$ must reduce the measured value and is, therefore, negative.  The $C_{pa}$ corrections are shown in Table~\ref{tab:pa_table}.  The values range from $-184$~ppb for \runonea\ to $-117$~ppb for \runonec.
The condition of the damaged resistors was worse for the \runoned dataset, which would generally lead to a larger $C_{pa}$ value.  In order to keep the correction for \runoned comparable in size to the other datasets, the fit start time was delayed from 30 to \SI{50}{\micro\second}, which reduces the correction by a factor of 1.8.

\begin{table}[htpb!]
\begin{ruledtabular}
\begin{tabular}{rrrrr}
Dataset        & \runonea & \runoneb & \runonec & \runoned \\
\hline
$C_{\text{pa}}$         & -184   & -165   & -117   & -164   \\
\hline
Stat. uncertainty       & 23     & 20     & 15     & 14     \\
\hline
Tracker and CBO         & 73     & 43     & 41     & 44     \\
Phase maps              & 52     & 49     & 35     & 46     \\
Beam dynamics           & 27     & 30     & 22     & 45     \\
\hline
Total uncertainty       & 96     & 74     & 60     & 80     \\
\end{tabular}
\end{ruledtabular}
\caption{Values of the phase-acceptance correction $C_{\text{pa}}$~(ppb) and their statistical, systematic, and total uncertainties for each of the \runone\ datasets.\label{tab:pa_table}}
\end{table}

The evaluations of statistical and systematic uncertainty are also detailed in Table~\ref{tab:pa_table}.  The statistical uncertainty originates from the limited number of tracks available to form $M^T(x,y,t)$ and ranges from 14 to 23~ppb depending on the size of the dataset.

The sources of systematic uncertainty can be grouped into three main areas with no single category dominating over the others.  First, there is imperfect knowledge of the tracker's alignment, resolution, and acceptance corrections, which all affect the measurement of $M^T(x,y,t)$.  Uncertainties are estimated by repeating the evaluation of $C_\text{pa}$ while varying the corrections within a reasonable range based on external constraints.  Additionally, there is a statistical tension in \runonea\ between the corrections extracted independently from each of the two tracking stations.  The uncertainty for this dataset is conservatively inflated to account for this.  As explained previously, the phase-acceptance effect originating from CBO decoherence will largely cancel when all the calorimeters are summed together.  The degree of this cancellation is dependent on the relative amplitudes of the CBO-modulated acceptance oscillation around the ring, which is strongly correlated with the tracker acceptance correction.  The uncertainty from the CBO decoherence cancellation is, therefore, combined linearly with those from the tracker corrections, resulting in values ranging from 41 to 73~ppb.

Second, uncertainties associated with the estimation of the phase, asymmetry, and acceptance maps in Eq.~\ref{eq:master}  are estimated using the \ringsim simulation.  These uncertainties are dominated by the knowledge of the phase map, which has a much stronger effect on $C_\text{pa}$ than either the asymmetry or acceptance maps.  The possible variation of the map is estimated by comparing measurements of the fitted phase in the actual \wam fits with predictions of $\varphi_{\text{pa}}^{\text{c}}$ from Eq.~\ref{eq:master}.  The time dependence of $M^c(x,y,t)$ in the latter is collapsed by taking data from all times in a single time bin.  The variation in the final fitted phase as a function of calorimeter number and as a function of calorimeter $y$ position matches the prediction to within 20\%.  This variation is used as an estimated uncertainty on the phase map, and the procedure is repeated to propagate this uncertainty to establish a contribution to $\delta C_\text{pa}$.  Interactions in material will also affect the detector acceptance and, therefore, change the phase map (Fig.~\ref{fig:VerticalProjGrid} in Appendix~\ref{ap:phase-acceptance-maps}).  The effect of material interactions in the simulation can be exaggerated to allow for quantification of an uncertainty due to mismodeling of material effects.  The total uncertainty associated with the map estimation ranges from 46 to 52~ppb.

Last, the procedure utilizes beam dynamics models to extract the calorimeter-specific  $M^c(x,y,t)$ distributions from the tracker-measured $M^T(x,y,t)$.  Uncertainties are estimated by calculating $C_\text{pa}$ while varying the beta functions and magnetic fields within constraints from measurements, varying the momentum distribution of the stored muons, and from a comparison between values extracted using either \COSY or \ringsim.  This uncertainty is smaller than those from the tracker and phase categories in Runs-1a, 1b and 1c (from 22 to 27\,ppb), but is comparable in size in \runoned\ ($45$~ppb). The larger uncertainty in \runoned\ is due to the worsening condition of the damaged resistors, which increases the azimuthal variation and, therefore, increases the reliance on the simulations.

Within each group of uncertainties, correlated effects are added linearly, and then all remaining effects are summed quadratically to give the total.  The total uncertainties (from 60 to 96\,ppb) range from 45\% to 52\% of the correction value.
 \section{\label{sec:conc} Conclusion}

In this paper, we described summaries and conclusions from in-depth studies of four beam dynamics systematic corrections that are required to adjust the measured muon precession frequency \wam (Ref.~\cite{\precession}) to its true physical value \wa.  The corrections for the vertical betatron motion---{\it{pitch}}---and the influence of the motional magnetic field on non-magic-momentum muons---{\it{electric field}}---are well understood and have been documented by previous generations of \gmtwo experiments.  Here, we have refined these studies and performed detailed uncertainty analyses.  The pitch correction requires knowledge of the vertical stored muon distribution from the {\it in situ} tracker system, which provides detailed time-dependent stored-muon spatial profiles in two areas of the storage ring.
The electric field correction requires knowledge of the stored muon momentum distribution (alternatively, radial distribution), which is deduced from studying the time evolution of the incoming muon bunch. Two distinct analysis methods are used.

Despite an initial scraping procedure meant to remove muons that will not remain stored in the ring, a small fraction that will eventually strike a fiducial-defining collimator do survive well into the measurement period.   A careful analysis demonstrated that those muons which are lost do not significantly alter the ensemble-averaged spin phase, but we apply a small data-driven correction.

Finally, owing to 2 of 32 damaged high-voltage resistors in the ESQ system that led to slower-than-designed charging times for two plates, the mean and rms of the stored muon distribution in \runone evolved throughout the first ${\sim}100$\,\SI{}{\micro\second} of the measuring period.   Investigation of this effect led to an extensive and new understanding of a subtle coupling of decay coordinate within the storage volume and acceptance versus spin phase.  The coupling to acceptance and phase is carried out by detailed \GEANT-based simulations, and the understanding of the beam motion around the ring through the evolution of the beta functions is simulated in detail using \COSY, whose key results are supported by storage simulations using \ringsim.  This ``phase-acceptance'' effect is time-dependent; its influence on the phase changes rapidly and faster than the muon population decays.  Therefore, in the \runoned period where the resistor time constants were longer than during periods 1a, 1b, and 1c, a fit start time in the precession fits of \SI{50}{\micro\second} is used to limit the systematic shift to the phase.

\subsection{Summary of \runone net corrections to \wam }

Table~\ref{tb:results} presents a summary overview of key findings discussed in this paper.  The statistical uncertainty from the asymmetry-weighted analyses of the muon precession data is provided for reference.   The four corrections discussed herein are multiplicative  adjustments that are applied to convert the raw extracted frequency \wam from the precession fits to the true \wa value that is needed to determine \amu as in Eq.~\ref{eq:amueq}.  The corrections are identified as electric field $C_{e}$, Sec.~\ref{sec:efield}; pitch $C_{p}$, Sec.~\ref{sec:pitch}; muon-loss-phase $C_{ml}$, Sec.~\ref{sec:muon_loss_phase}; and phase-acceptance $C_{pa}$, Sec.~\ref{sec:phase_acc}.  Accordingly,
\begin{equation}
a_\mu = \left(\frac{m}{q\tilde{B}}\right)\left[\wam \cdot (1 + C_{e}+C_{p}+C_{ml}+C_{pa})\right],
\end{equation}
where $\tilde{B}$ represents the muon-weighted average magnetic field that is discussed in Ref.~\cite{\field}.

\begin{table}[htbp]
\begin{ruledtabular}
\begin{tabular}{lcc}
& Correction (ppb) & Uncertainty (ppb)\\
\hline
\wam statistical & $\cdot\cdot\cdot$ &         434 \\
\hline
$C_{e}$  &        489&          53 \\
$C_{p}$  &        180&          13 \\
$C_{ml}$ &        $-11$&           5 \\
$C_{pa}$ &       $-158$&          75 \\
\hline
$C_{\text{total}}$ & 499 & 93 \\
\end{tabular}
\end{ruledtabular}
\caption{The \runone combined beam dynamics corrections to \wam from the four \runone datasets.  These values are computed with the full correlation matrix formalism used to provide the measured value for \wa. The values, thus, reflect fully correlated systematics and weighting from the statistical uncertainties from each dataset.}
\label{tb:results}
\end{table}

 \section{Acknowledgments}

We thank the Fermilab management and staff for their strong support of this experiment, as well as
the tremendous support from our university and national laboratory engineers, technicians, and workshops.
The Muon \gmtwo Experiment was performed at the Fermi National
Accelerator Laboratory, a U.S. Department of Energy, Office of
Science, HEP User Facility. Fermilab is managed by Fermi Research
Alliance, LLC (FRA), acting under Contract No. DE-AC02-07CH11359.
Additional support for the experiment was provided by the Department
of Energy offices of High Energy Physics and Nuclear Physics (USA), the National Science Foundation
(USA), the Istituto Nazionale di Fisica Nucleare (Italy), the Science
and Technology Facilities Council (UK), the Royal Society (UK), the
European Union's Horizon 2020 research and innovation programme under
the Marie Sk\l{}odowska-Curie Grant Agreements No. 690835 and
No. 734303, the National Natural Science Foundation of China
(Grants No. 11975153 and No. 12075151), MSIP, NRF and IBS-R017-D1 (Republic of Korea),
and the German Research Foundation (DFG) through the Cluster of
Excellence PRISMA+ (EXC 2118/1, Project ID No. 39083149).
 
\appendix
\section{Electric Quadrupole Nonlinearity}
\label{ap:Efield}
\begin{figure}[htbp]
      \includegraphics[width=\columnwidth]{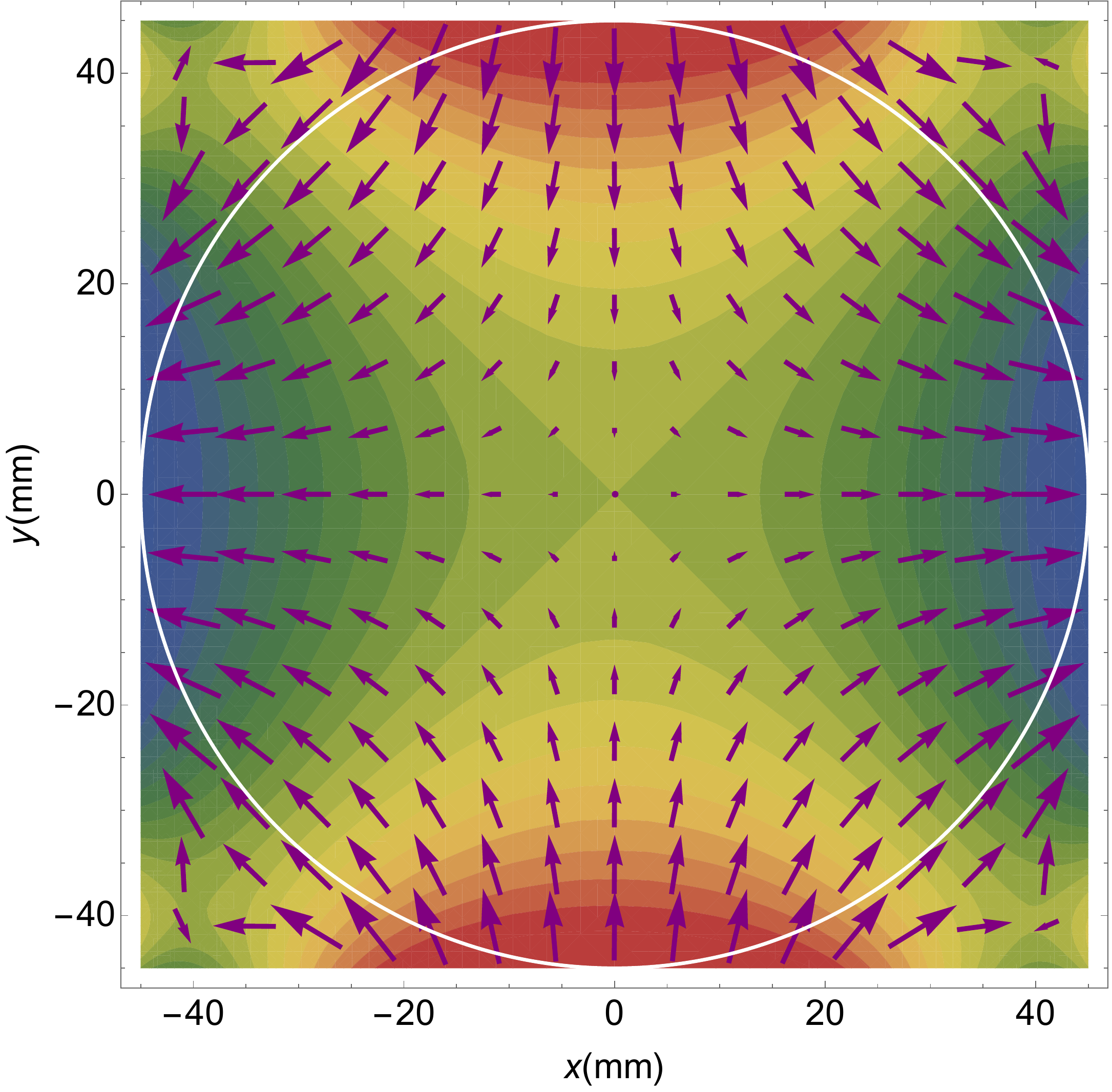}
\caption{A 2D representation of the equipotentials and electric field in the ESQ region.  The circle at radius 45\,mm represents the storage ring aperture defined by the collimators.}
\label{fig:Emap}
\end{figure}

Consider a 2D projection of the electric field in the ESQ region, as shown in Fig.~\ref{fig:Emap}.  Laplace's equation in two dimensions and Cartesian coordinates is
\begin{equation}
\nabla^2 V(x,y) = \frac{\partial^2 V}{\partial x^2}+\frac{\partial^2 V}{\partial y^2} =0.
\end{equation}
A potential corresponding to a pure quadrupole field can be written as $V = \frac{1}{2} k(x^2 -y^2)$ for some constant $k$.
The electric field is $E={\vec \nabla}V = k(x\hat x -y\hat y)$
with zero divergence, and it is linear in both $x$ and $y$.
Higher-order terms may appear owing to the plate geometry, but the
symmetry in the 2D Cartesian limit permits only those terms that are odd in $x$, $y$.
In particular there is no sextupole (${\sim}kx^2$) dependence of the field that would be symmetric in displacement.

In the limit of finite curvature, as employed in the storage ring geometry, it is more appropriate to represent the fields in cylindrical coordinates, where Laplace's equation is
\begin{equation}
\nabla^2 V(\rho,y,\phi) =  \left[\frac{1}{\rho}\frac{\partial}{\partial \rho}\left(\rho \frac{\partial}{\partial \rho}\right) + \frac{1}{\rho^2}\frac{\partial^2 }{\partial \phi^2}+\frac{\partial^2}{\partial y^2}\right]V = 0
\end{equation}
with $\rho = R_0+x$. Using the assumption---which is not accurate---that the ESQ plates are continuous around the ring so that there is no dependence on the angular coordinate $\phi$, the simplest possible solution is
\begin{equation}
V(\rho,y)= \tilde k\left[\frac{1}{2}\left(\frac{\rho^2}{R_0^2}-1\right)-\ln\frac{\rho}{R_0}-\left(\frac{y}{R_0}\right)^2\right] ,
\end{equation}
where $\tilde k$ is a constant.  The electric field is given by
\begin{equation}
  \vec E = -\vec\nabla V = -\frac{\tilde k}{R_0^2}\left[ \left(\rho-\frac{R_0^2}{\rho} \right)\hat\rho - 2y\hat y\right]
\end{equation}
and expanding in $x/R_0$:
\begin{equation}
  \vec E \approx k\left[\left(x-\frac{x^2}{2R_0}+\ldots\right){\hat \rho} -y\hat y\right] ,
  \label{eq:laplaciansolution}
\end{equation}
where $k = - 2 \tilde k / R_0^2$. Evidently, Maxwell's equations require a term quadratic in displacement. The quadratic term is equivalent to a sextupolelike
component that will affect the chromaticity, complicate the $E$-field correction, and possibly drive a third-order resonance.
The general conclusion is that there necessarily exists a symmetric component to the $E$-field that will contribute to the evaluation of the correction and its systematic uncertainty.
This uncertainty is included in Table \ref{Tab:EFieldCorrection} as part of the ESQ calibration.

\subsection{Path length correction owing to curved geometry}
The radial electric field along the trajectory of a muon with equilibrium radial offset $x_e$ and betatron amplitude $x_{\beta}$ is
\begin{equation}
E_r(s)  =  k(\eta\delta + x_\beta) - \frac{1}{2R_0}(\eta\delta +x_\beta)^2+\ldots ,
 \end{equation}
where $x_e=\eta\delta = \eta \frac{\Delta p}{p}$ and  the simple solution above to Laplace's equation is assumed.
The average electric field along the trajectory is
\begin{equation}
\langle E_r(s)\rangle = \frac{k}{L}\int_0 \left(\eta\delta + x_\beta - \frac{1}{2R_0}(\eta\delta +x_\beta)^2\right) (1+x_\beta/R_0){d s} ,
\end{equation}
where $L$ is the length of the trajectory.  Using $x_\beta = x_{\beta 0 }\cos\phi(s)$ with $d\phi = \frac{ds}{R_0}= \frac{dl}{(R_0+x)}$,  it can be shown that $\langle E_r\rangle$ is approximately
\begin{equation}
\langle E_r(s)\rangle  \sim k\left(\eta\delta - \frac{1}{2R_0}((\eta\delta)^2- \frac{1}{2} x_{\beta 0}^2) +\ldots    \right)\label{eq:avgefield} ,
\end{equation}
where $k = n\left(\frac{v_s B}{R_0}\right)$.
One observes that the contribution to the average electric field along a trajectory, due to the curvature of the ESQ plates, scales quadratically with the betatron amplitude.

Substitution of Eq.~\ref{eq:avgefield} into Eq.~\ref{eq:Efield_omega} gives the correction to $\omega_a$ due to fractional momentum offset $\delta$ and betatron amplitude $x_{\beta 0}$ as
\begin{eqnarray*}
C_e(\delta,x_{\beta 0})&\approx& -2\delta\frac{\beta k}{cB}\left(\eta\delta - \frac{1}{2R_0}((\eta\delta)^2- \frac{1}{2} x_{\beta 0}^2)\right)\\
&\approx& -2\frac{\beta k}{cB}\left(\eta\delta^2 - \frac{1}{2R_0}(\eta^2\delta^3- \frac{1}{2} x_{\beta 0}^2\delta)\right) .
\end{eqnarray*}

The next step requires averaging $C_e(\delta,x_{\beta 0})$ over the entire momentum and CBO distribution:
\begin{equation}
\begin{aligned}
\langle {C_e(x_e,x_{\beta 0})}\rangle
\approx & -2\frac{\beta k}{cB} \Bigg[\frac{\langle x_e^2\rangle}{\eta} \\
        & - \frac{1}{2R_0}\left(\frac{\langle x_e^3\rangle}{\eta}- \frac{1}{2} \langle{x_{\beta 0}^2}\rangle\frac{\vev{x_e}}{\eta}\right)\Bigg] ,
\end{aligned}
\end{equation}
where $\vev{\delta}=\vev{x_e}/\eta$.
If $\vev{x_e}/\eta =0$, and the betatron amplitude and momentum offset are assumed to be uncorrelated, and there is no contribution from the sextupole or path length terms, then
\begin{equation}
\langle C_e(\delta,x_{\beta 0})\rangle = -2\frac{\beta k}{cB}\frac{\vev{\delta^2}}{\eta}.
\end{equation}
In fact, momentum offset and betatron amplitude are strongly anticorrelated, which will be discussed below.
If $\langle x_e \rangle \ne 0$, that is, the muon momentum distribution mean is not centered at the magic momentum, then the fractional change to the $E$-field correction from the betatron amplitude consideration is
\begin{eqnarray*}
\frac{\vev{\Delta C_e(x_e,x_{\beta_0}}}{\vev{C_e}}&\sim& \frac{-\frac{1}{2R_0}\left[\frac{{x_e^3}}{\eta}-\onehalf\frac{\vev{x^2_{\beta_0}}\vev{x_e}}{\eta}\right]}{\frac{x_e^2}{\eta}}\\
&\sim& \frac{-\vev{x_e^3}+\vev{x_e}\vev{x^2_{\beta 0}}/2}{2\vev{x_e^2}R_0} .
\end{eqnarray*}
For the equilibrium radial distributions typical of \runone (see Fig.~\ref{fig:fourierdebunching}b),
$$\frac{\vev{\Delta C_e}}{\vev{C_e}} \sim   -1.1 \times 10^{-3} + (1.9\times 10^{-6}{\rm mm^{-2}})\vev{x_{\beta_0}^2}.$$
Conservatively estimating that $\vev{x_{\beta_0}^2}< (45{\rm mm)^2}$, the maximum radius of the storage volume, we find that
the change in the $E$-field correction, due to the nonlinearity associated with the curvature of the ESQ plates, is less than 0.5\%.

\subsection{Effect from the intrinsic quadrupole nonlinearity}
\label{sec:nonlinear}
\begin{figure}[htbp]
 \includegraphics[width=\columnwidth]{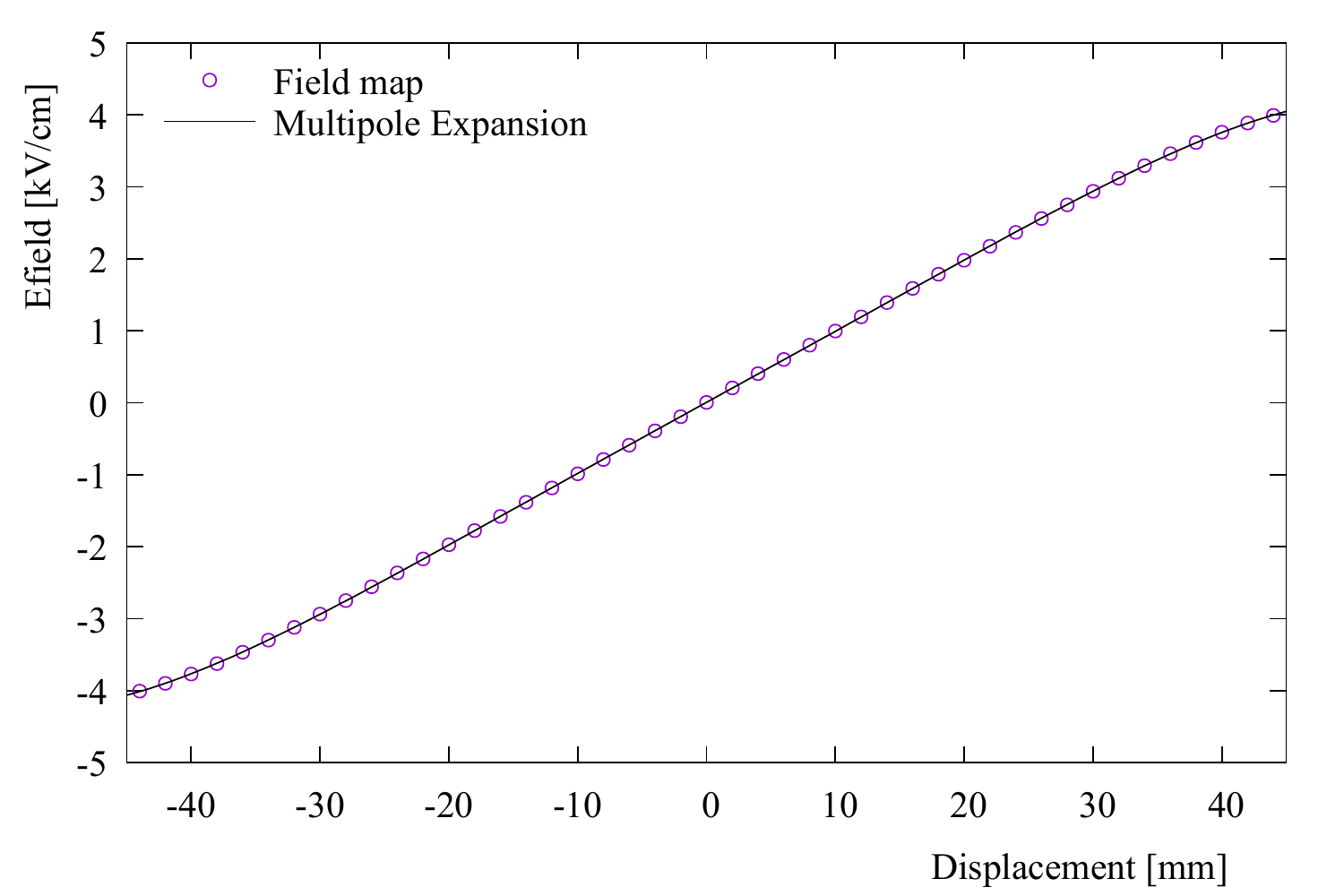}
\caption{The electric field along the $x$ axis in the midplane ($y=0$). The purple points are computed from the multipole expansion using Eq.~\ref{eq:mpoleexpansion}.
The black curve uses the values from the field map.}
\label{fig:Exprojection}
\end{figure}

The rectangular cross section of the ESQ geometry introduces a nonlinearity compared to the ideal case.
The electric field along the horizontal axis ($y=0$)  can be expressed by
\begin{equation}
E_x - iE_y = (b_n -ia_n)\frac{x^n}{r_0^n}\label{eq:mpoleexpansion} ,
\end{equation}
where $r_0=0.045$ m and $a_n$ and $b_n$ are  coefficients of the multipoles, which have been fit to an azimuthal slice of a \OPERATHREED~\cite{Opera3D} field map.
The fit is for a horizontally pure basis of McMillan functions~\cite{McMillan:1975yw}.
Figure~\ref{fig:Exprojection} shows the horizontal electric field in the midplane, with the evident nonlinear behavior at the extremes.  The values from the \OPERA map and from the multipole expansion are superimposed and are evidently in excellent agreement.

The fitted sextupolelike coefficient can be compared to the estimate discussed previously.
A  solution satisfying the Laplacian in the curved system was given in Eq.~\ref{eq:laplaciansolution}.
The ratio of the coefficients of
the quadratic and linear terms is $r_{hyp}=-\frac{1}{2R_0} = -0.0703{\rm m}^{-1}$, but that hypothetical solution is not unique.
Although it satisfies Maxwell's equations,  boundary conditions were not yet imposed.
The ratio based on the fit to the \OPERA field map, that satisfies both Maxwell and the boundary conditions,
is $r_{fit} = \frac{b_2}{b_1 r_0}= -2.71281\times 10^{3}/(1.01609\times 10^6\times 0.045{\rm m}) = -0.0593 {\rm m}^{-1}$, within 16\% of our guess.

As was shown above, the effect of the sextupolelike component of the ESQ field, and the asymmetry of the path length about the magic radius, is that the E-field correction
depends on betatron amplitude and nonlinearly on equilibrium radial offset.
The sextupolelike component and the path-length component contribute with opposite sign,
and the amplitude of the sextupolelike component is about 1/2 of the path-length piece.
Based on the measured equilibrium radial distribution, the
contribution of the sextupolelike component to the $E$-field correction is less than 1\%.
The potential effect from higher-order multipoles was explored in simulation.
 \section{$E$-field and Pitch Corrections Explored Numerically}
\label{ap:EpitchNumeric}
In simulation, $\omega_a$ can be determined directly by spin tracking. The trajectory of a muon is established by numerically integrating
the equations of motion, and its spin by integration of the Thomas-BMT equation along that trajectory.
In this section, the contribution to $\omega_a$ due to the electric field and the pitch correction (vertical betatron motion) is explored numerically, in particular, to study
the nonlinearity that arises from the ESQ geometry, mechanical alignment precision, and high-voltage uncertainty.
Spin tracking naturally incorporates the $E$-field and pitch effects; they cannot be trivially separated.

Full spin tracking verifies to high precision the conclusions for the $E$-field and pitch corrections as described in Eqs.~\ref{eq:Ce-equation} and \ref{eq:pitch_cor}, respectively. It is, however,
computationally intensive. Furthermore, for a trajectory that is both off momentum and oscillating vertically, spin tracking cannot distinguish the contributions from the electric field
and pitch.
In order to explore the full landscape of possible systematic uncertainties and to understand the contributions from pitch and $E$-field independently, an efficient and equivalently precise method was developed.

The anomalous precession frequency is commonly approximated as
\begin{equation}
\begin{aligned}
\vecwa = -\frac{q}{m}\Bigg[ a_{\mu} \vec{B} - \Big(a_{\mu} & - \frac{1}{\gamma^{2}-1} \Big) \frac{\vec{\beta} \times \vec{E}}{c} \\ 
                                                                 &   \left. -\, a_{\mu} \left(\frac{\gamma}{\gamma+1}\right)(\vec{\beta} \cdot \vec{B}) \vec{\beta} \right] ,
\label{eq:omega}
\end{aligned}
\end{equation}
where the experiment actually measures the scalar quantity \wam and not its vector approximation.  However, Eq.~\ref{eq:omega} can be used to produce spin dynamics simulations with
short run times that allow for the study of a large number of configurations.  First, consider the $\vec \beta  \times \vec E$ electric field term.
Define $\Phi_a = \int_0^T\omega_a dt$, there $T$ is the total integrated time that the muon is precessing.
In the absence of $\vec E$, $\Phi_a$ can be computed by integrating $\vec B_\perp$ along the muon trajectory.
The contribution from the electric field to $\phi_a$ is given by
\begin{equation}
\begin{split}
\frac{\Delta \Phi_E}{\Phi_a} &= \left(a_\mu-\frac{m^2 c^2}{p^2}\right)\frac{1}{c}\int_0^T\vec\beta\times\vec E dt \\
                             &\approx -2\frac{\Delta p}{p}\frac{1}{c}\int_0^T\vec\beta\times\vec E dt\label{eq:inte} ,
\end{split}
\end{equation}
where $\Delta p = p-\pmagic$.
In the tracking code, it is straightforward to compute the sums in Eq.~\ref{eq:inte}.
The electric field correction for the trajectory, assuming that $\vec B$ is parallel to $\vec\beta\times \vec E$, is
\begin{equation}
C_e(T) = -2\frac{\Delta p}{p}\frac{1}{T}\int^T\frac{\vec\beta\times\vec E}{Bc} dt.
\label{eq:efield_int}
\end{equation}

In a similar manner, the expressions for the pitch correction $C_p$ developed in Sec.~\ref{sec:pitch} can be rearranged to read
\begin{equation}
C_p(T)=\frac{\Delta \Phi_p}{\Phi_a} = \frac{1}{T} \int^T(1-|\bmbetahat\times \hat \bfB|) dt.
\label{eq:pitch_int}
\end{equation}

\begin{figure}[h]
\centering
\hspace*{5pt}
\includegraphics*[width=\columnwidth]{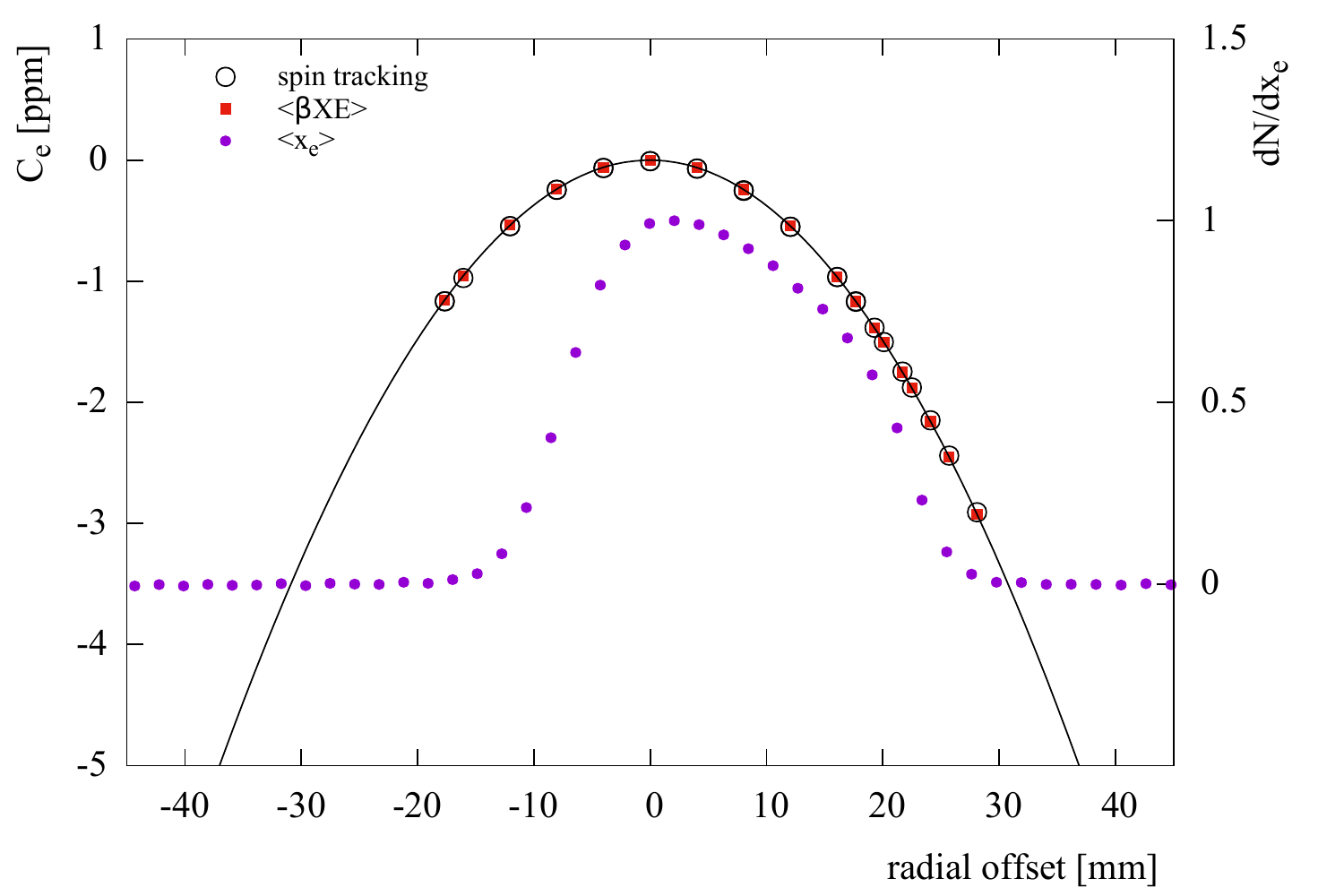}
\hspace*{-5pt}
\includegraphics*[width=\columnwidth]{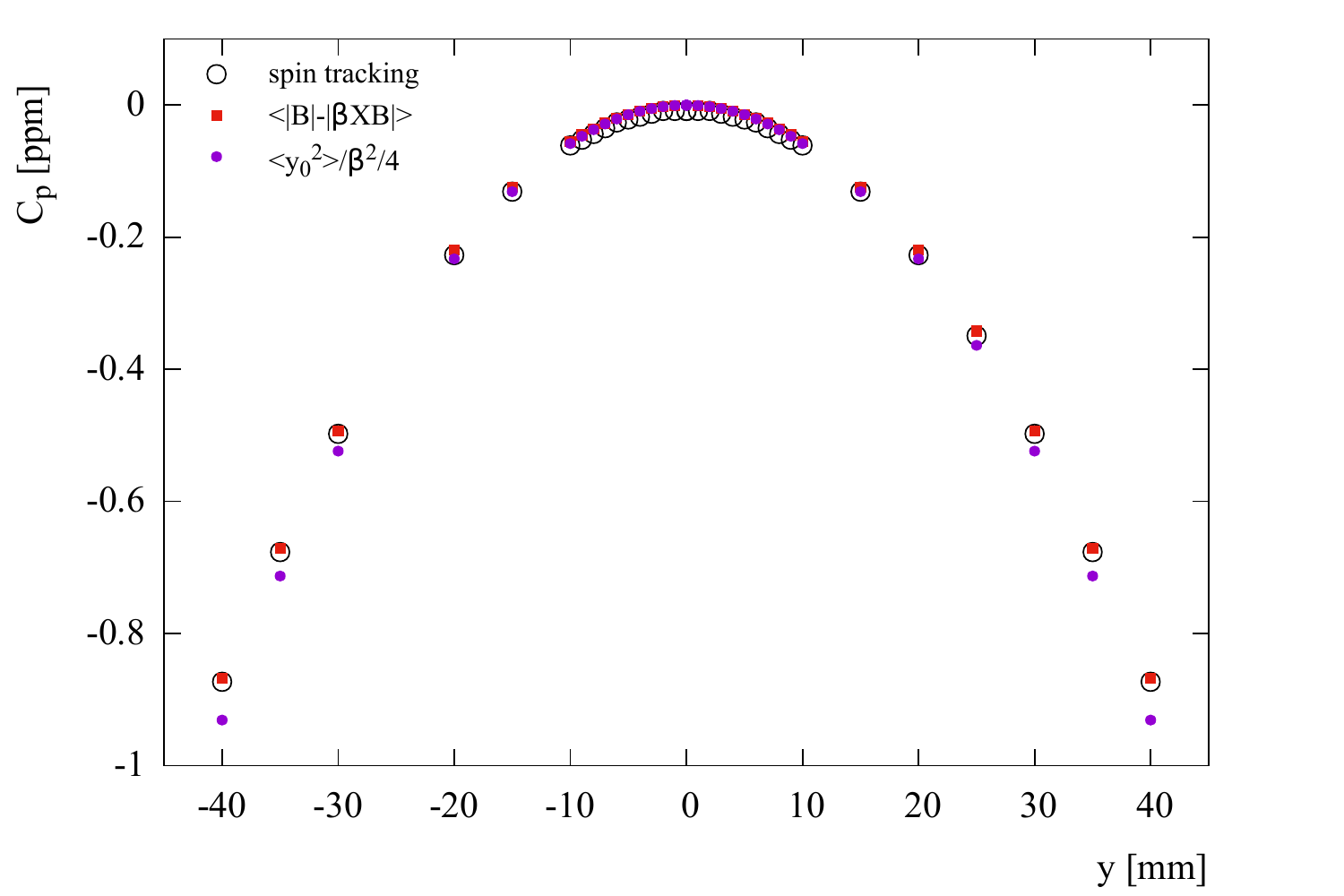}
\caption{Top: contribution to $\omega_a$ due to the electric field as computed by spin tracking (integration of the Thomas-BMT equation) and integration along trajectories (Eq.~\ref{eq:efield_int}) for closed orbits at varying radial offsets.
The vertical closed orbit is zero and there is no vertical betatron motion. A representative, measured radial distribution is superimposed for reference. Bottom: contribution to $\omega_a$ due to the vertical oscillation computed with spin tracking and integration along
the trajectory (Eq.~\ref{eq:pitch_int}) for muons with magic momentum and zero radial betatron amplitude. Note that the formula that assumes linearity overestimates the contribution at large amplitudes.}
 \label{fig:spintrackintlin}
 \end{figure}

To test the equivalence of the shift in precession frequency based on Eqs.~\ref{eq:efield_int} and~\ref{eq:pitch_int}, both were computed
simultaneously along a single trajectory. The muon is propagated for 1000 $2\pi/\omega_a$ in order to determine the spin tune by integrating the Thomas-BMT equation with sufficient accuracy, and $\vec\beta\times\vec E$ is summed along that same trajectory.
The trajectories chosen to test Eq.~\ref{eq:efield_int} have momentum offsets but no vertical motion.
To test Eq.~\ref{eq:pitch_int}, the tracked muon has magic momentum and zero radial betatron oscillation.
The agreement is excellent, as shown in Fig.~\ref{fig:spintrackintlin}, which shows $C_e$ and $C_p$ for both spin tracking and the integration method.

In the tests of systematics that are described next, the model of the quadrupole field is based on the sum of \OPERA-based field maps for each flat ESQ plate and, therefore, includes the nonlinearity associated with the geometry, as is evident in
Fig.~\ref{fig:Exprojection}.
At large amplitudes, the formulas that assume linearity of the electric field slightly overestimate the contributions of the electric field and pitch
to $\omega_a$.

The as-built ESQ plate alignment is known with a precision better than $\pm 1$\,mm, and the high-voltage set points to $\pm 0.01\%$.
To explore the uncertainty from unknown misalignments or voltage errors, 1024 configurations were analyzed, varying the conditions within the
specifications. In each case, the $E$-field and pitch corrections were computed and compared to the reference values.
From these studies, a systematic uncertainty of $\sigma(C_p) = 1.3$\,ppb is assigned for the contribution from pitching motion due to misalignment and voltage errors.
The distribution of perturbations to the electric field correction
is not Gaussianm and the standard deviation may underestimate the uncertainty. Indeed, the error distribution has a strong dependence on the
nature of the simulated misalignment and field-error configuration. The systematic uncertainty in the measurement of the contribution from the electric field due to field errors and misalignment is chosen conservatively to be the full width of distribution simulated values, yielding $\sigma(C_e) = 8.7$\,ppb.
 \section{Maps for the Phase-Acceptance Correction}
\label{ap:phase-acceptance-maps}

Estimation of the phase-acceptance correction relies on high-precision maps of detection acceptance, asymmetry, and phase as a function of muon decay coordinate $(x,y,\phi)$. In this section, we describe the procedure by which these maps are produced using our \ringsim\ \GEANT-based simulation in which
all detector systems (calorimeters and trackers) and vacuum chamber hardware of the experiment are fully modeled.
The simulation first determines the random time of a muon decay drawn from an exponential distribution. The muon four-momentum, position, and polarization direction are calculated based on the time spent in the storage ring. Finally, the muon is placed at the precalculated position and allowed to decay immediately. The orientation of the muon spin at the time of decay is determined from propagating the precession at the \wa frequency, which is known to sufficient precision from the E821 experiment.

\begin{figure}[tb]
\centering
\includegraphics[width=\columnwidth]{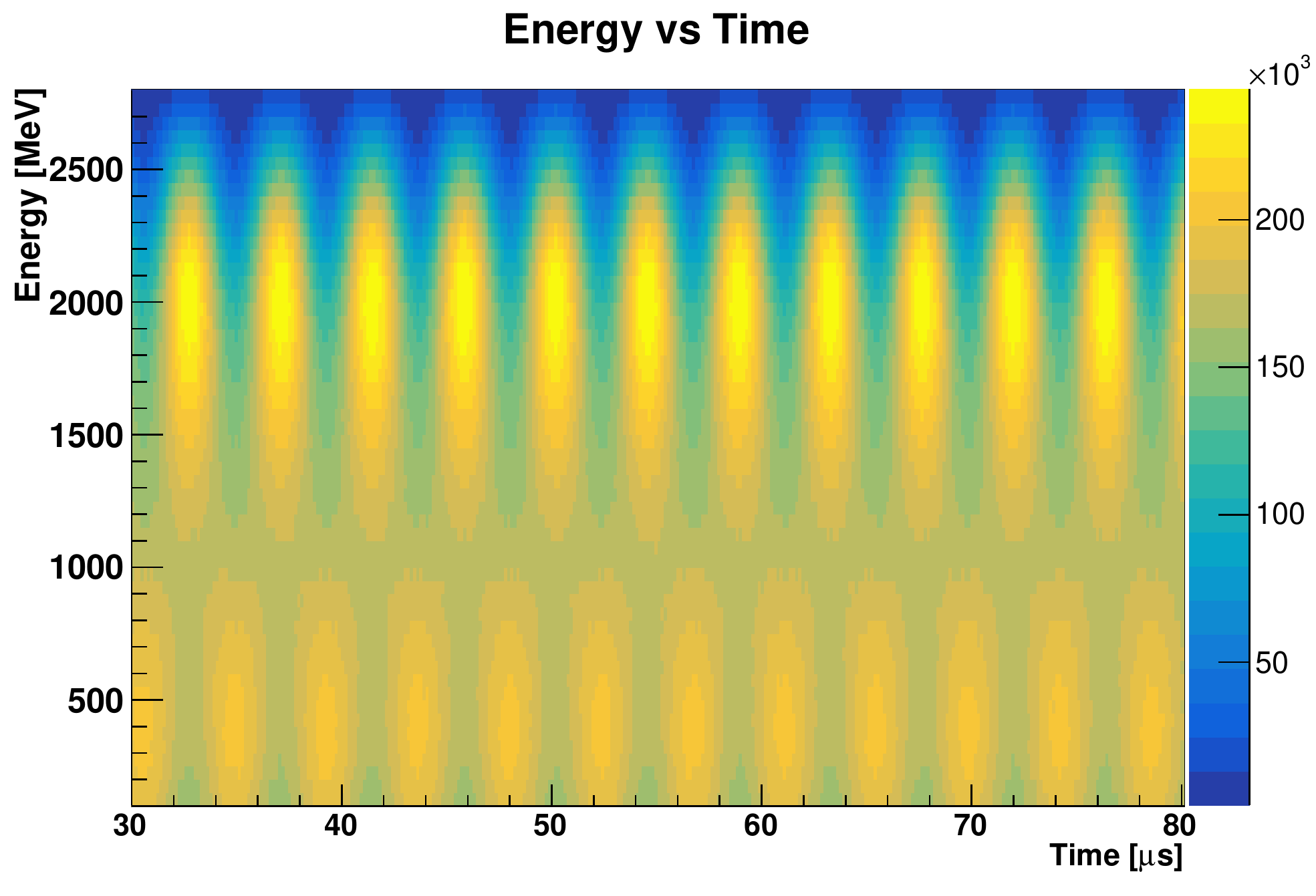}
\caption{\ringsim simulation of detected positron energy as a function of time, with muon decay disabled.}\label{fig:2DWiggle}
\end{figure}

The simulated time and energy distributions of the decay positron match well with the experimental measurement as shown in Fig.~\ref{fig:2DWiggle}.  This 2D intensity map---with the decaying exponential removed---shows positron energy versus time in fill from 30 to 80\,\SI{}{\micro\second}. The clear modulations are at the frequency \wa.  The asymmetry reverses around 1\,GeV, as expected from the kinematic boost of the Michel spectrum.
The simulation does not include CBO modulations because we are only interested in the average detection acceptance effect; the beam dynamic effects that are oscillating at a different frequency are strongly suppressed. Roughly 9 billion muon-decay events are simulated and analyzed to produce maps for three \wa analysis methods with different positron weightings (see Ref.~\cite{\precession}). In practice, the asymmetry-weighted method provides the most precise determination of \wa, and, therefore, we will show its maps.

The detection acceptance, asymmetry, and phase maps are produced in the form of 3D histograms $H^{c}(x,y,E)$, where $x$ and $y$ are the radial and vertical coordinates, respectively, in the storage ring volume, $E$ is the decay positron energy, and $c$ is a calorimeter index from 1 to 24.  For simplicity, we can express these maps as $H^{c}_{ijk}$, where $i$ and $j$ are $x$ and $y$ coordinates in 5\,mm bins, and $k$ is a 50\,MeV $E$ bin.

\subsection{Acceptance maps}

\begin{figure} [h]
\centering
\includegraphics[width=\columnwidth]{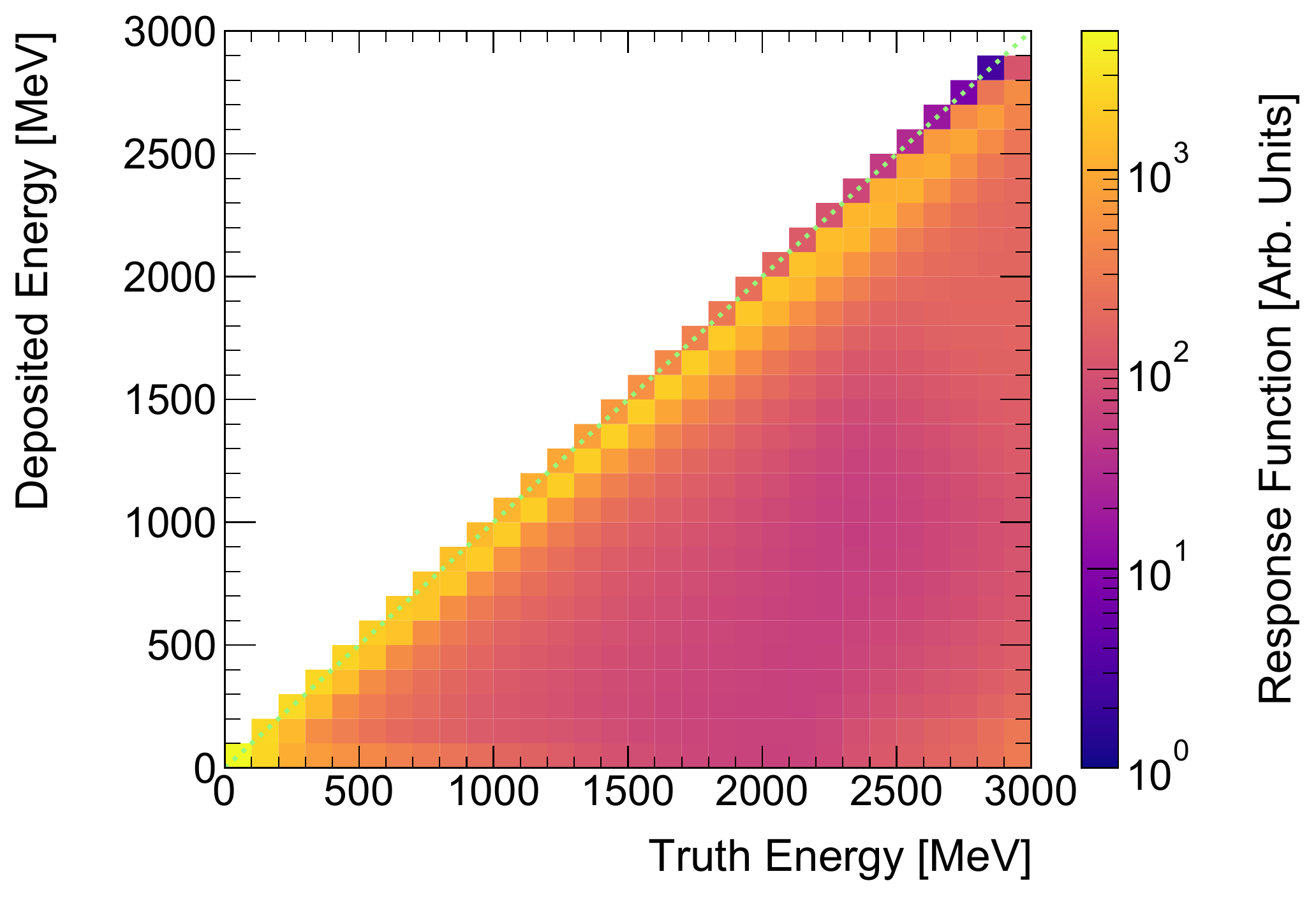}
\caption{The average calorimeter response function showing the deposited energy versus the truth energy, integrated over the decay volume.}\label{fig:CaloResponseFunction}
\end{figure}

The average acceptance $\bar{\varepsilon}_{ij}$ for a muon decaying in transverse position bin $ij$ is given by
\begin{equation}
\bar{\varepsilon}_{ij} = \sum_{k}f_{k}\varepsilon_{ijk} \sum_{k'>k_{th}'} R_{ijkk'} w_{k'} ,
\end{equation}
where the first sum is over positron truth energy bins $k$, the second sum is over positron deposited energy bins $k'$ above a threshold $k_{th}'$, $f_{k}$ is the fraction of events with energy $k$, $\varepsilon_{ijk}$ is the geometric acceptance for muon decays in bin $ij$ and positron energy $k$, $R_{ijkk'}$ is the calorimeter response function, and $w_{k'}$ is the analysis weighting factor~\cite{\precession}. $R_{ijkk'}$ represents the probability that a positron in position bin $ij$ and truth energy bin $k$ ends up in the $k'$ detected energy bin. This response function is illustrated in Fig.~\ref{fig:CaloResponseFunction}.

\begin{figure*}[tbh]
\centering
\includegraphics[width=\textwidth]{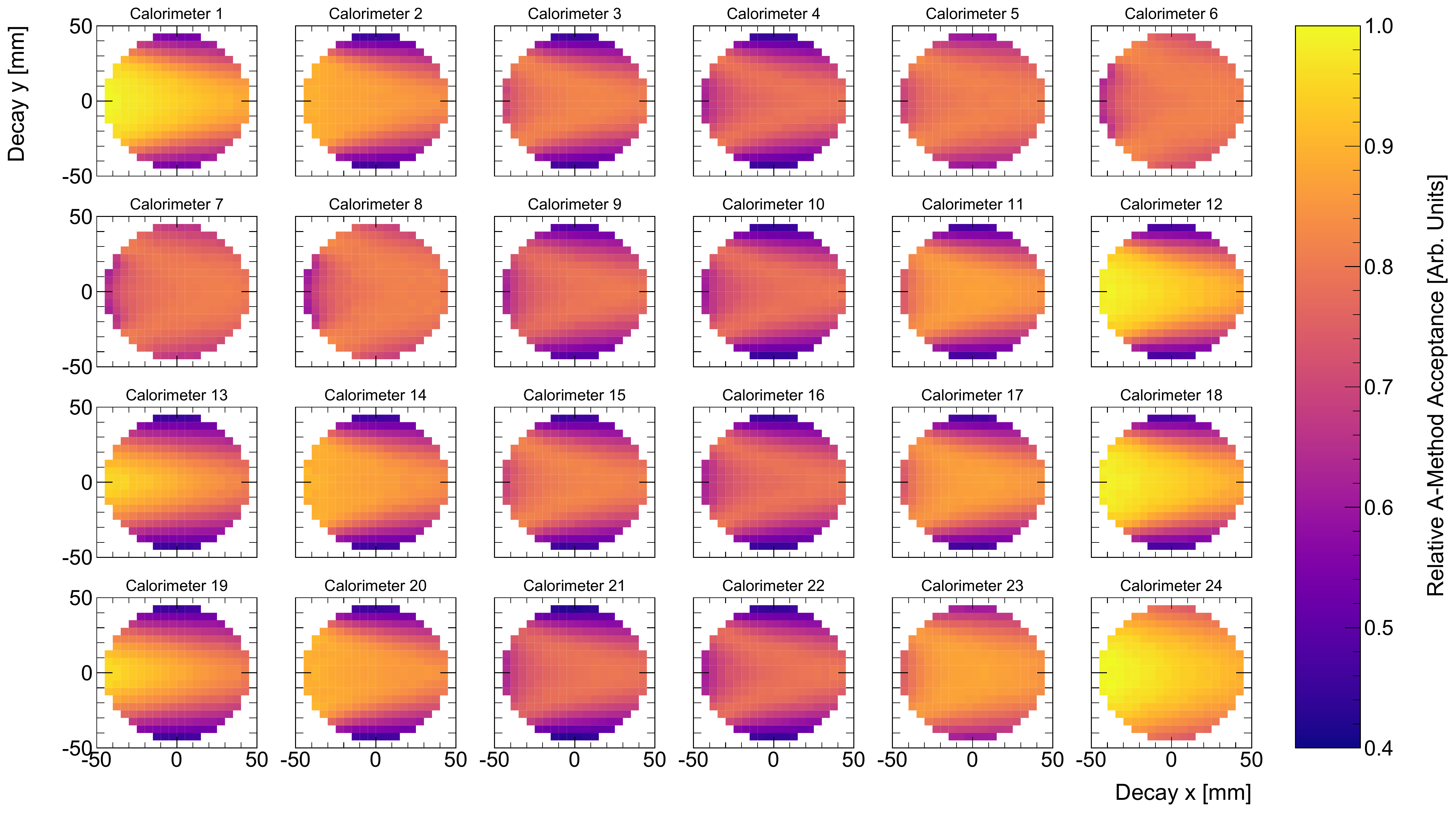}
\caption{The asymmetry-weighted relative acceptance maps for all calorimeters.  The differences in acceptance derive from the materials just upstream of each station; see Fig.~\ref{fig:VerticalProjGrid}.}\label{fig:acceptance-maps}
\end{figure*}

Figure \ref{fig:acceptance-maps} shows the relative acceptance function maps for each calorimeter, weighted for the asymmetry-method.  The variations among calorimeters are attributed to the material upstream of each, as is shown in Fig.~\ref{fig:VerticalProjGrid}.  We can loosely identify five categories:  i) behind ESQ plates;  ii) behind kicker plates; iii) downstream of the trackers; iv) next to the larger inflector vacuum chamber; and v) nothing in front.

\subsection{Phase and Asymmetry maps}

To produce the asymmetry and phase maps, two types of time spectra are generated: first, a time spectrum for all decay positrons (no matter whether the $e^+$ is detected or not) and, second, a decay $x-y$ grid of time spectra for each detected positron energy bin. Each time spectrum in the $ijk$ bin is fitted with the five-parameter function
$N(t)=N_{0}e^{-t/\tau}(1+A\cos{(\omega_{a}t+\varphi)}),$
where $\varphi$ is the fitted phase. The fitted phase for all decay positrons is denoted as $\varphi_{0}$, and the fitted phase for each decay $xy$ bin and energy bin is denoted as $\varphi_{ijk}$. The value of each $ijk$ bin in the phase map is then given by $\varphi_{ijk}-\varphi_{0}$
Similarly, the value of each $ijk$ bin in the asymmetry map is given by $A_{ijk}$.

\begin{figure*}[tbh]
\centering
\includegraphics[width=\textwidth]{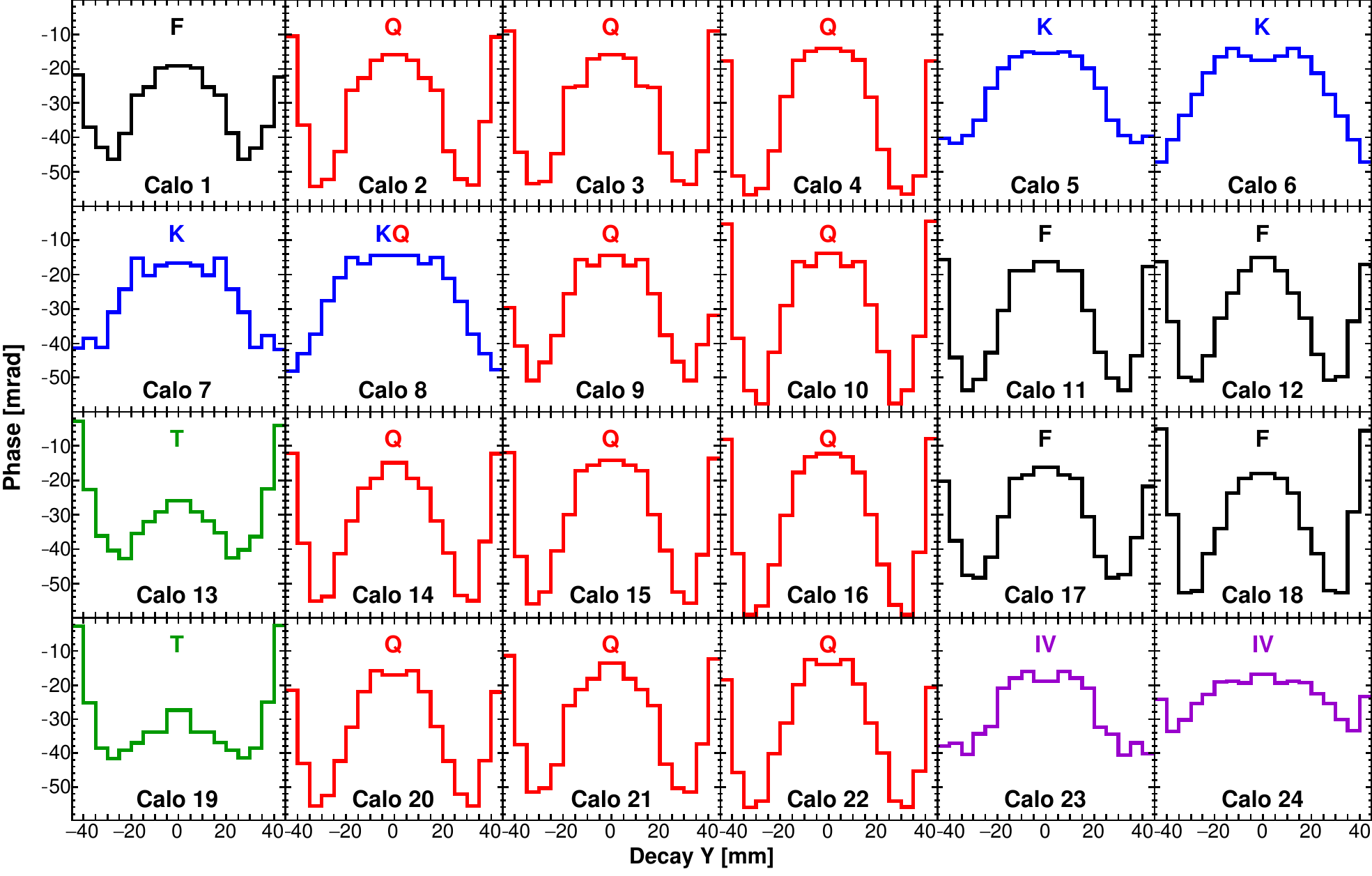}
\caption{The asymmetry-weighted vertical projections of the phase maps by calorimeter.  The calorimeters are ordered sequentially in azimuth with respect to the injection location.  The differences between stations is caused by material differences that impact the transmission of positron decays enroute to  detectors.  They also impact the acceptance (see Fig.~\ref{fig:acceptance-maps}) and the asymmetry maps. These indices represent:  Q, behind ESQ plates; K, behind kicker plates; T, behind tracker stations; IV, in the constrained inflector vacuum chamber; and, F, free space in front.}\label{fig:VerticalProjGrid}
\end{figure*}

Figure \ref{fig:VerticalProjGrid} shows beam-weighted vertical projections of the extracted phase maps.  Again, the variation from detector to detector shows a strong dependence on the material in front of the calorimeter.  The vertical width changes of the beam couple with the variation in these distributions such that some calorimeters, such as those behind ESQ plates, experience a much larger phase-acceptance effect than those behind the tracker stations.

 \section{Precession Fit Function}
\label{ap:fitfunction}
The complete fit function for the positron arrival time distribution described in Ref.~\cite{\precession} is

\begin{eqnarray}
F(t) = N_{0} \cdot N_{x} (t) \cdot N_{y} (t) \cdot \Lambda(t) \cdot
e^{-t/\gamma\tau_{\mu}} \cdot \nonumber \\
\left[1 + A_{0} \cdot A_{x}(t) \cdot \cos{(\wam t + \varphi_{0} \cdot \varphi_{x}(t))} \right].
\end{eqnarray}
Here, $\Lambda(t)$ is the muon loss function described in Sec.~\ref{sec:muonlossrate}.
The terms $N_x$, $N_y$, $A_x$ and $\varphi_x$ describe the interplay between acceptance
and beam dynamics that affect the overall calorimeter rate, the average asymmetry,
and the average phase.  They are defined as

\begin{alignat}{7}
N_x       (t) ={} 1 &{}+{} e^{-1 t/\tau_\text{CBO}}  A_{N,x,1,1}        \cos (1 \omega_\text{CBO} t + \varphi_{N,x,1,1}       )   \nonumber \\
                    &{}+{} e^{-2 t/\tau_\text{CBO}}  A_{N,x,2,2}        \cos (2 \omega_\text{CBO} t + \varphi_{N,x,2,2}       ) , \label{eq:nxt} \\
N_y       (t) ={} 1 &{}+{} e^{-1 t/\tau_y}           A_{N,y,1,1}        \cos (1 \omega_y          t + \varphi_{N,y,1,1}       )   \nonumber \\
                    &{}+{} e^{-2 t/\tau_y}           A_{N,y,2,2}        \cos (1 \omega_\text{VW}  t + \varphi_{N,y,2,2}       ) , \label{eq:nyt} \\
A_x       (t) ={} 1 &{}+{} e^{-1 t/\tau_\text{CBO}}  A_{A,x,1,1}        \cos (1 \omega_\text{CBO} t + \varphi_{A,x,1,1}       ) , \label{eq:axt} \\
\varphi_x (t) ={} 1 &{}+{} e^{-1 t/\tau_\text{CBO}}  A_{\varphi,x,1,1}  \cos (1 \omega_\text{CBO} t + \varphi_{\varphi,x,1,1} ) .
\end{alignat}

The parameters of the form $A_{N,x,i,j}$ and $\varphi_{N,x,i,j}$
correspond to the effect of the $i$th moment of the radial ($x$) beam distribution at the $j$th multiple of the relevant
fundamental frequency (in this case, $\omega_{\rm CBO}$)  on the rate normalization $N$.  Analogous parameters
model the modulation of the average asymmetry $A$ and phase $\varphi$, as well as the effect of moments of the vertical ($y$) beam distribution.
 
\providecommand{\noopsort}[1]{}\providecommand{\singleletter}[1]{#1}

\end{document}